\documentclass[12pt,preprint]{aastex}
\usepackage{amsmath}
\slugcomment{Submitted to {\it The Astrophysical Journal Supplements},\\   Accepted on 2013 January 15\\}
\shortauthors{G.N. Mace et al.}
\shorttitle{A Study of the Diverse T Dwarf Population Revealed by WISE}

\begin{document}

\title{A STUDY OF THE DIVERSE T DWARF POPULATION REVEALED BY WISE\\}

\author{Gregory N.\ Mace\altaffilmark{a,b},
J.\ Davy Kirkpatrick\altaffilmark{b},
Michael C.\ Cushing\altaffilmark{c},
Christopher R.\ Gelino\altaffilmark{b},
Roger L.\ Griffith\altaffilmark{b},
Michael F.\ Skrutskie\altaffilmark{d},
Kenneth A.\ Marsh\altaffilmark{e},
Edward L.\ Wright\altaffilmark{a},
Peter R.\ Eisenhardt\altaffilmark{f},
Ian S.\ McLean\altaffilmark{a},
Maggie A.\ Thompson\altaffilmark{g},
Katholeen Mix\altaffilmark{b},
Vanessa Bailey\altaffilmark{h},
Charles A.\ Beichman\altaffilmark{b},
Joshua S.\ Bloom\altaffilmark{i},
Adam J.\ Burgasser\altaffilmark{j,k},
Jonathan J.\ Fortney\altaffilmark{l},
Philip M.\ Hinz\altaffilmark{h},
Russell P.\ Knox\altaffilmark{h},
Patrick J.\ Lowrance\altaffilmark{b},
Mark S.\ Marley\altaffilmark{m},
Caroline V. Morley\altaffilmark{l},
Timothy J.\ Rodigas\altaffilmark{h},
Didier Saumon\altaffilmark{n},
Scott S.\ Sheppard\altaffilmark{o}, and
Nathan D.\ Stock\altaffilmark{h}\\}

\altaffiltext{a}{Department of Physics and Astronomy, UCLA, 430 Portola Plaza, Box 951547, Los Angeles, CA 90095-1547, USA;  gmace@astro.ucla.edu}
\altaffiltext{b}{Infrared Processing and Analysis Center, MS 100-22, California Institute of Technology, Pasadena, CA 91125, USA}
\altaffiltext{c}{Department of Physics and Astronomy, MS 111, University of Toledo, 2801 W. Bancroft St., Toledo, OH 43606-3328, USA}
\altaffiltext{d}{Department of Astronomy, University of Virginia, Charlottesville, VA 22904, USA}
\altaffiltext{e}{School of Physics and Astronomy, Cardiff University, Cardiff CF24 3AA, UK}
\altaffiltext{f}{NASA Jet Propulsion Laboratory, 4800 Oak Grove Drive, Pasadena, CA 91109, USA}
\altaffiltext{g}{Department of Astrophysical Sciences, Princeton University, Peyton Hall, 4 Ivy Lane, Princeton, NJ 08544-1001, USA\\}
\altaffiltext{h}{Steward Observatory, The University of Arizona, 933 N. Cherry Ave., Tucson, AZ 85721, USA}
\altaffiltext{i}{Department of Astronomy, University of California, Berkeley, CA 94720, USA}
\altaffiltext{j}{Department of Physics, University of California, San Diego, CA 92093, USA}
\altaffiltext{k}{Department of Physics, Massachusetts Institute of Technology, 77 Massachusetts Ave, Cambridge, MA 02139, USA}
\altaffiltext{l}{Department of Astronomy and Astrophysics, University of California, Santa Cruz, CA 95064, USA}
\altaffiltext{m}{NASA Ames Research Center, Moffett Field, CA 94035, USA}
\altaffiltext{n}{Los Alamos National Laboratory, Los Alamos, NM 87545, USA}
\altaffiltext{o}{Department of Terrestrial Magnetism, Carnegie Institution of Washington, 5241 Broad Branch Rd. NW, Washington, DC 20015, USA\\}

\begin{abstract}
We report the discovery of 87 new T dwarfs uncovered with the Wide-field Infrared Survey Explorer (WISE) and three brown dwarfs with extremely red near-infrared colors that exhibit characteristics of both L and T dwarfs.
Two of the new T dwarfs are likely binaries with L7$\pm$1 primaries and mid-type T secondaries. 
In addition, our follow-up program has confirmed 10 previously identified T dwarfs and four photometrically-selected
L and T dwarf candidates in the literature. This sample, along with the previous WISE discoveries, 
triples the number of known brown dwarfs with spectral types later than T5. 
Using the WISE All-Sky Source Catalog we present updated color-color and color-type diagrams for all the WISE-discovered T and Y dwarfs. 
Near-infrared spectra of the new discoveries are presented, along with spectral classifications.  
To accommodate later T dwarfs we have modified the integrated flux method of determining spectral indices to instead use the median flux.
Furthermore, a newly defined J-narrow index differentiates the early-type Y dwarfs from late-type T dwarfs based on the $J$-band continuum slope.
The K/J indices for this expanded sample show that 32\% of late-type T dwarfs have suppressed $K$-band flux and are blue relative to the spectral standards, while only 11\% are redder than the standards. 
Comparison of the Y/J and K/J index to models suggests diverse atmospheric conditions and supports the possible re-emergence of clouds after the L/T transition. 
We also discuss peculiar brown dwarfs and candidates that were found not to be substellar, including two young stellar objects and two active galactic nuclei. 
The substantial increase in the number of known late-type T dwarfs provides a population that will be used to test models of cold atmospheres and star formation. 
The coolest WISE-discovered brown dwarfs are the closest of their type and will remain the only sample of their kind for many years to come.
\end{abstract}


\section{Introduction} 
Brown dwarfs were initially theorized nearly 50 years ago as the lowest mass products of star formation \citep{kumar1963,hayashi1963}. 
Without the energy provided by hydrogen burning, brown dwarfs cool as they age and transition to later spectral types \citep[e.g.][Figure 2]{burgasser2004}. 
As a result, older field brown dwarfs can have higher masses than young brown dwarfs, even though they have the same effective 
temperature and spectral type \citep{baraffe2003, burrows1997}. Brown dwarfs are classified by their 
spectral morphology, which is primarily related to effective temperature, but is also modulated by metallicity, clouds, gravity, and binarity.
The cooling process results in a dearth 
of field L dwarfs and a surplus of T dwarfs relative to the initial mass function \citep{burgasser2004}. Since Gl 229B was discovered in 
1995 \citep{nakajima1995}, the T dwarf sample has increased in spurts thanks primarily to the Two Micron All Sky Survey 
\citep[2MASS;][]{skrutskie2006}, the Sloan Digital Sky Survey \citep[SDSS;][]{york2000}, the United Kingdom Infrared Telescope Infrared Deep Sky Survey \citep[UKIDSS;][]{lawrence2007},
and the Canada France Hawaii Telescope Legacy Survey \citep[CFHTLS;][]{delorme2008b}. 
T dwarfs are classified by increased methane, water and molecular hydrogen collision-induced absorption (CIA) 
in the near-infrared \citep{burgasser2002,burgasser2006,geballe2002}, and have inferred temperatures between $\sim$500 and $\sim$1300K \citep{cushing2008,cushing2011,stephens2009}. 

The cool brown dwarfs discovered by the Wide-field Infrared Survey Explorer \citep[WISE;][]{wright2010} will provide the most robust tests of cool atmospheric models. Recent analysis of the T9 spectral 
standard UGPS J072227.51$-$054031.2 \citep{lucas2010, bochanski2011,leggett2012}, and the Y dwarfs presented by \citet{cushing2011}, highlights the shortcomings of current models to match the low-flux 
near-infrared spectra of the coolest known objects. One promising avenue is the recent work of \citet{morley2012} to include sulfide clouds in modeling the coolest brown dwarf atmospheres.

Tight constraints on the field mass function can be imposed by a complete census of the low-mass component 
of the solar neighborhood, of which the local T-dwarf component is a vital part. Our knowledge of the functional form of the mass 
function has a compounding influence on models of galaxy, star and planet formation processes. 
Also, knowledge of the low-mass component of the solar neighborhood can be used to determine the minimum mass for star formation 
\citep{burgasser2004,allen2005}. To approach a complete volume-limited census, 
we have chosen brown dwarf candidates from the WISE All-Sky Source Catalog by imposing methane sensitive color constraints. Employing various near-infrared spectrographs, which provide 
diverse resolution and wavelength coverage, we have produced accurate classifications for these candidates through comparison to 
spectroscopic standards. Astrometric monitoring of the coolest objects provides trigonometric parallax 
measurements, as discussed by \citet{marsh2013} and \citet{beichman2013}. \citet{kirkpatrick2012} use this 
information to determine absolute magnitude relationships and constrain the mass function. 

In this paper we present the spectra of 100 T dwarfs, including 90 newly reported sources, which comprise approximately one-third of the present sample. 
Combined with the 89 T dwarfs reported by \citet{kirkpatrick2011}, WISE has more 
than tripled the number of brown dwarfs with spectral types later than T5. In \S 2 we discuss our photometric selection criteria and present WISE 
All-Sky photometry for all the objects in \citet{kirkpatrick2011} and the new brown dwarfs presented here. In \S 3 we present our follow-up 
photometry and spectroscopy for the new T dwarfs and a few objects in the literature. Spectral classification, indices, and further analysis
 are presented in \S 4, along with comments on noteworthy objects. In \S5 we present a dedicated discussion on the re-emergence of clouds in late-type T
 dwarfs. Interlopers, that mimic brown dwarfs in WISE color space, are discussed in \S6. Our results are summarized in \S7.
In the Appendix we discuss our motivation for redefining spectral indices as the median flux over a wavelength range rather than the integrated flux.

\section{Candidate Selection}

The WISE All-Sky Data Release \citep{cutri2012} is the culmination of a two-pass all-sky survey at 3.4, 4.6, 
12, and 22 $\mu$m, hereafter referred to as bands $W1$, $W2$, $W3$, and $W4$, respectively \citep{wright2010}. 
Designed as a variant of the methane imaging technique \citep{tinney2005}, the $W1-W2$ color compares the flux from the deep 3.3 $\mu$m CH$_4$ absorption 
band to the bright 4.6 $\mu$m flux in T and Y dwarfs \citep{mainzer2011}. Cooler T and Y dwarf atmospheres possess redder $W1-W2$ colors than M and L dwarfs. 
In $W3$ and $W4$, past the blackbody peak for the coolest T dwarf, the flux decreases and is further reduced by NH$_3$ absorption \citep{saumon2003,burrows2003}. 
Objects with $W2-W3$ $\ge$ 4.0 can be disregarded as brown dwarf candidates, and only two of thirteen candidates with $W2-W3$ $>$ 3.0 are T dwarfs (see \S 6).

Initial candidate selection, as described by \citet{kirkpatrick2011}, was made using the WISE Preliminary Data 
Release\footnote{See http://wise2.ipac.caltech.edu/docs/release/prelim/expsup/ .}.  Our search required $W1-W2$ colors (or limits) greater than 
1.5 mag, which corresponds roughly to types $\ge$T5, and $W2-W3$ $<$ 3.0 mag, which reduces the number of very red extragalactic contaminants and 
sources embedded in star formation regions. To identify brown dwarfs earlier than T5 we searched for objects with $W1-W2$ $>$ 0.4 mag and no 
association with a 2MASS source (implying that the $J-W2$ color is either very red or the object has moved). Extragalactic contaminants were culled 
from the bright source candidates using the following criteria: $W1-W2$ $>$ 0.96($W2-W3$) $-$ 0.96 (as shown in Figure 3 of \citet{kirkpatrick2011}). 
Spectroscopic follow-up of our initial candidates confirmed $\sim$100 new brown dwarfs. \citet{kirkpatrick2011} provide photometry from the WISE Preliminary Data Release for
these new objects and the objects in \citet{mainzer2011} and \citet{burgasser2011}.

Making use of the WISE All-Sky Data Release\footnote{See http://wise2.ipac.caltech.edu/docs/release/allsky/.}, specifically the database of 
extractions from the atlas images that is a union of the WISE All-Sky Source Catalog and the WISE All-Sky Reject Table, we modified our 
selection criteria to identify the reddest candidates. The details of our query method can be found in \citet{kirkpatrick2012}. Briefly, 
we required $W1-W2$ $>$ 2.0 mag, $W2-W3$ $<$ 3.5 mag, signal-to-noise ratio (SNR) $>$ 3 per frame at $W2$, that the source is not blended with another source in visual inspection, 
that the source is point-like, and that the absolute Galactic latitude be greater than three degrees if its Galactic longitude falls within twenty degrees of the Galactic Center. 
To aid in follow-up, finder charts of a 2$\arcmin$x2$\arcmin$ field around each candidate were constructed using DSS2 $BRI$ (epoch $\sim$1980s), SDSS $ugriz$ 
(where available; epoch $\sim$2000), 2MASS $JHK_s$ (epoch $\sim$2000), and WISE four-band data (epoch $\sim$2010). Prioritization was accomplished 
by visually inspecting each of these finder charts\footnote{In lieu of creating $\sim$90 additional figures for the finder charts of the new objects, we point the readers 
to the excellent DSS+SDSS+2MASS+WISE Finder Chart Service at IRSA (http://irsa.ipac.caltech.edu/).}. 
T dwarf candidates that were detected at visual wavelengths, and showed no clear proper motion, were 
excluded from further follow-up.

Comparison of the WISE photometry in the Preliminary Data Release and the All-Sky Data Release reveals measurements biases of $\sim$0.3, 0.05, and 0.4 
magnitudes in $W1$, $W2$ and $W3$, respectively. This is a known difference in the catalogs and is further discussed in Section 6.3 of the WISE Explanatory Supplement\footnote{See http://wise2.ipac.caltech.edu/docs/release/allsky/expsup/.}.
The Pass-2 source extraction, which produced the All-Sky Source Catalog, implemented a number of improvements listed in section 4.1 of the Explanatory Supplement. 
Because the WISE All-Sky Data Release is the premier data product from the WISE mission, we present this new photometry in Table~\ref{WISE_all_sky_phot_100} for the brown dwarfs in \citet{kirkpatrick2011} 
and in Table~\ref{WISE_all_sky_phot_Ts} present photometry for our newly confirmed brown dwarfs. 

\begin{figure}
\epsscale{0.9}
\figurenum{1}
\plotone{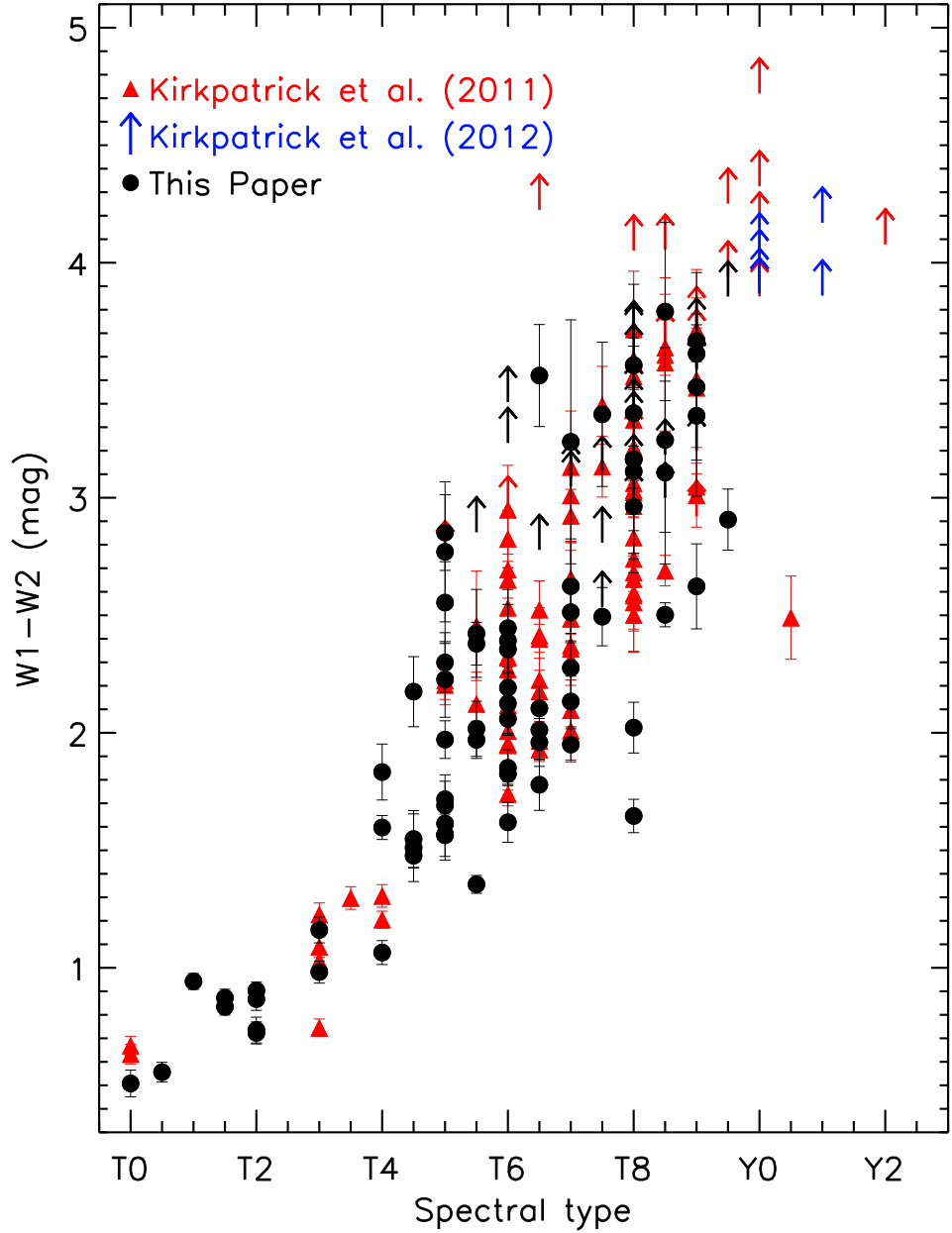}
\caption{WISE $W1-W2$ color versus spectral type. Objects from \citet{kirkpatrick2011} (including those announced by \citet{mainzer2011}, \citet{cushing2011}, and \citet{burgasser2011}) are marked with red triangles and limit arrows. The Y dwarfs from \citet{kirkpatrick2012} are marked with blue squares and limit arrows. T dwarfs announced in this paper are marked with black circles and limit arrows.
\label{W1W2_type}}
\end{figure}

\begin{figure}
\epsscale{0.9}
\figurenum{2}
\plotone{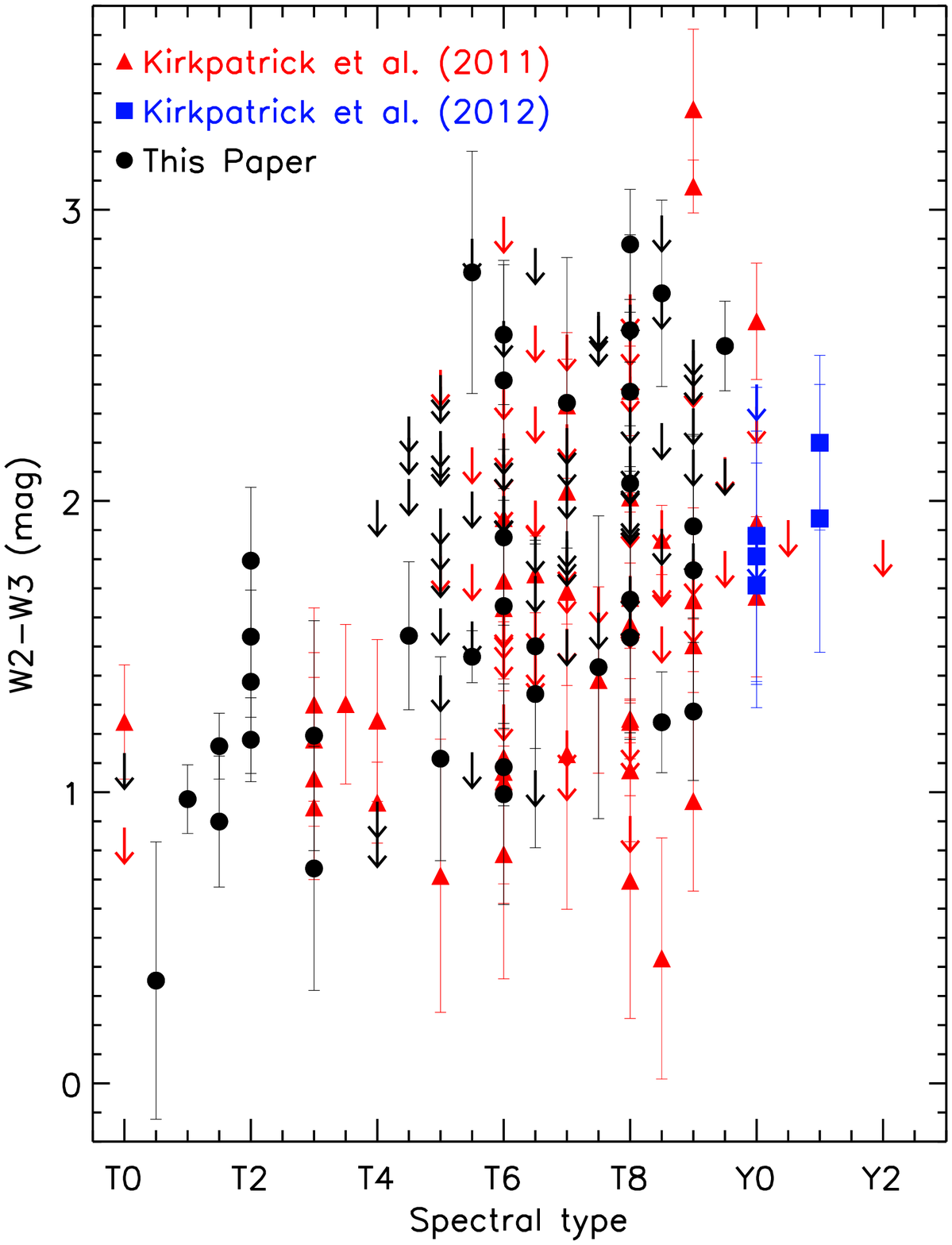}
\caption{WISE $W2-W3$ color versus spectral type. Same symbols as in Figure~\ref{W1W2_type}.
\label{W2W3_type}}
\end{figure}

Figures~\ref{W1W2_type} \& \ref{W2W3_type} present the updated $W1-W2$ and $W2-W3$ colors as a function of spectral type for objects in
Tables~\ref{WISE_all_sky_phot_100} \& \ref{WISE_all_sky_phot_Ts} and for Y dwarfs from \citet{kirkpatrick2012}. 
Similar plots were included in \citet{kirkpatrick2011} for M, L, T, and Y dwarfs, but here we focus on the T dwarf sequence.
The gradual onset of methane absorption at 3.3 $\mu$m \citep{noll2000} complicates the functional form of the $W1-W2$ color around the L-T transition.
Although the dispersion in the $W1-W2$ color for the late-type T dwarfs is large ($\sim$1.5 mag), early-type T dwarfs display a considerably smaller dispersion until T4 or T5. 
Most objects later than T9 only have $W2$ and $W3$ detections, resulting in $W1-W2$ lower limits. The gap between the limits and the verified detections of the late-type sources implies that the $W1$ limit is an overestimate and the resultant $W1-W2$ color limits are artificially high.  
Upper limits in $W2-W3$ are a result of non-detections in $W3$, which is the case for many of our new T dwarfs. 
The dispersion in the $W1-W2$ and $W2-W3$ colors is likely caused by clouds and/or non-equilibrium chemistry in T dwarf atmospheres \citep{saumon2003}, and are largest for spectral types later than T4.  

\section{Follow-up Photometry and Spectroscopy}

\begin{figure}
\epsscale{0.9}
\figurenum{3}
\plotone{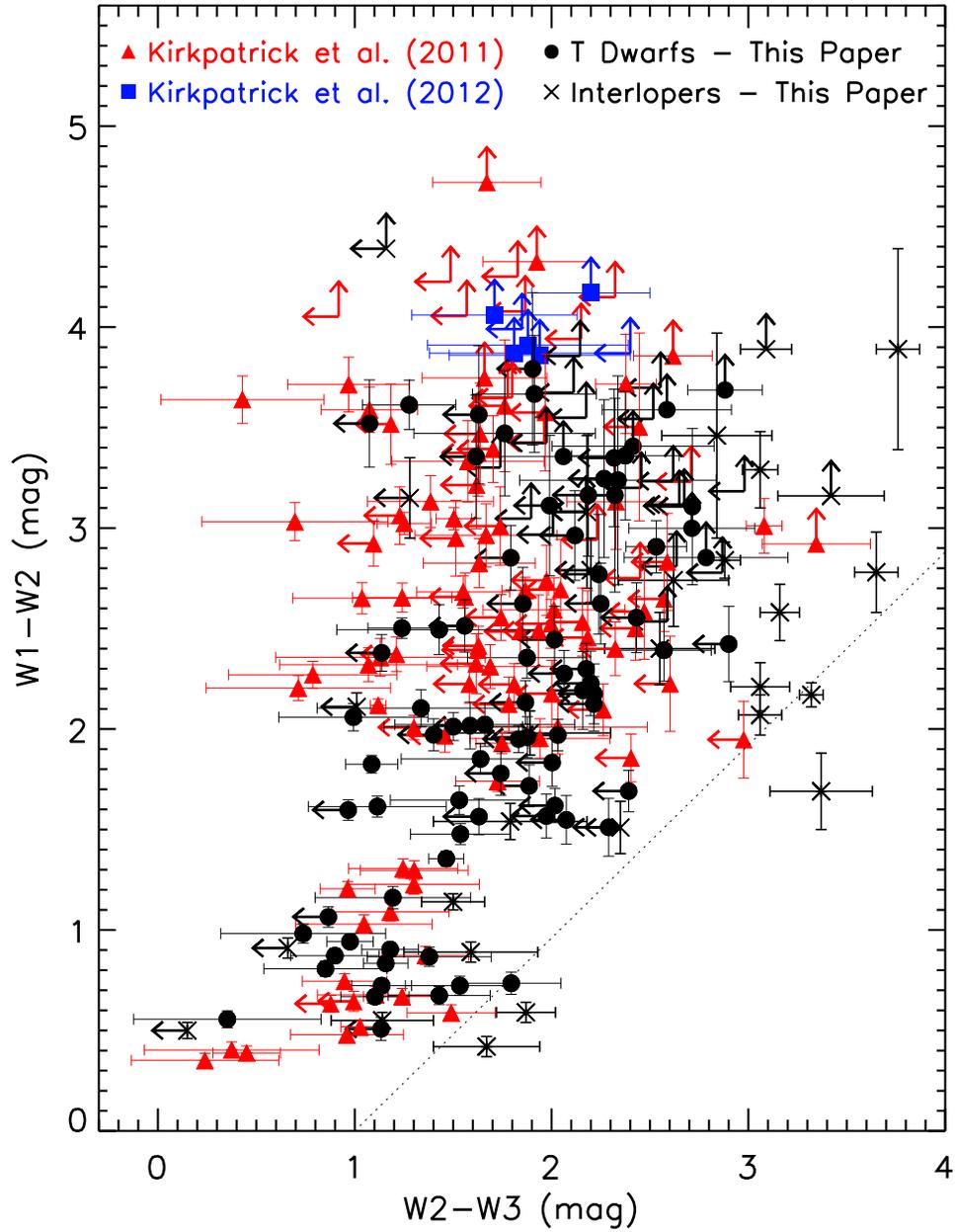}
\caption{WISE $W1-W2$ versus $W2-W3$ color-color diagram. Same symbols as in Figure~\ref{W1W2_type}. Interlopers are marked by black x's and limits. The dotted line identifies our $W1-W2$$>$0.96($W2-W3$)$-$0.96 selection criterion. 
\label{W1W2_W2W3}}
\end{figure}

T dwarf candidates selected by WISE colors alone are reliable when they fall within the core of the T dwarf $W1-W2$ versus $W2-W3$ color-color diagram in Figure~\ref{W1W2_W2W3}.  However,
near the periphery of this region the number of non-stellar interlopers increases.  Additional photometry of WISE candidates 
allows us to identify sources that have near-infrared colors characteristic of brown dwarfs, and provides near-IR magnitudes that guide facility selection for 
spectroscopic observation. 

\subsection{Follow-up Photometric Observations}

Most of our photometric observations are described in \citet{cushing2011} and \citet{kirkpatrick2011,kirkpatrick2012}.
We have separated the 2MASS and Mauna Kea Observatories Near-Infrared \citep[MKO;][]{simons2002,tokunaga2002} systems since care should be taken when making comparisons between the two 
filter systems whose different bandpasses produce slightly different magnitudes. Some of our photometry is provided by the 2MASS All-Sky Point Source Catalog and the 
Reject Table\footnote{See http://www.ipac.caltech.edu/2mass/releases/allsky/doc/explsup.html for details.}. Other follow-up ground-based photometry on the 2MASS system is from the 2MASS camera on the 1.5~m Kuiper Telescope at Steward Observatory 
and the Peters Automated Infrared Imaging Telescope \citep[PAIRITEL;][]{bloom2006} on the 1.3~m telescope at Fred Lawrence Whipple Observatory. 
Photometry on the MKO system is from the Wide-field Infrared Camera \citep[WIRC;][]{wilson2003} at the 5~m Hale Telescope at Palomar Observatory, the Persson's Auxilliary Nasmyth Infrared Camera \citep[PANIC;][]{martini2004} on the 6.5~m Magellan telescope at Las Campanas Observatory, 
and the NOAO Extremely Wide Field Infrared Imager \citep[NEWFIRM;][]{autry2003} on the 4~m Victor M. Blanco Telescope at Cerro Tololo Inter-American Observatory (CTIO). 
Ten of our candidates are available in UKIDSS (the UKIRT Infrared Deep Sky Survey) Data Release 8 \citep{hewett2006,lawrence2007,casali2007,hambly2008,hodgkin2009}. 
Photometric filters installed in the Ohio State Infrared Imager/Spectrometer \citep[OSIRIS;][]{depoy1993} use Barr filters which produce T dwarf photometry most similar to the MKO system \citep[][Section 2.2.5]{kirkpatrick2012}, 
so they are listed with the MKO measurements. All images were acquired between 2010 August and 2012 August, and source extractions from our observations use source apertures that are 1.5$\times$FWHM of the source point-spread function.

Figures 5, 7, and 8 of \citet{kirkpatrick2011} specify the $J-H$, $J-W2$, and $H-W2$ colors of M, L, T, and Y dwarfs. Objects lacking photometry in Table 3 of that paper, but having follow-up photometry now, are presented in Table~\ref{first100_more_followup_photometry} of this paper. 
Photometry of new T dwarf candidates is listed in 
Table~\ref{Ts_followup_photometry}.  For the figures in this paper we use magnitudes on the 2MASS system. MKO magnitudes are depicted in \citet{kirkpatrick2011,kirkpatrick2012}.

\begin{figure}
\epsscale{0.9}
\figurenum{4}
\plotone{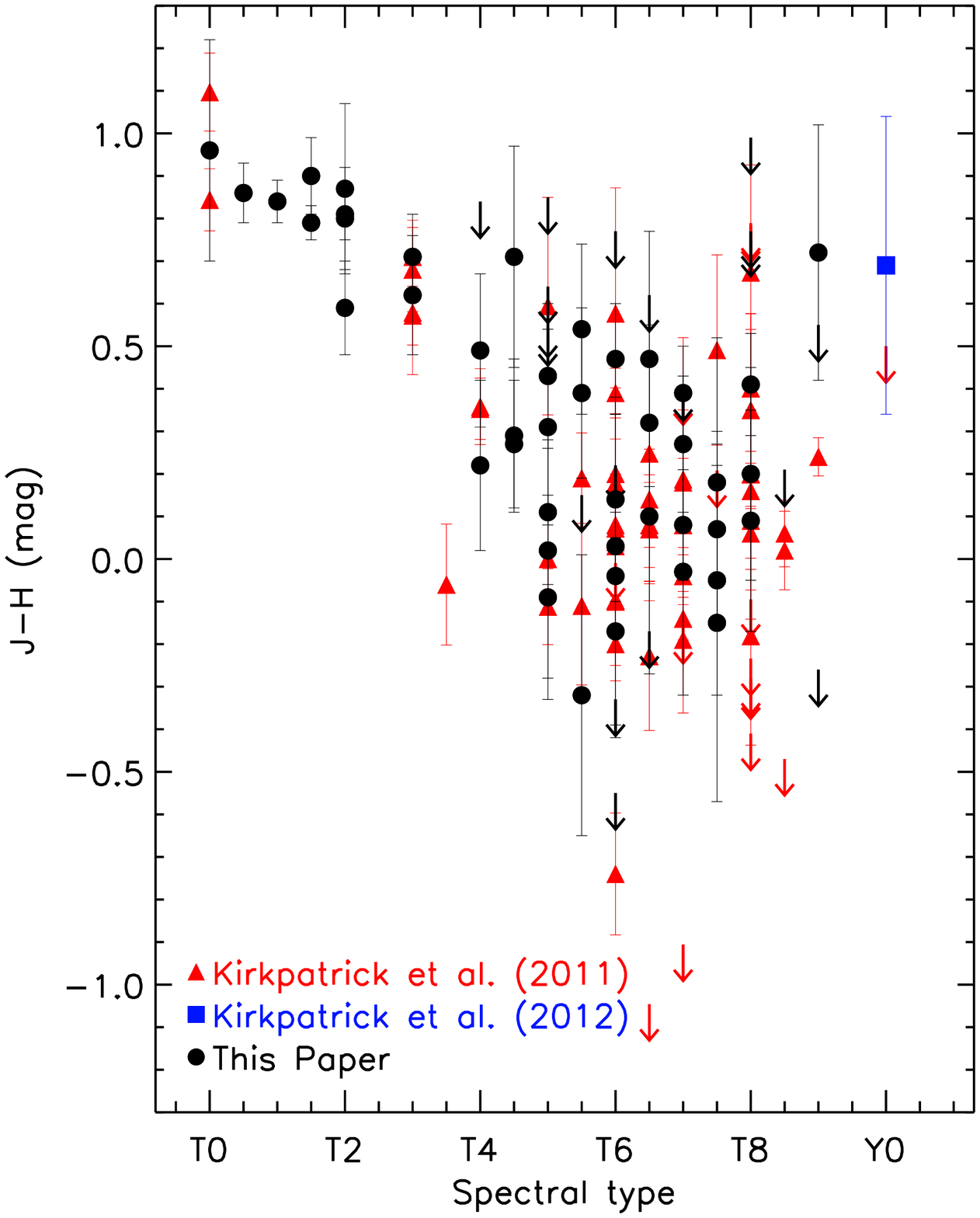}
\caption{$J-H$ color versus spectral type. Same symbols as in Figure~\ref{W1W2_type}. A turnaround in the $J-H$ color at T7$\pm$1 is seen here, but it is tentative based on the large dispersion and uncertainties in our photometry.
\label{JH_type}}
\end{figure}

The $J-H$ color-type diagram for WISE-discovered T and Y dwarfs is shown in Figure~\ref{JH_type}. The reversal in the $J-H$ color shown in Figure 7 of \citet{kirkpatrick2012} can also be seen in Figure~\ref{JH_type} around T7$\pm$1. 
However, the turnaround in the $J-H$ color is driven by only a few objects later than T8 and there are large uncertainties in our colors, making a clear identification of the reversal location difficult.

\begin{figure}
\epsscale{0.9}
\figurenum{5}
\plotone{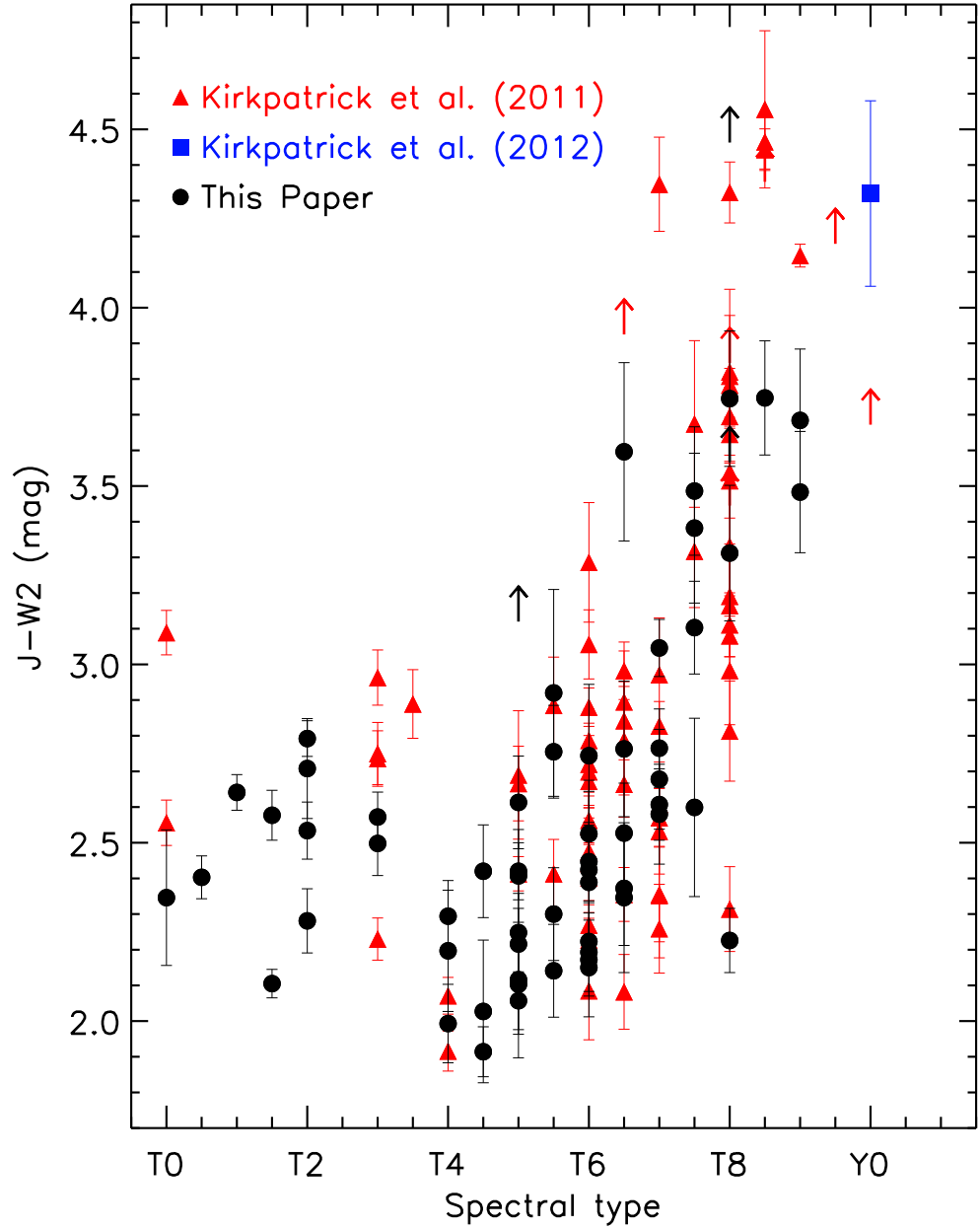}
\caption{$J-W2$ color versus spectral type. Same symbols as in Figure~\ref{W1W2_type}.
\label{JW2_type}}
\end{figure}

\begin{figure}
\epsscale{0.9}
\figurenum{6}
\plotone{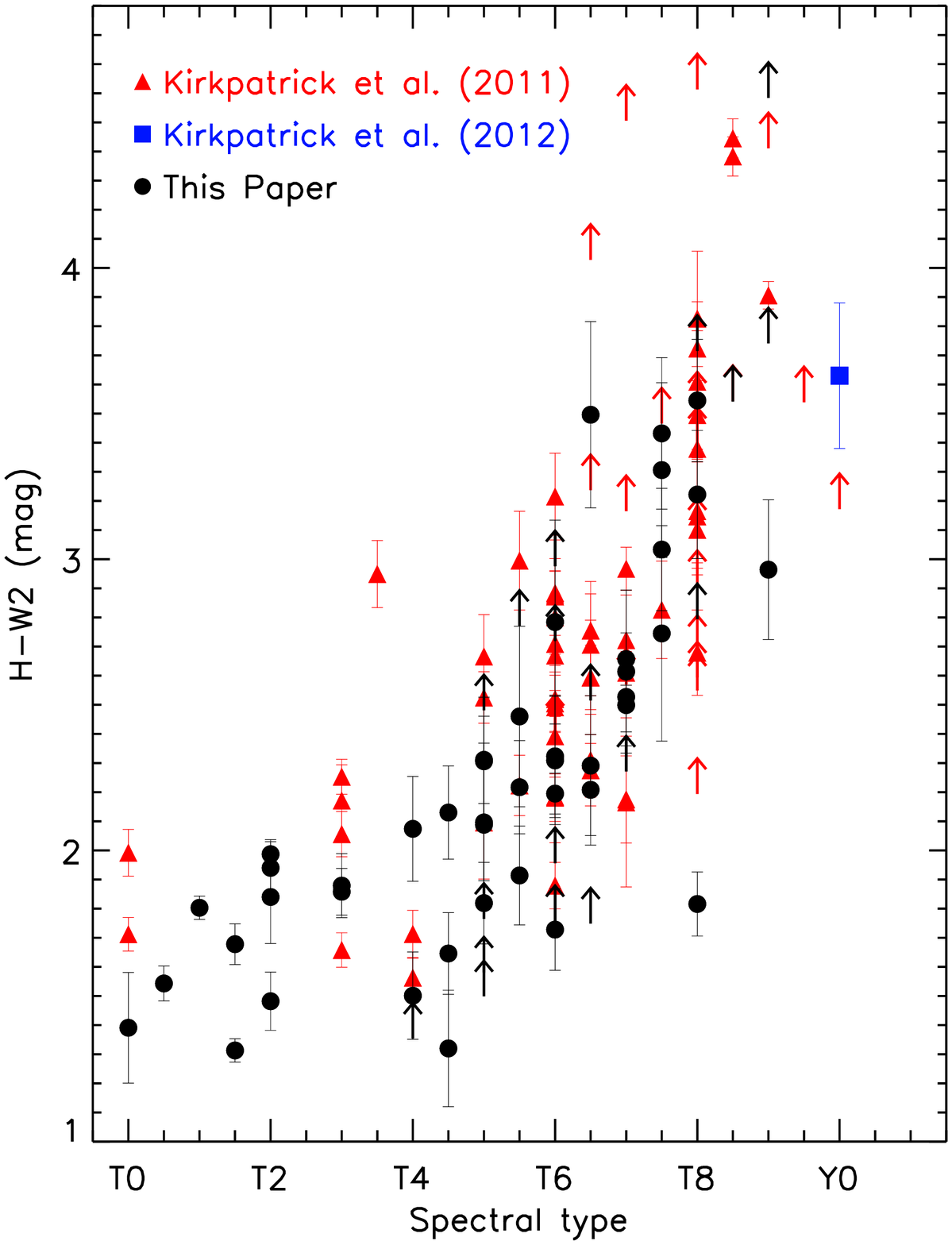}
\caption{$H-W2$ color versus spectral type. Same symbols as in Figure~\ref{W1W2_type}.
\label{HW2_type}}
\end{figure}

The $J-W2$ and $H-W2$ colors, shown in Figures~\ref{JW2_type} \& \ref{HW2_type}, of early-type T dwarfs are slightly bluer (or flat) for successive types before they become redder for late-type T dwarfs.  
These colors are complicated as H$_2$ CIA, non-equilibrium CO and CO$_2$ absorption, and the shift of the Planck peak to longer wavelengths influence 
the emergent flux \citep{linsky1969,borysow1997,frommhold2010,saumon2012,abel2012}.

\begin{figure}
\epsscale{0.9}
\figurenum{7}
\plotone{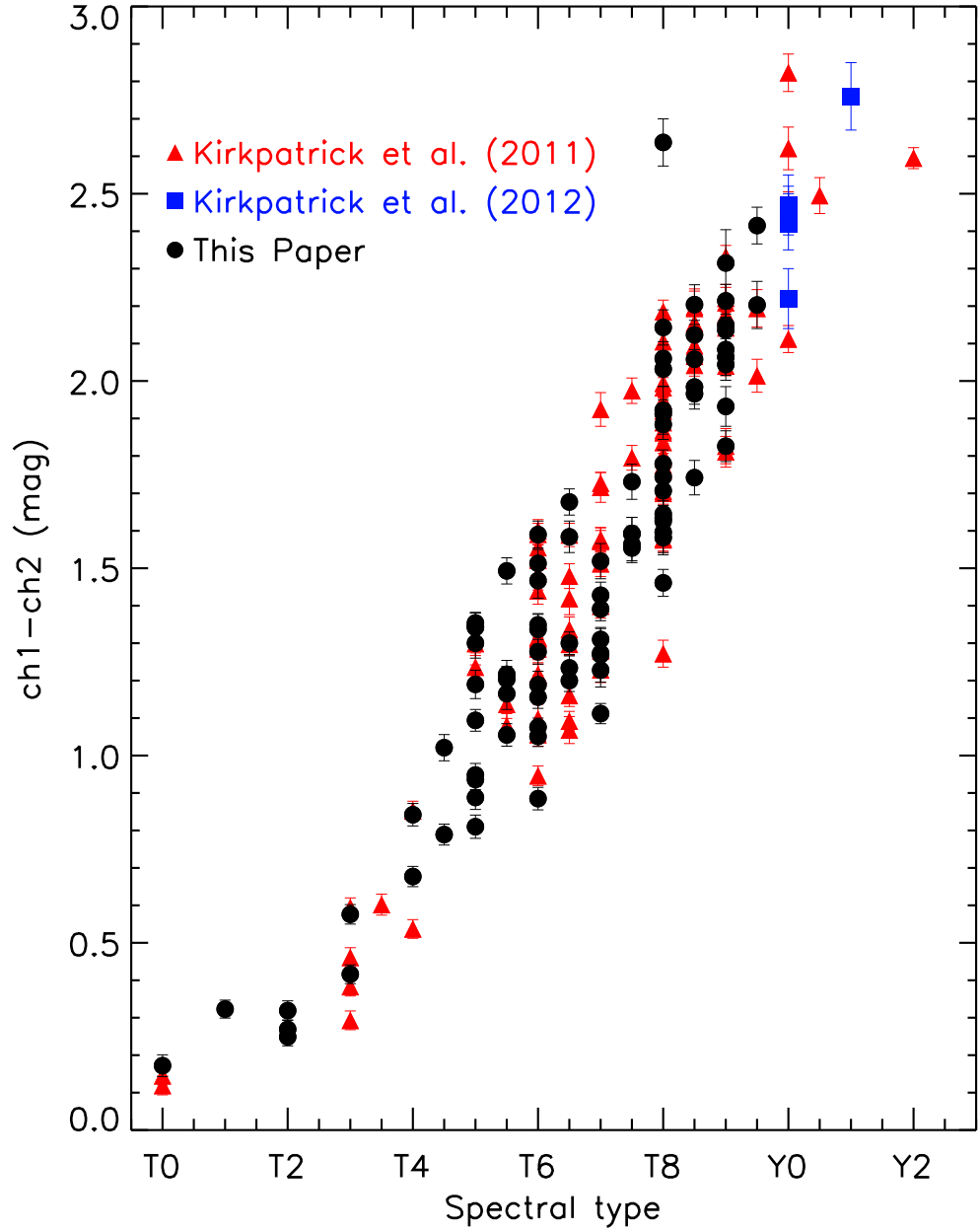}
\caption{{\it Spitzer} $ch1-ch2$ versus spectral type. Same symbols as in Figure~\ref{W1W2_type}. These colors can nearly be fit by a straight line, but the early-type T dwarfs are all bluer than the fit.
\label{ch1ch2_type}}
\end{figure}

\begin{figure}
\epsscale{0.9}
\figurenum{8}
\plotone{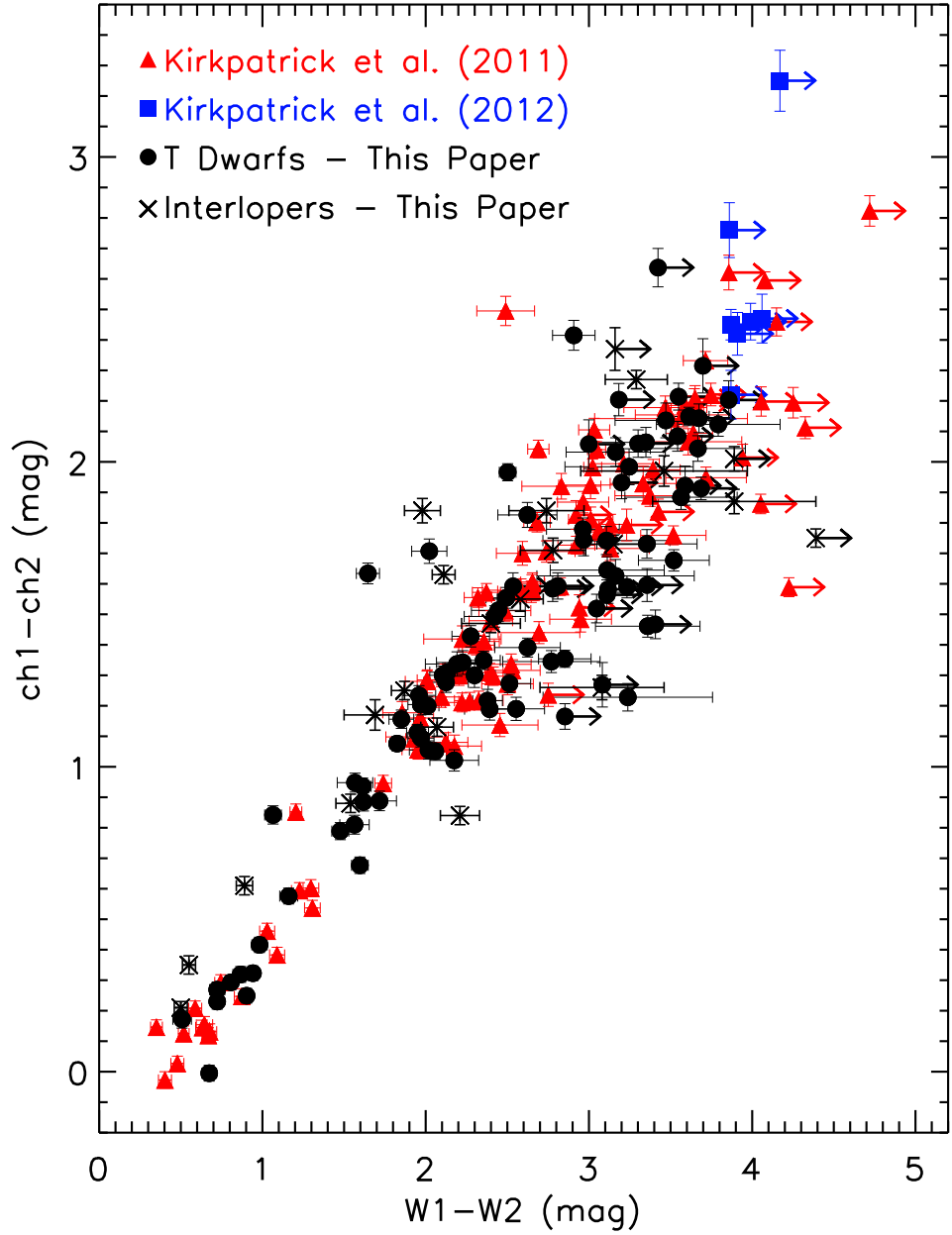}
\caption{{\it Spitzer} $ch1-ch2$ versus WISE $W1-W2$ color-color diagram. Same symbols as in Figure~\ref{W1W2_type}. Interlopers are marked by black x's and limits.
\label{ch1ch2_W1W2}}
\end{figure}

The Infrared Array Camera (IRAC; \citealt{fazio2004}) onboard the {\it Spitzer Space Telescope}  (\citealt{werner2004})
was used during the warm {\it Spitzer} mission to obtain deeper 
photometry in its 3.6 and 4.5 $\mu$m channels (hereafter, $ch1$ and $ch2$, respectively) allowing the WISE $W1$ and $W2$ filters to be robustly measured. These observations were made as part of Cycle 7 and Cycle 8 programs 70062 and 80109 
(Kirkpatrick, PI). {\it Spitzer}/IRAC photometry is listed in Tables~\ref{first100_more_followup_photometry} \&~\ref{Ts_followup_photometry} along with our ground-based photometry. 
The $ch1-ch2$ color, as a function of spectral type, is shown in Figure~\ref{ch1ch2_type}.
The trend is similar to the WISE $W1-W2$ colors (Figure~\ref{W1W2_type}), but differences in the WISE and {\it Spitzer}/IRAC passbands \citep{mainzer2011} 
result in a shallower slope in the $ch1-ch2$ color, as a function of spectral type, than the $W1-W2$ color. Figure~\ref{ch1ch2_W1W2} directly compares the 
$ch1-ch2$ and $W1-W2$ colors for T dwarfs. Figure 1 of \citet{griffith2012} shows 
the $ch1-ch2$ and $W1-W2$ colors of many WISE candidate and confirmed brown dwarfs, clearly identifying the dip in $ch1-ch2$ at the L/T transition and the 
dispersion in colors at later spectral types that are also seen in Figure~\ref{ch1ch2_W1W2}.

\subsection{Follow-up Spectroscopic Observations}
Summaries  of the spectroscopic follow-up observations of the 90 new brown dwarfs and fourteen known objects from the literature are given in Tables~\ref{spec_log} \& \ref{repeat_spec}, respectively. 
Our observation and data reduction procedures are discussed in detail in \citet{kirkpatrick2011,kirkpatrick2012}.  The analysis by \citet{kirkpatrick2012} of the low-mass end of the field brown dwarf 
mass function used objects observed prior to 2012 May 1 (UT) and spectral types for WISE J152305.10+312537.6, WISE J180901.07+383805.4, WISE J201404.13+042408.5, and 
WISE J201920.76$-$114807.5 have been adjusted by half spectral types in this paper. An improved spectrum of the T9 dwarf WISE J033515.01+431045.1 
has increased the precision of the spectral type provided by \citet{kirkpatrick2012}.

No changes were made in our procedures for spectra acquired with SpeX \citep{rayner2003} at NASA's Infrared Telescope Facility (IRTF), the Triple Spectrograph \citep[TSpec;][]{herter2008} at 
Palomar Observatory, or the Wide Field Camera 3 \citep[WFC3;][]{kimble2008} onboard the {\it Hubble Space Telescope} (HST). Non-peculiar T dwarf spectra from SpeX and TSpec are sorted by 
spectral type (see \S4.1 for classification procedure) and shown in Figures~\ref{SpeX_1} and \ref{FGT_1}, respectively. Figure~\ref{HST} shows the WFC3 confirmed T9.5 dwarf WISE J154214.00+223005.2.

We spectroscopically confirmed seven of our candidates employing the Gemini Near-InfaRed Spectrograph \citep[GNIRS;][]{elias2006} at the Gemini-North Telescope on Mauna Kea, HI.  We used the cross-dispersed 
mode with the 32 lines/mm grating and a 1$\arcsec$-wide slit which provided an R of 540 over the 0.8 to 2.5 $\mu$m wavelength range.  A series of 300 sec exposures were obtained at two positions along the 7$\arcsec$-long slit.  
An A0 V standard star was obtained either before or after each science target for telluric correction and flux calibration purposes. The data were reduced with a modified version of the Spextool data reduction 
package \citep{cushing2004}.  The raw frames were first cleaned of fixed patterns using the \texttt{cleanir.py} routine provided by Gemini Observatory.  A normalized flat-field image was then generated using the 
nightly calibration frames.  Pairs of images taken at two different positions along the slit were then pair-subtracted and flat-fielded.  The spectra were then extracted after subtracting off residual background and 
wavelength calibrated using the night sky OH emission lines.  The raw spectra were corrected for telluric absorption using the spectra of the associated A0 V standard star and the technique described in \citet{vacca2003}.  
Finally the spectra from the individual orders were merged to produce a continuous 0.8 to 2.4 $\mu$m spectrum.
The GNIRS spectra of our new T dwarfs are shown in Figure~\ref{FGT_1} and are colored blue to differentiate them from the other instruments.

Since the publication of \citet{kirkpatrick2011}, the reduction procedure has changed for spectra from the Folded-port Infrared Echellette (FIRE; \citealt{simcoe2008}, \citealt{simcoe2010}) instrument at the 6.5m 
Walter Baade Telescope on Cerro Manqui at the Las Campanas Observatory, Chile. FIRE reductions in this paper make use of the FIREHOSE package for low-dispersion data 
following the procedure in the online ``cookbook."\footnote{See http://www.mit.edu/people/rsimcoe/FIRE/ob\_data.htm for details.} The FIREHOSE package uses NeAr arc lines for the calibrator, and OH sky emission lines for target, wavelength solutions. 
For our faintest objects, pair subtraction of the 2-D spectra prior to insertion into the pipeline greatly improved the accuracy of the bspline sky-line fitting procedure. 
Although the automated continuum fitting worked well for most of our targets, the faint sources required manual selection of the continuum position when sky-lines were poorly subtracted.
Spectra were extracted using the optimal spectral extraction process, where an iterative, local background model is fit to improve estimates of the sky flux for subtraction.
The combined spectrum was then corrected for telluric absorption and flux calibrated using the observations of an A0 V star and the technique described in \citet{vacca2003} 
and the XTELLCOR program from SpeXtool (see \citealt{cushing2004}). FIRE spectra are presented in Figure~\ref{FGT_1} and are colored red to differentiate them from the other instruments.

Observations with the Near-Infrared Spectrometer (NIRSPEC, \citealt{mclean1998,mclean2000}) at the 10~m 
W.~M.~Keck Observatory on Mauna Kea, Hawai'i, were made with slit widths of 0${\farcs}$38, 0${\farcs}$57, or 0${\farcs}$76 to reduce slit-loss for specific seeing conditions.  
These slit widths result in spectral resolutions at 1.27$\mu$m of R$\sim$ 2250, 1500, and 1100, respectively. After being confirmed in our $J$-band (N3) observations, two of our new 
T dwarfs were observed with the $H$-band (N5) filter with a 0${\farcs}$76 slit (R$\sim$1400 at 1.58$\mu$m). Observations from multiple epochs used the same instrument configurations and
were combined into a single spectrum.
Data reduction made use of the publicly available REDSPEC package, with modifications to remove residuals from the sky-subtracted pairs prior to 1-D spectral extraction. Normalized 
Keck/NIRSPEC $J$-band spectra are shown in Figure~\ref{NIRSPEC_N3_1}, and objects with both $J$- and $H$-band observations are shown in Figure~\ref{NIRSPEC_N3N5}. 
The $H$-band spectra were scaled relative to the $J$-band data using the ratio of peak fluxes produced by REDSPEC \citep{mclean2003}. 

\section{T Dwarf Spectral Analysis}

After collecting near-IR spectra for each candidate we determined spectral types through comparison to the spectral standards (\citealt{burgasser2006} and \citealt{cushing2011}). 
Some of the objects are noteworthy for their peculiarity. Fourteen of our candidates have been published by others since we first identified or confirmed them, and we discuss each one
individually. For all the T dwarfs presented here and in \citet{kirkpatrick2011}, as well as the Y dwarfs from \citet{cushing2011} and \citet{kirkpatrick2012}, we derive spectral indices using the median flux method.

\subsection{Spectral Classification}

We derived the spectral types for each of our near-IR spectra by visual comparison to the T0-to-T8 spectral standards of \citet{burgasser2006} and the T9 and Y0 standards defined by \citet{cushing2011}.
Classifications are based on the complete near infrared flux when it is available. Normalization at $\sim$1.27~$\mu$m makes the $J$-band morphology the easiest to compare to the standards and 
emphasizes peculiarities in the $Y$, $H$, and $K$ bands. Half spectral types are assigned where the spectrum falls consistently between two adjacent spectral standards. In cases where the spectrum 
deviates from the standards in a way that is inconsistent with the spectral sequence (e.g. the $J$ band is narrower than a T7 and wider than a T8, but the $H$- and $K$-band fluxes are consistent with a T6) we 
have identified these objects as peculiar. We identify spectral types of low SNR spectra with `:' to indicate that they are more uncertain than the standard $\pm$0.5 type uncertainty. 
The extremely red brown dwarfs defy classification and are discussed in detail in \S4.2.
Our binary candidates are given spectral types and marked as `sb' to denote that they are peculiar and best matched with combined spectral templates.
Assigned spectral types for all the objects in Table~\ref{WISE_all_sky_phot_Ts} are listed in column 2 of Tables~\ref{spec_log} and~\ref{repeat_spec}. 

\subsection{Are the Extremely Red Brown Dwarfs L or T Type?}

The T dwarf spectral standards from \citet{burgasser2006} were chosen for their progressively stronger H$_2$O and CH$_4$ absorption features throughout the near infrared. It is the CH$_4$ features in 
T dwarfs like the prototype Gl 229B \citep{geballe1996} that makes these objects stand apart from L dwarfs.
\citet{kirkpatrick2010} outlined the procedure for defining a near infrared L dwarf sequence that varies smoothly between spectral types and is tied closely with the optical spectral types. For this L dwarf sequence,
spectra are normalized at the $J$-band flux peak ($\sim$1.27$\mu$m) and the type is determined from only the 0.9-1.4$\mu$m spectral shape. From this initial classification, the spectral morphology from 1.4-2.5$\mu$m identifies the source as 
either ``red" or ``blue" relative to the spectral standard. Based on these systems, we have identified three objects (WISE J064205.58+410155.5, WISE J075430.95+790957.8, and WISE J173859.27+614242.1) that defy classification. 

As the number of red L dwarfs with near-IR spectra has increased, classification and physical interpretation of the spectra has become more complicated. Many of the known red L dwarfs are early-type and 
have accompanying optical spectra to facilitate classification and analysis \citep{looper2008, kirkpatrick2006, kirkpatrick2010}. 
At later spectral types, L dwarf $J-K$ colors redden by nearly 1.5 magnitudes relative to L0's (as cloud opacity increasingly dominates at shorter wavelengths) and the red L dwarfs mark the upper boundary of these colors \citep[][Figure 4]{gizis2012}. 
Recently, as a result of 2MASS proper motion surveys and the WISE mission, the sample of red, late-type L dwarfs has been expanded \citep{kirkpatrick2011, gizis2012}.
Since the reddest L dwarfs are a few magnitudes fainter at visible wavelengths, high-quality optical spectra of these objects are rare. 

Infrared spectra of red L dwarfs are primarily influenced by temperature, metallicity and gravity differences, but also by possible binarity and cloud morphology.
\citet{looper2008} discuss the influence of low surface gravity and high metallicity on reddening near-IR colors, and the unlikely reddening from binarity and interstellar extinction. 
\citet[][Figure 3]{allers2010} show how dust and low surface gravity alters near-IR spectra for adjacent spectral types.  
Other extremely red objects, like 2MASS J035523.37+113343.7 \citep{faherty2013}, 2MASS 1207b \citep{patience2010, barman2011b} and the HR 8799 planets \citep{bowler2010, barman2011a, marley2012} 
have low surface gravity, appear to be dusty, resemble the L dwarf standards, and lack the methane features of T dwarfs.
\citet{gizis2012} present an extremely red L dwarf (WISEP J004701.06+680352.1) that they can best classify as an L7.5pec based on rough comparison to the spectral standards and other red L dwarfs in the literature, 
but they also discuss the difficulty in determining spectral types and estimating surface gravities for such red sources, which may not be very young. 
The extremely red objects shown in this paper don't show the `peakiness' associated with youth, but instead display a broader plateau in the $H$ band, and are most similar to WISEP J0047+6803 and 2MASSW J2244316+204343 \citep{dahn2002,mclean2003,looper2008}.

WISE J0754+7909 is the least red of the three new objects. Figure~\ref{ERGs1} shows the Palomar/TSpec spectrum of this object along with the L9 and T2 spectral standards. Comparison to the L9 standard produces the 
best match to the integrated flux, but poorly matches the H$_2$O and CH$_4$ absorption depths. Classification based solely on the $J$-band peak, as was done by \citet{kirkpatrick2010}, pushes the classification of this object into the 
T sequence. However, given the noise of this spectrum it is difficult to fully identify CH$_4$ features, ruling out its classification as a T dwarf using the scheme of \citet{burgasser2006}. Near-IR colors for this object are consistent with the T2 classification
and WISE colors are similar to sources at the L/T transition. However, this object is blended in the {\it Spitzer} images with a source that is not visible in our near-IR imaging, so WISE and {\it Spitzer} colors should be used cautiously.

\citet{burgasser2010} used the SpeX Prism Library to identify possible binary systems based on the composite of L and T dwarf spectra. 
Using the same analysis, we find that WISE J0754+7909 is poorly matched by composite spectra and is likely not a binary. 
The spectrum of SDSS J1516+3053 is also included in Figure~\ref{ERGs1} since it matches the spectrum of WISE J0754+7909 well at shorter wavelengths.
SDSS J151643.01+305344.4 has been classified as a T0.5$\pm$1 \citep{chiu2006} and T1.5$\pm$2 \citep{burgasser2010}, and is a weak binary candidate based on their analysis. The large uncertainties in the SDSS J1516+3053 classifications are
due to the difficulty in classifying this source using the methods in the literature.

The Palomar/TSpec spectrum of WISE J0642+4101 is shown in Figure~\ref{ERGs2} with the L9 and T2 spectral standards. Just like WISE J0754+7909, classification of this object based on the $J$ band only requires that we classify this object as a T dwarf. However,
there is a clear absence of CH$_4$ in this spectrum. Compared to SDSS J1516+3053, WISE J0642+4101 is significantly redder. The red L7.5$\pm$2 dwarf 2MASS J2244+2043 is one of the reddest late-type 
L dwarfs and is very similar to WISEP J0047+6803 \citep[L7.5pec;][]{gizis2012}. Figure~\ref{ERGs2} shows that WISE J0642+4101 is nearly as red as 2MASS J2244+2043 in the $H$ band, but has a significantly lower $K$-band flux.

The reddest object that we have identified, WISE J1738+6142, is shown in Figure~\ref{ERGs3}. We observed this object with Palomar/TSpec on 2011 July 13 (UT) and with IRTF/SpeX on 2012 July 23 (UT). 
Although the Palomar/TSpec spectrum is noisy, comparison of these two spectra reveals that the near-infrared flux of WISE J1738+6142 is extremely variable.\footnote{The airmass of each A0 standard was less than 0.05 away from the mean value for WISE J1738+6142.} 
Photometry of this object from PAIRITEL produces $J-K_s$=2.55$\pm$0.16, which is equivalent with the extremely red L dwarf WISE J0047+6803. The WISE $W1-W2$=0.72$\pm$0.04 
color for this source is slightly redder than the similar red L dwarfs WISEP J0047+6803 ($W1-W2$=0.65$\pm$0.03) and 2MASS J2244+2043 ($W1-W2$=0.67$\pm$0.03).
Comparison of the WISE J1738+6142 IRTF/SpeX spectrum to 2MASS J2244+2043 reveals excess flux on the blue side of the $H$ and $K$ bands, possibly hinting at the presence of a 
T dwarf secondary. Since the red L dwarf phenomenon isn't entirely understood, having a coeval T dwarf companion with similar metallicity or low gravity indicators will help reveal the cause of the cloudiness in the L dwarfs, 
should any of these extremely red objects prove to be an L+T binary.

Although gravity, metallicity, and temperature changes are the primary variables in modeling these red objects, the variability seen in WISE J1738+6142 hints at considerably rapid changes in atmospheric conditions. 
If we assume that the source of reddening is common to all these objects, then surface gravity and metallicity are not likely the mechanisms at work.
The proper motions of these three objects, determined by comparing 2MASS and WISE coordinates [WISE J0642+4101 (0.37 $\pm$ 0.03~$\arcsec$/yr), WISE J0754+7909 (0.43 $\pm$ 0.04~$\arcsec$/yr), and WISE J1738+6142 (0.28 $\pm$ 0.04~$\arcsec$/yr)] 
is consistent with the red L dwarfs from \citet{kirkpatrick2010} that are likely part of an older field population. That is inconsistent with either high metallicity or low gravity.
Although analysis of the underlying physical mechanisms will require detailed analysis and modeling of higher resolution optical and near-IR spectroscopy, identifying companions to these red objects and disentangling their spectral features are good initial tests. 
We believe that further analysis (i.e. optical spectroscopy, high-resolution imaging, and time domain observations) are warranted before final classifications are made for these extremely red objects at the L/T transition.

\subsection{Comments on Interesting T Dwarf Discoveries}

In the following subsection we discuss the peculiar objects individually, along with noteworthy objects that warrant separate discussion.

\subsubsection{WISE J0336$-$0143 - T8:}
This object is a clear outlier in Figure~\ref{ch1ch2_type} ($ch1-ch2$ =  2.64 $\pm$ 0.06) and is undetected in the WISE $W1$ filter ($W1-W2$ $>$ 3.42). These colors are more consistent with a Y dwarf than a T8, but the Keck/NIRSPEC $J$-band spectrum is broad.
WISE J0336$-$0143 may be a T+Y binary, with the Y dwarf contributing more flux at the WISE passbands than in the near-IR bands. Follow-up photometry and broader near-IR spectroscopy of this source are warranted.

\subsubsection{WISE J0413$-$4750 - T9}
The A nod position of the pair subtracted 2-D spectra for this source was contaminated by a galaxy. The spectrum of the galaxy was extracted from the B-position images and fit with a line. This linear fit was then subtracted from the final extracted spectrum of WISE J0413$-$4750, which is shown in Figure~\ref{FGT_1}.

\subsubsection{WISE J0629+2418 - T2 sb}
The IRTF/SpeX spectrum for this object is presented in Figure~\ref{sbsolutions} along with the T2 standard, which is the best overall match to the CH$_4$ and H$_2$O absorption troughs. 
Relative to the T2 standard, the $H$ band is wider and displays greater CH$_4$ absorption redward of the flux peak and the $K$-band flux is high. 
Using the statistical method described by \citet{burgasser2010} we identify WISE J0629+2418 as a strong binary candidate with a L7$\pm$1 primary and T5.5$\pm$0.5 secondary ($\Delta$$J_{MKO}$=0.47$\pm$0.28, $\Delta$$H_{MKO}$=1.67$\pm$0.30).
The best composite match to WISE J0629+2418 is also shown in Figure~\ref{sbsolutions}.

\subsubsection{WISE J1523+3125 - T6.5 pec}
Figure~\ref{pec} shows the Gemini/GNIRS spectrum for this object. Comparison to the T6 standard produces the best match to the $H$-band flux, while the $J$ band is better matched by the T7 standard.
The $Y$- and $K$-band fluxes are inconsistent with the standards. These features are similar to the metal-poor T dwarfs 2MASSI J0937347+293142 \citep[T6pec;][]{burgasser2002} and BD +01$\degr$ 2920B \citep[T8pec;][]{pinfield2012}. 
However, we also observed BD +01$\degr$ 2920B (\S4.4.5) and find that, other than a suppressed $K$ band, it matches the T8 standard well and we do not classify it as peculiar. 

\subsubsection{WISE J1542+2230 - T9.5}
The HST/WFC3 spectrum for this object is shown in Figure~\ref{HST}, compared to the T9 and Y0 spectral standards from \citet{cushing2011}. 
WISE J1542+2230 has a measured F140W(Vega) magnitude of 20.3. Located 2$\farcs$0 west is a neighbor that has a measured F140W(Vega) magnitude of 22.8.
The slitless grism image of this field revealed that the neighbor has a flux peak coincident with the $H$-band flux peak in cool brown dwarfs, however we detect no flux in the $J$-band opacity window at 1.27 $\mu$m.
Additional high resolution images of WISE J1542+2230 were obtained using the Keck II
LGS-AO system \citep{wizinowich2006,vandam2006} with NIRC2 on 2012 April 14 (UT).
Very good seeing ($\approx$0$\farcs$3) and clear skies allowed us to use the
relatively faint ($R$=18.5) USNO-B star 1124-0287738 
\citep{monet2003} as the tip-tilt reference star.  Observations were obtained 
in the MKO $H$ filter with the wide camera (nominal plate scale 
40mas/pixel) on NIRC2.  Twenty-one frames with durations of 120sec 
each were collected using a 3-point dither pattern that was repeated seven times
with position offsets 1.5--5$\arcsec$.  The images were dark subtracted, 
flat-fielded, shifted to a common grid, and median averaged to make a 
single mosaic image with a total exposure time of 42 minutes in the central 
region. From this imaging we measured the proper motion of WISE J1542+2230 to be $\mu_{\alpha}$ = $-$0.82 $\pm$ 0.07 arcsec/yr and $\mu_{\delta}$ = $-$0.41 $\pm$ 0.06 arcsec/yr, while the neighbor remained stationary, thus ruling out binarity.

\subsubsection{WISE J1730+4207 - T0 sb}
This object's spectrum is very similar to WISE 0629+2418 (\S4.3.3), with the integrated flux most similar to the T0 standard (see Figure~\ref{sbsolutions}).
This source is matched well by a composite of L7.5$\pm$1 and T3$\pm$1 spectra. With $\Delta$$J_{MKO}$=-0.31$\pm$0.27 and $\Delta$$H_{MKO}$=0.38$\pm$0.36, this system is likely a flux-reversal binary.
The best composite match to WISE J1730+4207 is also shown in Figure~\ref{sbsolutions}, and is well within the uncertainties of the average fit L7.5+T3.

\subsubsection{WISE J2014+0424 - T6.5 pec}
Much like WISE J1523+3125 (\S4.3.4), comparison to the T6 standard produces the best match to the $H$-band flux, while the $J$ band is better matched by the T7 standard.
The $Y$- and $K$-band fluxes are inconsistent with the standards. Figure~\ref{pec} shows the Gemini/GNIRS spectrum for this object. 

\subsection{Comments on Known and Candidate Brown Dwarfs in the Literature}

Some of our candidates were identified and/or confirmed but subsequently published by others as independent discoveries. Table~\ref{repeat_spec} lists these objects along with their discovery papers and previous spectral classifications. Spectra for these objects are shown in Figure~\ref{WISERedo1} unless noted otherwise.

\subsubsection{WISE J0920+4538 - L9 sb?}
\citet{aberasturi2011} performed a cross correlation of the 2MASS Point Source, SDSS Data Release 7, and WISE Preliminary Release Catalogs to identify six brown 
dwarf candidates based on photometric colors and proper motion. Based on a temperature derived from spectral energy distribution fitting, they estimated the spectral type of this object 
as L4 or L5. However, comparison to spectral standards reveals that this object is best classified as an L9 and a weak binary candidate. Figure~\ref{sbsolutions} shows the Palomar/TSpec spectrum for this source and 
the best fitting composite spectrum with a L7.5$\pm$1.5 primary and a T1.5$\pm$1.5 secondary ($\Delta$$J_{MKO}$=-0.17$\pm$0.33 and $\Delta$$H_{MKO}$=0.22$\pm$0.33).
The composite fit to WISE J0920+4538 is highly uncertain, but fitting to a single object spectrum is also a poor fit and binarity may be the source of the discrepant photometric colors and spectral features of this source. 

\subsubsection{ULAS J095047.28+011734.3 - T8}
Photometry, and a T8 spectral type, for this object was reported by \citet{leggett2012}. Our Magellan/FIRE spectrum for this source is consistent with the T8 standard, but 
displays a slightly higher flux in the $Y$ and $H$ band, and is slightly suppressed in the $K$ band. However, these deviations are within our noise estimates and so we do not classify this object as peculiar. 

\subsubsection{WISE J1246$-$3139 - T1}
This source was first identified in the literature by \citet{andrei2011} as one of the targets in their parallax program. Based on preliminary photometry they estimated the spectral type as T1.0.
The Palomar/TSpec spectrum for this source is shown in Figure~\ref{WISERedo1} and we classify this object as a T1.

\subsubsection{LHS 2803B - T5.5}
This wide separation, common proper motion companion to the M4.5$\pm$0.5 dwarf LHS 2803 was reported by \citet{deacon2012} and classified as a T5.5. We produce the same T5.5 classification from our IRTF/SpeX spectrum.

\subsubsection{BD +01\degr 2920B - T8}
\citet{pinfield2012} identify BD +01$\degr$ 2920B as a T8pec dwarf companion to BD +01$\degr$ 2920 (HIP 70319), a metal-poor ([Fe/H] = $-$0.38 $\pm$ 0.06) G1 dwarf at a distance of 17.2 pc. 
We observed BD +01$\degr$ 2920B and find that, other than a suppressed $K$ band, the object matches the T8 standard. 
Since our analysis (\S4.5.5) shows that suppressed $K$-band flux is fairly common in late-type T dwarfs we do not classify it as peculiar.

\subsubsection{WISE J1449+1147 - T5 pec}
In a search for high proper motion T dwarf candidates between the UKIDSS DR6 and SDSS DR7 databases, \citet{scholz2010} identified this source as a candidate T7$\pm$2 (ULAS J144901.90+114711.3).
This object was observed with two instruments, Gemini/GNIRS and Palomar/TSpec. Figure~\ref{pec} shows the near-IR spectra from each instrument along with the T5 spectral standard.
The $J$- and $H$-band flux match the T5 standard, but the $Y$ band displays a flux excess and the $K$ band is more heavily suppressed than the standard. 
These are the same low-metallicity features seen in the other peculiar objects.

\subsubsection{Repeats from \citet{deacon2011}}
\citet{deacon2011} identified four new T dwarfs by their proper motions between the PanSTARRS 1 commissioning data and 2MASS. We independently identified three of these objects in the WISE catalog and they 
were observed prior to publication by \citet{deacon2011}; PSO J226.2599$-$28.8959 (WISE J1505-2853), PSO J246.4222+15.4698 (WISE J1625+1528), and PSO J247.3273+03.5932 (WISE J1629+0335). 
Our derived spectral types are consistent with their visually derived types, but with higher uncertainties. The fourth object observed by \citet{deacon2011}, PSO J201.0320+19.1072, is not included in the partial
sky coverage of the WISE Preliminary Release Source Catalog and was not observed by us after it was published by \citet{deacon2011}.

\subsubsection{VHS J154352.78$-$043909.6 - T5:}
Our IRTF/SpeX spectrum for this source is fairly low signal-to-noise. \citet{lodieu2012} photometrically selected this object from the first data release of the VISTA Hemisphere Survey and the WISE Preliminary Data Release.
They derive a spectral type of T4.5 based on a Magellan/FIRE spectrum of the source, which is constant with the uncertain T5: classification that we produce.

\subsubsection{Repeats from \citet{kirkpatrick2011}}
Two objects from \citet{kirkpatrick2011}, WISE J1717+6129 and WISE J2344+1034, were observed again with Gemini/GNIRS to provide more wavelength coverage than the original Keck/NIRSPEC observations.
The classifications from \citet{kirkpatrick2011} remain unchanged at T8 and T9, respectively.

\subsubsection{WISE J1809+3838 - T7.5}
\citet{luhman2012} used the WISE All-Sky Source Catalog along with the 2MASS Point Source Catalog to search for proper motion companions. 
From their IRTF/SpeX spectrum they produce a type of T7$\pm$1, while we classify the object as T7.5$\pm$0.5.  

\subsubsection{WISE J2342+0856 - T6.5}
This object was also identified by \citet{scholz2010} as a candidate T7$\pm$2 (ULAS J234228.96+085620.1). We classify this object as a T6.5 based on the IRTF/SpeX spectrum for this object.

\subsection{Median Derived Spectral Indices}

\citet{burgasser2006} defined a number of primary spectral indices that compare the ratio of the integrated $J$- and $H$-band flux peaks to the blueward H$_2$O and 
redward CH$_4$ absorption features; H$_2$O$-$J, CH$_4$$-$J, H$_2$O$-$H, and CH$_4$$-$H. $K$-band indices were also defined, but $K$-band fluxes reduced 
by H$_2$ CIA results in low SNR $K$-band spectra and highly uncertain $K$-band indices. The Y/J and K/J color ratios, defined by \citet{burgasser2004b,burgasser2006BBK}, 
serve as proxies for the broadband $Y-J$ and $J-K$ colors.
Two additional indices that correlate well with the T dwarf spectral sequence are W$_J$ \citep{warren2007} and NH$_3$$-$H  \citep{delorme2008}, 
which approximate the slope of the $J$- and $H$-band flux peaks, respectively.

In addition to these indices, we have derived an H/J color ratio as a proxy for the $J-H$ color,
\begin{equation} \label{indxH/J}
\textnormal{H/J} = \frac{\textnormal{$\widetilde{F}$$_{1.560-1.600 \mu m}$}}{\textnormal{$\widetilde{F}$$_{1.250-1.290\mu m}$}}  ,\\
\end{equation}

and the J-narrow index which is designed to differentiate Y dwarfs from T dwarfs based on their narrow $J$-band flux peaks,
\begin{equation} \label{indxJnarrow}
\textnormal{J-narrow} = \frac{\textnormal{$\widetilde{F}$$_{1.245-1.260\mu m}$}}{\textnormal{$\widetilde{F}$$_{1.260-1.275\mu m}$}}  ,\\
\end{equation}
where $\widetilde{F}$ is the \emph{median} flux between the subscripted wavelengths in microns. Figure~\ref{indexwaves} depicts the wavelengths over which these indices are derived for a series of Palomar/TSpec 
spectra from this paper.

Although indices in the literature are defined as the integrated flux over the desired wavelength range, this method is unreliable when trying to compute indices for low-SNR spectra. Even with 
moderate SNR spectra, the flux in the numerator of most indices is minuscule for late-type T dwarfs and easily modulated by noisy pixels and poor telluric correction. For this reason, we have computed all of our 
indices \emph{over the classical wavelengths} but use the \emph{median} flux instead of the integrated flux. This method reduces the scatter in index values as a function of spectral type, especially at later types.
Further discussion about using median fluxes to compute indices is deferred to the Appendix, where we also include the integrated flux indices for all our WISE discoveries. 

We computed the index values using a Monte Carlo model of each spectrum. This was accomplished by randomly drawing from a normal distribution with means 
given by the flux densities at each wavelength and standard deviations given by the uncertainty in the flux densities. For 1000 realizations of each spectrum, the index is 
given as the mean and standard deviation of the distribution of index values. Table~\ref{indices} lists the index values for the T and Y dwarf spectra in \citet{kirkpatrick2011,kirkpatrick2012} and \citet{cushing2011}, the new WISE T dwarf 
discoveries in this paper, and for objects from Table~\ref{repeat_spec}. 

Spectral indices for the WISE-discovered T and Y dwarfs are shown in Figures~\ref{Six_Indices_Ref} to~\ref{model_indices} along with the T0-to-Y0 spectral standards. 
Because our sample is mostly T dwarfs later than T5, indices for the T dwarfs in the SpeX Prism Library (Burgasser et al., in prep.) are also included in our figures. The average index value and standard deviation, 
for the $\sim$320 unique T and Y dwarfs shown in the figures, are listed in Table~\ref{avg_indices} for each half spectral type from T0 to Y0. In the subsections below we discuss each index in detail.

\subsubsection{Primary Indices from \citet{burgasser2006}}
The primary classification indices for T dwarfs were designed by \citet{burgasser2002} and \citet{geballe2002}, and then revised by \citet{burgasser2006}. We modify these as follows to incorporate median fluxes:
\begin{equation}
\textnormal{H$_2$O$-$J} = \frac{\textnormal{$\widetilde{F}$$_{1.140-1.165 \mu m}$}}{\textnormal{$\widetilde{F}$$_{1.260-1.285 \mu m}$}}\\
\end{equation}
\begin{equation}
\textnormal{CH$_4$$-$J} = \frac{\textnormal{$\widetilde{F}$$_{1.315-1.340 \mu m}$}}{\textnormal{$\widetilde{F}$$_{1.260-1.285 \mu m}$}}\\
\end{equation}
\begin{equation}
\textnormal{H$_2$O$-$H} = \frac{\textnormal{$\widetilde{F}$$_{1.480-1.520 \mu m}$}}{\textnormal{$\widetilde{F}$$_{1.560-1.600 \mu m}$}}\\
\end{equation}
\begin{equation}
\textnormal{CH$_4$$-$H} = \frac{\textnormal{$\widetilde{F}$$_{1.635-1.675 \mu m}$}}{\textnormal{$\widetilde{F}$$_{1.560-1.600 \mu m}$}} . \\
\end{equation}
These indices each show a general trend of having a maximum value at the earliest T spectral type and a steady decrease at later spectral types. However, by the end of the T sequence theses indices have saturated and are less useful for classification. For Y dwarfs, these indices have highly uncertain and spurious values since the numerator in each case has essentially no flux and is dominated by background noise. 

\subsubsection{W$_J$ and NH$_3$$-$H Indices for Late-type T Dwarfs}
The W$_J$ index was defined by Warren et al. (2007) to characterize the $J$ band width. The original integrated flux definition required that 
the denominator be multiplied by a factor of two in order to compensate for the larger width numerator. When modified to use the median flux, the index becomes,
\begin{equation}
\textnormal{W$_J$} = \frac{\textnormal{$\widetilde{F}$$_{1.180-1.230 \mu m}$}}{\textnormal{$\widetilde{F}$$_{1.260-1.285 \mu m}$}} ,\\
\end{equation}
and it is no longer necessary to keep this factor in the denominator. We have chosen to remove it in our analysis. This produces index values between 0.0 and 0.9, which is consistent with the integrated flux results in the literature. The flux from 1.180-1.230 $\mu$m gradually and consistently decreases with spectral type, producing unique index values all the way to the T/Y boundary, as seen by \citet{cushing2011}.  

\citet{delorme2008} defined the NH$_3$$-$H index to probe the blue wing of the $H$ band in later T dwarfs.
We modify this index to incorporate median fluxes,
\begin{equation}
\textnormal{NH$_3$$-$H} = \frac{\textnormal{$\widetilde{F}$$_{1.530-1.560 \mu m}$}}{\textnormal{$\widetilde{F}$$_{1.570-1.600 \mu m}$}} . \\
\end{equation}
This index is not useful at identifying T dwarfs earlier than T7, but has a large dynamic range for for the latest T dwarfs and the Y0s. \citet{cushing2011} identified 
the break in the NH$_3$$-$H index at Y0, suggesting that near infrared NH$_3$ absorption begins at the T/Y boundary. This break is still clearly identified in our median flux index.

\subsubsection{Y/J Ratio}

The Y/J index was defined by \citet{burgasser2006} as a possible tracer of variations in metallicity. We modify this index to incorporate median fluxes,
\begin{equation}
\textnormal{Y/J} = \frac{\textnormal{$\widetilde{F}$$_{1.005-1.045 \mu m}$}}{\textnormal{$\widetilde{F}$$_{1.250-1.290 \mu m}$}} . \\
\end{equation}
The $Y$-band spectra that we have acquired are from the low-resolution, multi-band spectrographs IRTF/SpeX, Palomar/TripleSpec, Magellan/FIRE, and Gemini/GNIRS. From lower resolution spectra we derive indices with larger uncertainties. This index shows significant scatter compared to the H/J and K/J ratios and the spectral standards produce a trend with redder Y/J flux ratios relative to the rest of the population.  The dispersion in this index is smallest for T4's and later objects show the largest scatter. 

\subsubsection{H/J Ratio}

\citet{burgasser2002} defined a broader version of the H/J index that encompasses most of the T dwarf flux contributing to the broadband $J-H$ color, and noted a possible reversal at later types. 
It is not one of the indices revised in \citet{burgasser2006} and so we have constructed the H/J index, presented in Equation~\ref{indxH/J},
employing the wavelengths used by \citet{burgasser2006} to define other indices.
This index shows a tight trend with an average dispersion of 0.1 and shows a similar reversal as the 2MASS $J-H$ colors in Figure~\ref{JH_type}.
As discussed in \S3.1, the turnaround of the $J-H$ color is evident before the T/Y boundary, and appears to occur around T6$\pm$1 for the H/J index. 
However, Y dwarfs show a significant scatter in H/J. The Y0s break the trend of the T dwarfs, and the H/J index drops, possibly due to the onset of NH$_3$ absorption. WISE J182831.08+265037.8 has been typed as $\ge$ Y2 \citep{kirkpatrick2012} based on the near-equal heights of the $J$- and $H$-band flux peaks and has an index value equivalent to the earliest T dwarfs. This dichotomy of H/J flux ratios hints at drastically different atmospheric conditions for Y0s and later type Y dwarfs.

\subsubsection{K/J Ratio}

The wavelengths used for the K/J ratio were initially defined by \citet{burgasser2002} to compare the $K$-band flux peak to the $J$-band flux peak, and modified here to incorporate median fluxes,
\begin{equation}
\textnormal{K/J} = \frac{\textnormal{$\widetilde{F}$$_{2.060-2.100 \mu m}$}}{\textnormal{$\widetilde{F}$$_{1.250-1.290 \mu m}$}} . \\
\end{equation}
Models have long predicted dominant H$_2$ CIA of the $K$-band flux in cool brown dwarf atmospheres, \citep{saumon1994, saumon2012, marley1996, allard1996} which has been observationally verified and is evident in 
Figure~\ref{Indices_Bands_Ref_Med}. The large dispersion in the late-type T K/J index was pointed out by \citet{burgasser2006} and we find that the dispersion is smallest at T4 and flares at both earlier and later types. 
Although the early-type T dwarf standards fall near the mean of each spectral bin, the late-type T spectral standards are on the redder side of the mean. Figure~\ref{SpTKJ}
shows the K/J indices for T dwarfs, sorted by instrument.
We find that the average K/J index for the T5-T9 spectral standards is 0.147$\pm$0.013 and only $\sim$11\% (18 out of 170; 8 WISE discoveries and 10 from the SpeX Prism Library) are redder than the 3$\sigma$ upper limit. 
However, 47 of the 126 WISE-discovered and 7 of the 44 SpeX Prism Library late-type T dwarfs are below the 3$\sigma$ lower limit (shown in Figure~\ref{SpTKJ}). In total 54 out of 
170, \emph{approximately one third}, of the T5 or later dwarfs are blue relative to the standards. 

The largest dispersion ($\sim$1.5 mags) in the K/J ratio is in the T7-T8 spectral bins, in which the number of known T dwarfs has been quintupled by WISE. The T dwarf sample prior to WISE peaked at T5 
\citep[][Figure 28]{kirkpatrick2011}, and WISE discoveries shift this peak to T6, with the number of T8's nearly as bountiful. Figure~\ref{SpTKJ} shows that, although the spectral standards all have similar K/J ratios, 
the value of the index for late-type T dwarfs gradually decreases at later types. The small number of late-type T dwarfs in the SpeX Prism Library, and the steady decrease in the K/J index with spectral type, results in a smaller fraction of blue T dwarfs from the SpeX Prism Library ($\sim$16\%) relative to the WISE discoveries ($\sim$37\%).

Although $\sim$32\% of the late-type T dwarfs are blue ($K$ band suppressed) and $\sim$11\% are red, the current spectral standards still provide the best overall match to the bulk of observed spectra.
The spectral standards match the mean H/J indices very well.
It is possible that the standards, which were chosen to be bright in order to be easily observed, are also younger. If younger than the mean of the sample, the standards would generally be more luminous for their effective temperature (larger), 
have a lower surface gravity (reduced collision induced absorption), and be metal-rich. All of these factors would result in the redder colors seen in the K/J index and bluer Y/J index values.
Objects with blue or red K/J index values, relative to the $J$ band normalized standards, are most commonly called peculiar in the literature. 
Because this is fairly common in our sample, the objects we classify as peculiar deviate from the standards in more than one band.

\subsubsection{J-narrow Index}

The J-narrow index is defined in Equation~\ref{indxJnarrow} and also presented in \citet{kirkpatrick2012}. This index was derived empirically from the spectra 
presented in this paper in order to produce a useful index for classifying Y dwarfs using a spectral index.  As discussed above, many of the 
indices derived for T dwarfs have saturated before the T/Y boundary and are no longer useful for Y dwarfs.
The KI doublet at 1.243 and 1.254 $\mu$m is a feature present in M though T dwarfs down to mid-type T. The numerator in this index encloses the red 
member of the $J$-band neutral potassium doublet and the denominator captures the $J$-band flux peak. Variations in the KI feature are what produce 
the dispersion and undulation seen in Figure~\ref{Jnarrow} for the T dwarf sequence. Most of our spectra are lower resolution IRTF/SpeX and 
Palomar/TSpec spectra and the KI 1.254$\mu$m line is barely resolved. In these data, there is no significant difference in the J-narrow index when 
using the median or the integrated flux. In higher resolution spectra the integrated flux produces slightly lower index values than the median method 
while increasing the dispersion in each spectral bin by including more spectral variation. To minimize this variability we use the median flux to derive 
this index. \citet[][Figure 15]{mclean2003} shows how the KI lines broaden towards later spectral types, before also decreasing in strength around T6 \citep{mclean2007}. 
It is this broadening that causes the J-narrow index to decrease from T0-T6, and the disappearance of the line in late-type T dwarfs that causes the slight 
increase in the index before falling for the Y dwarfs as a result of the narrowing $J$-band flux peak. 

As discussed in \citet{kirkpatrick2012}, because our by-eye classification of Y dwarfs requires that 
the $J$-band peak be narrower than the Y0 standard, WISE J173835.52+273258.9, all the other Y0s have smaller J-narrow indices. Also, this index is only useful for 
distinguishing very early-type Y dwarfs from late-type T dwarfs since it does not appear to work well for WISE J1828+2650, which is either much later in type or 
spectroscopically peculiar.
Future use of the J-narrow index for higher-resolution spectra may provide a means of sorting each spectral bin by surface gravity by employing the KI feature \citep{mclean2003}, 
a pursuit that may be aided by computing the index using the integrated flux.

\section{Clouds and Metallicity at the T/Y Boundary}

\citet{burrows2006} developed a set of brown dwarf models that combined cloudy and cloudless models to reproduce 
the observed $J-K$ colors of L and T dwarfs, respectively. Similar synthetic colors were derived by \citet{saumon2008} that matched the overall brown dwarf sequence. Both models
failed to match the late-type T dwarf dispersion, which they could modulate through variations in metallicity and binarity \citep[see Figure 15 in][]{saumon2008}. 
\citet{burgasser2011} utilized the \citet{saumon2008} models to determine the best fit parameters for their five WISE-discovered T dwarfs and found that cloudy, low surface gravity models provided the best match to observed spectra.

The addition of sulfide clouds to the brown dwarf models by \citet{morley2012} reproduces the dispersion of the late-type T dwarf photometry.
The flux reversal we see in the $J-H$ color (and the H/J index) around T7 corresponds to the turn-off of the late-type T dwarfs from the cloudless models of \citet{morley2012}. 
Based on this turn-off, the spectral sequence beyond T7 (Teff $\le$800K) is possibly dominated by cloudy atmospheres that have redder H/J and K/J indices. 
This includes the T dwarf spectral standards and the Y dwarfs later than Y0. The bluest H/J and K/J indices continue the trend of the early-type T dwarfs and are still consistent with cloudless atmospheres. 
The Y0s in this case stand-out in the H/J index as breaking the trend, and may be less cloudy than the later Y dwarfs. 
Thus, the spectral sequence of the late-type T dwarfs and the early-type Y dwarfs may no longer be correlated directly to temperature, but also to the atmospheric conditions.

The top panel of Figure~\ref{model_indices} shows the objects that we have computed indices for, sorted by spectral type. Early-type T dwarfs are uniquely identified by their K/J index, while 
late-type T dwarfs have a larger Y/J dispersion. The T5-T9 spectral standards all have similar index values, and the T8 dwarfs show the largest spread in index values. 

The middle panel in Figure~\ref{model_indices} depicts the index values for the \citet{morley2012} sulfide cloud models with solar metallicity, and various gravity and condensate sedimentation efficiencies ($f_{sed}$), between 400K and 1300K. 
These models do not include silicate clouds, and so they do not replicate the early-type T dwarfs.
The cloudless models predict a range of Y/J index values for the coolest objects when gravity is varied, but relatively little difference for most T dwarfs. Both the $Y$ and $J$ bands are impacted by the alkali opacity and small variations to the broadening
produces large changes in the Y/J index, so our interpretation here is limited.
For $f_{sed}$=5, the models predict a slight increase in both the Y/J and K/J indices before they both decrease, and this matches our measured index values. 
However, for $f_{sed}$=4 only the log g=4.5 model remains consistent with our measurements. Smaller $f_{sed}$ values have Y/J $\leq$ 0.5 and large K/J values that are inconsistent with our measurements. 

In the bottom panel of Figure~\ref{model_indices} we show the index values for the cloudy (Case E) and cloudless models presented in \citet{burrows2006}, 
for log g=5.0, T$_{eff}$s from 800K to 1700K, and various metallicities. 
The highest temperature is marked with a square and the lowest temperature is marked with a circle in the figure. Here we see that the K/J index produced for the cloudless models, for a specific temperature, do not match the early-type T dwarfs.
This figure shows that the cloudy models trace the early-type T dwarfs well and identifies a similar cloudy late-type T population as the \citet{morley2012}
models. Although both the cloudy and cloudless -0.5 dex tracks enclose the increasing Y/J index values, the cloudless sub-solar metallicity model best matches our measurements. 

Overall, the T dwarf spectral standards occupy a minimum in the Y/J index for the bulk of the late-type T dwarfs. Objects with Y/J indices below the spectral standards are best matched to cloudy models
 and possibly oversolar metallicities. Objects with a Y/J index greater than the standards are best matched to cloudless models and sub-solar metallicities. 
 The K/J index can be modulated by temperature, metallically, and surface gravity \citep{leggett2007}. However, objects with the lowest K/J values, also have the highest Y/J indices, 
 which is consistent with many $K$ band suppressed late-type T dwarfs that are best matched to low-metallicity models.

\section{Interlopers in WISE Color Space}

Some of the brown dwarf candidates survived the photometric selection process but were spectroscopically identified as contaminants in our sample.
In cases where the acquired spectrum did not match a brown dwarf, we saved images at the telescope to later verify pointing. In the few cases where pointing was questioned, the candidate was added 
back to the candidate list and the observations were attempted again. If pointing was confirmed, and the observed flux was not that of a brown dwarf, we logged the object as an interloper. 

WISE All-Sky photometry for the interlopers is listed in Table~\ref{WISE_all_sky_phot_duds}. 
Figure~\ref{W1W2_W2W3} shows where these sources encroach on the WISE $W1-W2$ versus $W2-W3$ color space of T and Y dwarfs. Although many are located on the periphery, some look like legitimate brown dwarf candidates because they fall squarely in the brown dwarf color-color space.
Interlopers that were observed with {\it Spitzer} are logged in Table~\ref{Spitzer_phot_duds} and shown in Figure~\ref{ch1ch2_W1W2}. 
Only a few of the interlopers are distinguishable from brown dwarfs in this figure.

Spectroscopic observations are listed in Table~\ref{duds} along with descriptions of the spectra. In most cases we observe faint near-IR continuum and no discernible spectral features.
However, some objects are unique and we discuss those sources in the following subsections.

\subsection{Young Stellar Objects}

Two of the interlopers, WISE J0421+3721 and WISE J0546+2639, were selected from candidates in our L-dwarf color selection process ($W1-W2$ $>$ 0.4 mag and no association with a 2MASS source).
Both objects are missing from the 2MASS All-Sky PSC, but are listed in the 2MASS Survey Point Source Reject Table. WISE J0421+3721 has 2MASS magnitudes 
$J$=17.0$\pm$0.2, $H$=15.0$\pm$0.1, and $K_s$=14.1$\pm$0.1 mag. WISE J0546+2639 is flagged as an artifact contaminated source with $J$=15.88$\pm$0.07, $H$=14.50$\pm$0.06, $K_s$=13.26$\pm$0.03 mag. These two sources have
very similar near-IR colors, and consistent $W1-W2$ colors. The IRTF/SpeX spectra of these sources are shown in Figure~\ref{YSOs} with prominent spectral features marked.

We identify these sources as YSOs based on comparisons to the survey of 110 young stellar objects (YSOs) by \citet{connelley2010}. We find that a few of their YSO spectra have similar morphologies to our 
newly identified YSOs in that they that lack emission features, peak around 1.8$\mu$m, and show weak H and CO absorption features. The best spectral match Elias 2-25 (16:26:34.167 $-$24:23:28.26~J2000)
 is a T Tauri star with similar near-IR colors to the new YSOs ($J-H$=1.598, $H-K_s$=0.943, and $J-K_s$=2.542). 
 The other T Tauri objects from \citet{connelley2010} with similar colors and spectral morphologies  are V806 Tau, FS Tau A, and LZK 12. 
 
 Employing the $\alpha$ spectral index, defined as the slope of log($\lambda$F$_\lambda$) versus log($\lambda$) \citep{lada1987}, over the WISE passbands we find $\alpha$=-1.63 and -1.42 for WISE J0421+3721 and WISE J0546+2639, respectively.
 Inclusion of the $K_s$ and $W3$ magnitudes for WISE J0421+3721 produces  $\alpha$=-1.45. These values are consistent with evolved T Tauri stars, between Class II and Class III, in the proximity of Taurus-Auriga.

\subsection{Active Galactic Nuclei}

Although many of the interlopers show faint continuum flux, two of them show spectral features indicative of AGN. 
WISE J1530$-$2617 and WISE J1754+3233 have {\it Spitzer} and WISE colors consistent with reddened extragalactic sources \citep{wright2010,eisenhardt2012,stern2012}.

Figure~\ref{WISE1530_NIRSPEC} shows the Keck/NIRSPEC $J$-band spectrum of WISE J1530$-$2617. We identify H$\alpha$, [NII]$\lambda$6583, and broad [SII]$\lambda$$\lambda$6716, 6731 
at 1.3136, 1.3158 and 1.3313, 1.3334 $\mu$m, respectively. This produces a redshift z=1.00 and line flux ratio N[II]/H$\alpha$ = 1.314, which is typical for an AGN \citep{veilleux1987}. Also, $W1-W2$=3.89$\pm$0.50 for this source, 
which is even redder than the $W1W2$-dropouts shown in Figure 1 of \citet{eisenhardt2012}.

The HST/WFC3 spectrum of WISE J1754+3233 is shown in Figure~\ref{WISE1754_HST}. We determine that z=1.92 for H$\beta$ and [O III]$\lambda$$\lambda$4959, 5007 at 1.4192, 1.4471 and 1.4611$\mu$m, respectively.
Although [O III]$\lambda$5007/H$\beta$ = 1.129 is not unique to AGN, its WISE colors identify it as an AGN \citep{wright2010}. This object is one that was initially identified as a candidate in the Preliminary Data Release, 
but its updated All-Sky Data Release photometry moved it outside our selection criteria (see Figure~\ref{W1W2_W2W3}).

\section{Conclusions}

We present the spectra of $\sim$100 T dwarfs uncovered with the Wide-field Infrared Survey Explorer (WISE).
Spectral types were derived by visual comparison to spectral standards, and modified median-flux spectral indices were computed for WISE brown dwarf discoveries.
We also define the J-narrow index to help better identify future early-type Y dwarf discoveries.
Along with the 89 T dwarfs reported by \citet{kirkpatrick2011}, our follow-up triples the number of known brown dwarfs with spectral types later than T5.

Using the classification schemes in the literature, we identify three extremely red brown dwarfs that can not easily be classified as L or T type.
These objects display $Y$- and $J$-band spectral morphologies most similar to T dwarfs, have red $J-K$ colors, and lack distinct CH$_4$ absorption features.
Although gravity, metallicity, and temperature variations can all contribute to the reddening in these objects, variability seen in WISE J1738+6242 hints at rapid (on the order of a year) atmospheric changes.
Perhaps the discrepant features we see in these extremely red objects are due to an unresolved redL + T binary. 
If the T dwarf component of this system constrains other physical parameters, such as the age or metallicity of the system, this would shed light on the underlying cause of the red L dwarf phenomenon.  
Alternatively, these objects may bridge the L dwarf sequence to the HR 8799 planets and provide a greater understanding of both with further investigation.

Along with the spectral indices, we derive the Y/J, H/J and K/J peak flux ratios for $\sim$320 T and Y dwarfs.
The H/J ratio traces the $J-H$ color and shows a reversal around T7.  
The Y0s stand-out in the H/J index as breaking the trend, and may be less cloudy than the later Y dwarfs. 
Thus, the spectral sequence of the late-type T dwarfs and the early-type Y dwarfs may no longer be correlated directly to temperature, but also to other atmospheric conditions like cloud fraction, metallicity, grain size, and surface gravity.
The K/J index shows a large dispersion of $\sim$1.5 mag, which can be matched by the sulfide cloud models presented by \citet{morley2012} and not by previous cloudless models used for late-type T dwarfs.
We find that although most late-type T dwarfs match the $J$- and $H$-band morphology of the spectral standards, one-third show significantly suppressed $K$-band flux relative to the standards (smaller K/J values), 
which implies higher surface gravities, cloudless atmospheres, and/or low metallicities.
Comparison to atmospheric models reveals that the small Y/J peak flux ratios of late-type T dwarfs are best matched to cloudy models and/or oversolar metallicities. 
Objects with a Y/J index greater than the standards are best matched to cloudless models and/or sub-solar metallicities.
However, we can not disentangle the physical parameters modulating T dwarf colors until model grids are extended to include various cloud species at lower temperatures and non-solar metallicities.

\section{Acknowledgments}

This publication makes use of data products from the Wide-field Infrared Survey Explorer, which is a 
joint project of the University of California, Los Angeles, and the Jet Propulsion Laboratory/California 
Institute of Technology, funded by the National Aeronautics and Space Administration.
We thank the Infrared Processing and Analysis Center at Caltech for funds provided by the Visiting Graduate Fellowship.
Observations presented here were obtained under programs GN-2011A-Q-67 and GN-2011B-Q-7 at the Gemini Observatory, which is operated by the Association of Universities for Research in Astronomy, Inc., 
under a cooperative agreement with the NSF on behalf of the Gemini partnership: the National Science Foundation (United States), the Science and Technology Facilities Council (United Kingdom), the National 
Research Council (Canada), CONICYT (Chile), the Australian Research Council (Australia), Minist\'{e}rio da Ci\^{e}ncia, Tecnologia e Inova\c{c}\~{a}o (Brazil) and Ministerio de Ciencia, Tecnolog\'{i}a e Innovaci\'{o}n Productiva (Argentina).
Some photometry in this paper is based on observations obtained at the Southern Astrophysical Research (SOAR) telescope, which is a joint project of the Minist\'{e}rio da Ci\^{e}ncia, Tecnologia, e Inova\c{c}\~{a}o (MCTI) da Rep\'{u}blica Federativa do Brasil, the U.S. National Optical Astronomy Observatory (NOAO), the University of North Carolina at Chapel Hill (UNC), and Michigan State University (MSU).
This publication also makes use of data products from 2MASS, SDSS, and UKIDSS. 2MASS
is a joint project of the University of Massachusetts and the Infrared Processing and Analysis 
Center/California Institute of Technology, funded by the National Aeronautics and Space Administration 
and the National Science Foundation. SDSS is funded by the Alfred P. Sloan Foundation, the Participating 
Institutions, the National Science Foundation, the U.S. Department of Energy, the National Aeronautics 
and Space Administration, the Japanese Monbukagakusho, the Max Planck Society, and the Higher Education 
Funding Council for England.
UKIDSS uses the Wide Field Camera at the United Kingdom Infrared Telescope atop Mauna Kea, Hawai'i. We
are grateful for the efforts of the instrument, calibration, and pipeline teams that have made the 
UKIDSS data possible.
We acknowledge use of the DSS, which were produced at the Space Telescope Science Institute under
U.S.\ Government grant NAG W-2166. The images of these surveys are based on photographic data
obtained using the Oschin Schmidt Telescope on Palomar Mountain and the UK Schmidt Telescope.
This research has made use of the NASA/IPAC Infrared Science Archive (IRSA),
which is operated by the Jet Propulsion Laboratory, California Institute of Technology, under contract
with the National Aeronautics and Space Administration. Our research has benefited from the M, L, and
T dwarf compendium housed at DwarfArchives.org, whose server was funded by a NASA Small Research Grant, 
administered by the American Astronomical Society.
The Brown Dwarf Spectroscopic Survey (BDSS) provided and essential comparison library for our moderate-resolution spectroscopy.
We are also indebted to the SIMBAD database, operated at CDS, Strasbourg, France. 
This work is based in part on observations made with the {\it Spitzer Space Telescope} , which is
operated by the Jet Propulsion Laboratory, California Institute of Technology, under a contract with
NASA. Support for this work was provided by NASA through an award issued to program 70062 by JPL/Caltech. This work
is also based in part on observations made with the NASA/ESA {\it Hubble Space Telescope}, obtained
at the Space Telescope Science Institute, which is operated by the Association of Universities for
Research in Astronomy, Inc., under NASA contract NAS 5-26555. These observations are associated with 
program 12330. Support for program 12330 was provided by NASA through a grant from the Space
Telescope Science Institute.
Some of the spectroscopic data presented herein were obtained at 
the W.M. Keck Observatory, which is operated as a scientific partnership among 
the California Institute of Technology, the University of California and the 
National Aeronautics and Space Administration. The Observatory was made 
possible by the generous financial support of the W.M. Keck Foundation.
In acknowledgement of our observing time at Keck and the IRTF, 
we further wish to recognize the very significant 
cultural role and reverence that the summit of Mauna Kea has always had within the indigenous Hawai'ian 
community. We are most fortunate to have the opportunity to conduct observations from this mountain.  
We acknowledge use of PAIRITEL, which is operated by the Smithsonian Astrophysical Observatory (SAO) 
and was made possible by a grant from the Harvard University Milton Fund, the camera loaned from the 
University of Virginia, and the continued support of the SAO and UC Berkeley. The PAIRITEL project is 
supported by NASA Grant NNG06GH50G. This paper also includes data gathered with the 6.5 m Magellan
Telescopes located at Las Campanas Observatory, Chile. 
We acknowledge fruitful discussions with Tim Conrow, Roc Cutri, and Frank Masci, and
acknowledge assistance with Magellan/FIRE observations by Emily Bowsher.
We thank the anonymous referee for detailed and thoughtful recommendations for improving this paper prior to publication.

\clearpage

\appendix{\bf{Appendix A:} Integrated Flux Indices for WISE Discovered Brown Dwarfs}

Spectral indices from the literature are defined as integrated flux ratios such that,

\begin{center}
integrated flux index = $\int$F$_{\lambda_1 - \lambda_2}$ d$\lambda$ / $\int$F$_{\lambda_3 - \lambda_4}$ d$\lambda$,\\
\end{center}
where the numerator, integrated between $\lambda_1 - \lambda_2$, covers a specified absorption region and the denominator, 
integrated between $\lambda_3 - \lambda_4$, includes the corresponding peak flux. This results in index values generally less than unity and greater than zero.
In this paper we redefine the spectral indices using the median flux ratios which when defined as,
\begin{center}
median flux index = $\widetilde{F}$$_{\lambda_1 - \lambda_2}$ / $\widetilde{F}$$_{\lambda_3 - \lambda_4}$,\\
\end{center}
for the same wavelengths produce consistent index values with the integrated flux method, but decreased dispersion per spectral type bin.

For each of the WISE discovered brown dwarfs listed in Table~\ref{indices} we first derived the integrated flux ratios from 
\citet{burgasser2006}, \citet{warren2007}, and \citet{delorme2008}. These are listed in Table~\ref{int_indices}. Indices were derived as described in \S4.5, 
using a Monte Carlo method with 1000 realizations of the spectrum. Figure~\ref{Six_Indices_Inst_A} shows the median flux and integrated flux indices 
plotted as a function of spectral type, like Figure~\ref{Six_Indices_Ref}, but color coded by spectrograph. The Y/J, H/J, and K/J flux ratios computed using the median 
flux and integrated flux methods are shown in Figure~\ref{Indices_Bands_Inst_A}, also color coded by spectrograph. 

The integrated flux indices show a clear instrumental bias, where the indices from IRTF/SpeX spectra sit distinctly above the others. 
This offset is the result of differences in spectral resolution between instruments, and between near-infrared bands for the same instrument. 
To properly compare integrated flux spectral indices between instruments we would first need to interpolate the spectra onto similar wavelength scales.
Even for the same instrument, changes in the resolution as a function of wavelength would need to be removed by interpolation.
Using the median flux, rather than the integrated flux, sidesteps this issue and reduces the likelihood of errors when comparing spectra from numerous instruments, like we have done here.
Another, less significant, source of this offset is a negative noise bias of all non-IRTF/SpeX spectra reduced with Spextool. This is clearly seen in Figure~\ref{pec}, where the scatter 
in the per pixel flux is more often below the mean. This bias in the reduced spectrum results in smaller integrated fluxes and consequently, smaller index values. 
Employing the \emph{median} flux instead of the integrated flux reduces data reduction and instrumental influences on the computed indices.

\clearpage
																											
\begin{deluxetable}{lcllllll}																																
\tabletypesize{\tiny}																																																			
\tablewidth{6.4in}																																
\tablenum{1}																																
\tablecaption{Updated WISE All-Sky Photometry for Discoveries from \citet{kirkpatrick2011}\label{WISE_all_sky_phot_100}}																																
\tablehead{																																
\colhead{Object Name} &																																
\colhead{NIR Sp.} &																																
\colhead{$W1$} &																																
\colhead{$W2$} &																																
\colhead{$W3$} &																																
\colhead{$W4$} &																																
\colhead{$W1-W2$} &																																
\colhead{$W2-W3$} \\																																
\colhead{} &																																
\colhead{Type} &																																
\colhead{(mag)} &																																
\colhead{(mag)} &																																
\colhead{(mag)} &																																
\colhead{(mag)} &																																
\colhead{(mag)} &																																
\colhead{(mag)}\\																																
\colhead{(1)} &																																
\colhead{(2)} &																																
\colhead{(3)} &																																
\colhead{(4)} &																																
\colhead{(5)} &																																
\colhead{(6)} &																																
\colhead{(7)} &																																
\colhead{(8)}																																
}																																
\startdata																																
WISE J000849.75$-$173922.9	&	  T6            	&		16.41	$\pm$	0.09	&	14.56	$\pm$	0.07	&	$>$	12.15			&	$>$	8.49			&		1.86	$\pm$	0.12	&	$<$	2.40			\\
WISE J003119.76$-$384036.1	&	  L2 pec (blue) 	&		12.40	$\pm$	0.03	&	12.01	$\pm$	0.02	&		11.56	$\pm$	0.17	&	$>$	8.76			&		0.39	$\pm$	0.04	&		0.45	$\pm$	0.17	\\
WISE J004928.48+044059.9	&	  L9            	&		13.45	$\pm$	0.03	&	12.97	$\pm$	0.03	&		12.01	$\pm$	0.29	&		8.39	$\pm$	0.27	&		0.48	$\pm$	0.04	&		0.96	$\pm$	0.29	\\
WISE J010637.07+151852.8	&	  M8 pec        	&		13.08	$\pm$	0.03	&	12.73	$\pm$	0.03	&		12.49	$\pm$	0.37	&	$>$	8.81			&		0.35	$\pm$	0.04	&		0.24	$\pm$	0.38	\\
WISE J012333.23+414203.8	&	  T7            	&		17.04	$\pm$	0.11	&	15.03	$\pm$	0.07	&		12.99	$\pm$	0.45	&	$>$	9.20			&		2.01	$\pm$	0.14	&		2.03	$\pm$	0.46	\\
WISE J013836.58$-$032221.2	&	  T3            	&		14.43	$\pm$	0.03	&	13.40	$\pm$	0.03	&		12.35	$\pm$	0.35	&	$>$	9.30			&		1.03	$\pm$	0.05	&		1.05	$\pm$	0.35	\\
WISE J014807.25$-$720258.7	&	  T9.5          	&	$>$	18.94			&	14.69	$\pm$	0.05	&	$>$	12.86			&	$>$	9.52			&	$>$	4.25			&	$<$	1.83			\\
WISE J015010.89+382724.1	&	  T0            	&		13.65	$\pm$	0.03	&	12.98	$\pm$	0.03	&		11.74	$\pm$	0.19	&	$>$	9.04			&		0.67	$\pm$	0.04	&		1.24	$\pm$	0.20	\\
\enddata		
\tablecomments{\citet{scholz2011} independently discovered the T9 dwarf WISE J1741+2553, first announced in \citet{gelino2011}, and also noted WISE J0254+0223 as a proper motion object of presumably late type.}		
\end{deluxetable}																																																																																						

\clearpage

\begin{deluxetable}{lllllll}																														
\tabletypesize{\tiny}																																														
\tablewidth{5.7in}																														
\tablenum{2}																														
\tablecaption{WISE All-Sky Photometry for Brown Dwarf Discoveries from This Paper\label{WISE_all_sky_phot_Ts}}																														
\tablehead{																														
\colhead{Object Name} &																														
\colhead{$W1$} &																														
\colhead{$W2$} &																														
\colhead{$W3$} &																														
\colhead{$W4$} &																														
\colhead{$W1-W2$} &																														
\colhead{$W2-W3$} \\																														
\colhead{} &																														
\colhead{(mag)} &																														
\colhead{(mag)} &																														
\colhead{(mag)} &																														
\colhead{(mag)} &																														
\colhead{(mag)} &																														
\colhead{(mag)}\\																														
\colhead{(1)} &																														
\colhead{(2)} &																														
\colhead{(3)} &																														
\colhead{(4)} &																														
\colhead{(5)} &																														
\colhead{(6)} &																														
\colhead{(7)}																														
}																														
\startdata																														
WISE J000517.48+373720.5   	&		16.88	$\pm$	0.12	&	13.27	$\pm$	0.03	&		11.99	$\pm$	0.24	&	$>$	9.23			&		3.61	$\pm$	0.12	&		1.28	$\pm$	0.24	\\
WISE J003829.05+275852.1   	&		18.04	$\pm$	0.29	&	14.38	$\pm$	0.05	&		12.46	$\pm$	0.31	&	$>$	8.78			&		3.67	$\pm$	0.29	&		1.91	$\pm$	0.32	\\
WISE J003830.54+840517.7	&		17.53	$\pm$	0.17	&	15.34	$\pm$	0.09	&	$>$	13.18			&	$>$	9.53			&		2.19	$\pm$	0.19	&	$<$	2.16			\\
WISE J004024.88+090054.8   	&		16.06	$\pm$	0.08	&	13.92	$\pm$	0.07	&	$>$	12.06			&	$>$	8.65			&		2.13	$\pm$	0.11	&	$<$	1.87			\\
WISE J004542.56+361139.1   	&		15.34	$\pm$	0.04	&	13.73	$\pm$	0.03	&		12.61	$\pm$	0.35	&	$>$	9.28			&		1.61	$\pm$	0.05	&		1.12	$\pm$	0.35	\\
WISE J004945.61+215120.0   	&		15.47	$\pm$	0.04	&	12.97	$\pm$	0.03	&		11.73	$\pm$	0.17	&	$>$	9.21			&		2.50	$\pm$	0.05	&		1.24	$\pm$	0.17	\\
WISE J013525.64+171503.4   	&		17.19	$\pm$	0.15	&	14.74	$\pm$	0.07	&	$>$	12.73			&	$>$	8.86			&		2.45	$\pm$	0.17	&	$<$	2.02			\\
WISE J021010.25+400829.6   	&		16.29	$\pm$	0.09	&	14.74	$\pm$	0.08	&	$>$	12.66			&	$>$	9.17			&		1.55	$\pm$	0.12	&	$<$	2.08			\\
\enddata																														
\tablenotetext{a}{~L dwarf candidate, \citet{aberasturi2011}}																														
\tablenotetext{b}{ULAS J095047.28+011734.3, \citet{leggett2012} }																														
\tablenotetext{c}{IR source, \citet{andrei2011} }																														
\tablenotetext{d}{~LHS 2803B, \citet{deacon2012} }																														
\tablenotetext{e}{BD +01\degr 2920B, \citet{pinfield2012} }																														
\tablenotetext{f}{T7 candidates, \citet{scholz2010}}																														
\tablenotetext{g}{PSO J226.2599$-$28.8959, PSO J246.4222+15.4698, PSO J247.3273+03.5932, \citet{deacon2011}}																														
\tablenotetext{h}{VHS J154352.78$-$043909.6, \citet{lodieu2012} }																														
\tablenotetext{i}{WISE J180901.07+383805.4, \citet{luhman2012} }																														
\tablenotetext{j}{A nearby object caused source confusion in the pipeline used to generate the All-Sky Catalog, resulting in a non-detection of this object. The magnitudes reported here were taken from the first-pass processing operations co-add Source Working Database.}																														
\end{deluxetable}																																																												
\clearpage

\begin{deluxetable}{lcccccccccc}																																											
\tabletypesize{\tiny}																																											
\rotate																																											
\tablewidth{7.7in}																																											
\tablenum{3}																																											
\tablecaption{Additional Follow-up 2MASS, MKO and IRAC Photometry for Discoveries from \citet{kirkpatrick2011}\label{first100_more_followup_photometry}}			
\tablehead{
\colhead{Object} &																																											
\colhead{Obs.} &																																											
\multicolumn{3}{c}{Two Micron All-Sky Survey filter system} &																																											
\multicolumn{2}{c}{Mauna Kea Observatories filter system} &																																											
\multicolumn{3}{c}{{\it Spitzer}/IRAC observations} \\																																											
\colhead{Name} &																																											
\colhead{Ref.} &																																											
\colhead{$J$} &																																											
\colhead{$H$} &																																											
\colhead{$K_s$} &																																											
\colhead{$J$} &																																											
\colhead{$H$} &																																											
\colhead{$ch1$} &																																											
\colhead{$ch2$} &																																											
\colhead{$ch1-ch2$} \\																																											
\colhead{} &																																											
\colhead{} &																																											
\colhead{(mag)} &																																											
\colhead{(mag)} &																																											
\colhead{(mag)} &																																											
\colhead{(mag)} &																																											
\colhead{(mag)} &																																											
\colhead{(mag)} &																																											
\colhead{(mag)} &																																											
\colhead{(mag)} & \\																																											
\colhead{(1)} &																																											
\colhead{(2)} &																																											
\colhead{(3)} &																																											
\colhead{(4)} &																																											
\colhead{(5)} &																																											
\colhead{(6)} &																																											
\colhead{(7)} &																																											
\colhead{(8)} &																																											
\colhead{(9)} &																																											
\colhead{(10)}																																											
}																																											
\startdata																																											
WISE J0448$-$1935	&	---	&	---				&	---				&	---				&	---				&	---				&		15.37	$\pm$	0.02	&		14.16	$\pm$	0.02	&		1.21	$\pm$	0.03	\\
WISE J0656+4205	&	---	&	---				&	---				&	---				&	---				&	---				&		13.64	$\pm$	0.02	&		13.26	$\pm$	0.02	&		0.38	$\pm$	0.03	\\
WISE J0744+5628	&	---	&	---				&	---				&	---				&	---				&	---				&		16.27	$\pm$	0.03	&		14.55	$\pm$	0.02	&		1.71	$\pm$	0.04	\\
WISE J0759$-$4904 	&	3	&	---				&	---				&	---				&		17.38	$\pm$	0.05	&		17.41	$\pm$	0.04	&	---				&	---				&	---				\\
WISE J0821+1443	&	---	&	---				&	---				&	---				&	---				&	---				&		15.59	$\pm$	0.03	&		14.51	$\pm$	0.02	&		1.08	$\pm$	0.03	\\
WISE J0836$-$1859 	&	1	&		18.99	$\pm$	0.22	&		18.79	$\pm$	0.26	&	$>$	18.48			&	---				&	---				&	---				&	---				&	---				\\
WISE J1018$-$2445	&	1	&		17.82	$\pm$	0.12	&		18.00	$\pm$	0.23	&		17.81	$\pm$	0.31	&	---				&	---				&	---				&	---				&	---				\\
WISE J1042$-$3842	&	4	&	---				&	---				&	---				&		18.98	$\pm$	0.09	&		19.08	$\pm$	0.11	&	---				&	---				&	---				\\
\enddata																																											
\tablecomments{Ref.\ for JHK instrument or catalog: 																																											
2MASS filter system: 																																											
(1) Bigelow/2MASS,																																											
(2) PAIRITEL;																																											
MKO filter system:																																											
(3) Magellan/PANIC,																																											
(4) CTIO-4m/NEWFIRM}																																											
\end{deluxetable}																																																																																																																																																																		

\clearpage

\begin{deluxetable}{lccccccccccc}																																											
\tabletypesize{\tiny}																																											
\rotate																																											
\tablewidth{7.7in}																																											
\tablenum{4}																																											
\tablecaption{Follow-up 2MASS, MKO and IRAC Photometry for New Brown Dwarf Discoveries\label{Ts_followup_photometry}}																																											
\tablehead{																																											
\colhead{Object} &																																											
\colhead{Obs.} &																																											
\multicolumn{3}{c}{Two Micron All-Sky Survey filter system} &																																											
\multicolumn{2}{c}{Mauna Kea Observatories filter system} &																																											
\multicolumn{3}{c}{{\it Spitzer}/IRAC observations} \\																																											
\colhead{Name} &																																											
\colhead{Ref.} &																																											
\colhead{$J$} &																																											
\colhead{$H$} &																																											
\colhead{$K_s$} &																																											
\colhead{$J$} &																																											
\colhead{$H$} &																																											
\colhead{$ch1$} &																																											
\colhead{$ch2$} &																																											
\colhead{$ch1-ch2$} \\																																											
\colhead{} &																																											
\colhead{} &																																											
\colhead{(mag)} &																																											
\colhead{(mag)} &																																											
\colhead{(mag)} &																																											
\colhead{(mag)} &																																											
\colhead{(mag)} &																																											
\colhead{(mag)} &																																											
\colhead{(mag)} &																																											
\colhead{(mag)} & \\																																											
\colhead{(1)} &																																											
\colhead{(2)} &																																											
\colhead{(3)} &																																											
\colhead{(4)} &																																											
\colhead{(5)} &																																											
\colhead{(6)} &																																											
\colhead{(7)} &																																											
\colhead{(8)} &																																											
\colhead{(9)} &																																											
\colhead{(10)}}																																											
\startdata																																											
WISE J0005+3737   	&	9	&	---				&	---				&	---				&		18.33	$\pm$	0.12	&		18.10	$\pm$	0.08	&		15.43	$\pm$	0.02	&		13.28	$\pm$	0.02	&		2.15	$\pm$	0.03	\\
WISE J0038+2758   	&	5	&	---				&	---				&	---				&		18.61	$\pm$	0.02	&		18.92	$\pm$	0.04	&		16.45	$\pm$	0.04	&		14.41	$\pm$	0.02	&		2.04	$\pm$	0.04	\\
WISE J0038+8405   	&	4	&		17.53	$\pm$	0.06	&		17.06	$\pm$	0.11	&	$>$	16.50			&	---				&	---				&		16.33	$\pm$	0.04	&		14.99	$\pm$	0.02	&		1.34	$\pm$	0.04	\\
WISE J0040+0900 	&	1	&		16.50	$\pm$	0.12	&		16.54	$\pm$	0.27	&	$>$	15.74			&	---				&	---				&		15.07	$\pm$	0.02	&		13.76	$\pm$	0.02	&		1.31	$\pm$	0.03	\\
	&	7\tablenotemark{a} 	&	---				&	---				&	---				&	---				&		16.56		0.02	&	---				&	---				&	---				\\
WISE J0045+3611	&	1	&		16.15	$\pm$	0.08	&		16.04	$\pm$	0.15	&		15.80	$\pm$	0.20	&	---				&	---				&		14.68	$\pm$	0.02	&		13.75	$\pm$	0.02	&		0.94	$\pm$	0.03	\\
WISE J0049+2151   	&	1	&		16.72	$\pm$	0.16	&	$>$	16.51			&	$>$	15.86			&	---				&	---				&		15.01	$\pm$	0.02	&		13.04	$\pm$	0.02	&		1.97	$\pm$	0.03	\\
WISE J0135+1715   	&	---	&	---				&	---				&	---				&	---				&	---				&		16.08	$\pm$	0.03	&		14.57	$\pm$	0.02	&		1.51	$\pm$	0.04	\\
\enddata
\end{deluxetable}																																																																																																																																																																																																																

\clearpage

\begin{deluxetable}{lccccc}												
\tabletypesize{\scriptsize}												
\tablewidth{5.6in}												
\tablenum{5}												
\tablecaption{Follow-up Near-IR Spectroscopy and Spectral Classifications for T Dwarf Discoveries \label{spec_log}}												
\tablehead{												
\colhead{Object Name} &												
\colhead{NIR Sp.} &												
\colhead{Spectrograph} &												
\colhead{Int.\ Time} &												
\colhead{Tell.\ Corr.} &												
\colhead{Obs.\ Date} \\												
\colhead{and J2000 Coordinates} &												
\colhead{Type} &												
\colhead{} &												
\colhead{(s)} &												
\colhead{Star} &												
\colhead{(UT)} \\												
\colhead{(1)} &												
\colhead{(2)} &												
\colhead{(3)} &												
\colhead{(4)} &												
\colhead{(5)} &												
\colhead{(6)}}												
\startdata												
WISE J0005+3737   	&	T9	&	 Keck/NIRSPEC-J  	&	2400	&	HD 9711	&	 2011 Oct 07              	\\	
WISE J0038+2758   	&	T9	&	 Keck/NIRSPEC-J  	&	4800	&	HD 7215	&	 2011 Sep 08              	\\	
WISE J0038+8405   	&	T6	&	 Keck/NIRSPEC-J  	&	3600	&	HD 33541	&	 2011 Aug 09              	\\	
WISE J0040+0900   	&	T7	&	 IRTF/SpeX     	&	1200	&	HD 7215	&	 2011 Nov 30 	\\	
WISE J0045+3611   	&	T5	&	 IRTF/SpeX     	&	1200	&	HD 9711	&	 2011 Aug 25 	\\	
WISE J0049+2151   	&	T8.5	&	 IRTF/SpeX     	&	1200	&	HD 9711	&	 2011 Aug 25 	\\	
WISE J0135+1715   	&	T6	&	 IRTF/SpeX     	&	1200	&	HD 19600	&	 2011 Aug 25 	\\	
WISE J0210+4008   	&	T4.5	&	 IRTF/SpeX     	&	1200	&	HD 19600	&	 2011 Nov 30 	\\
\enddata												
\tablecomments{Classification notes: `pec' - peculiar, `sb' - spectral binary, `:' - low SNR, `extremely red' - discussed in \S 4.2.}												
\end{deluxetable}																																																						

\clearpage

\begin{deluxetable}{lccccccl}															
\tabletypesize{\scriptsize}															
\rotate															
\tablewidth{8in}															
\tablenum{6}															
\tablecaption{New Near-IR Spectroscopy and Spectral Classifications for Other Brown Dwarfs from the Literature\label{repeat_spec}}															
\tablehead{															
\colhead{Object Name} &															
\colhead{NIR Sp.} &															
\colhead{Spectrograph} &															
\colhead{Int.\ Time} &															
\colhead{Tell.\ Corr.} &															
\colhead{Obs.\ Date} &															
\colhead{Published} &															
\colhead{Discovery}\\															
\colhead{} &															
\colhead{Type} &															
\colhead{} &															
\colhead{(s)} &															
\colhead{Star} &															
\colhead{(UT)} &															
\colhead{NIR Sp. Type} &															
\colhead{Paper}\\															
\colhead{(1)} &															
\colhead{(2)} &															
\colhead{(3)} &															
\colhead{(4)} &															
\colhead{(5)} &															
\colhead{(6)} &															
\colhead{(7)} &															
\colhead{(8)}}															
\startdata															
WISE J0920+4538   	&	L9 sb?	&	 Palomar/TSpec 	&	1440	&	HD 89239	&	 2012 Jan 07	&	L4/5 Candidate	&	\citet{aberasturi2011}	\\
ULAS J095047.28+011734.3	&	T8	&	 Magellan/FIRE 	&	1080	&	HD 75159	&	 2012 May 05	&	T8	&	\citet{leggett2012}	\\
WISE J1246$-$3139 	&	T1	&	 Palomar/TSpec 	&	1440	&	HD 109134	&	 2012 Jan 07	&	IR Source	&	\citet{andrei2011}	\\
LHS 2803B	&	T5.5	&	 IRTF/SpeX     	&	1440	&	HD 134013	&	 2012 May 13	&	T5.5	&	\citet{deacon2012}	\\
BD +01\degr 2920B	&	T8	&	 Magellan/FIRE 	&	1440	&	HD 113147	&	 2012 May 05	&	T8p	&	\citet{pinfield2012}	\\
WISE J1449+1147   	&	T5 pec	&	 Palomar/TSpec 	&	3600	&	HD 131951	&	 2011 May 12 	&	T7 Candidate	&	\citet{scholz2010}	\\
	&	T5 pec	&	 Gemini/GNIRS  	&	2400	&	HD 147295	&	 2011 Jul 04	&	---	&	---	\\
PSO J226.2599$-$28.8959	&	T2:	&	 Palomar/TSpec 	&	2400	&	HD 133652	&	 2011 May 13              	&	T1.5	&	\citet{deacon2011}	\\
\enddata															
\tablecomments{Classification notes: `pec' - peculiar, `sb' - spectral binary, `:' - low SNR.}															
\end{deluxetable}																																																				

\clearpage

\begin{deluxetable}{lccccccccc}																																	
\tabletypesize{\tiny}																																	
\rotate																																	
\tablewidth{8.2in}																																	
\tablenum{7}																																	
\tablecaption{Modified Spectral Indices Based on Median Fluxes \label{indices}}																																	
\tablehead{																																	
\colhead{Object Name} &																																	
\colhead{Instrument} &																																	
\colhead{NIR Sp.} &																																	
\colhead{H$_2$O$-$J} & 																																	
\colhead{CH$_4$$-$J} &																																	
\colhead{H$_2$O$-$H} &																																	
\colhead{CH$_4$$-$H} &																																	
\colhead{W$_J$} &																																	
\colhead{NH$_3$$-$H} &																																	
\colhead{J-narrow}\\																																	
\colhead{J2000 Coordinates} &																																	
\colhead{} & 																																	
\colhead{Type} &																																	
\colhead{} & 																																	
\colhead{} &																																	
\colhead{} &																																	
\colhead{} &																																	
\colhead{} &																																	
\colhead{} &																																	
\colhead{} \\																																	
\colhead{(1)} &																																	
\colhead{(2)} &																																	
\colhead{(3)} &																																	
\colhead{(4)} &																																	
\colhead{(5)} &																																	
\colhead{(6)} &																																	
\colhead{(7)} &																																	
\colhead{(8)} &																																	
\colhead{(9)} &																																	
\colhead{(10)}}																																	
\startdata																																	
\\                                                                                                                                                   																																	
\multicolumn{10}{c}{T and Y Dwarf Standards} \\  																																	
\\  																																	
SDSS J1207+0244	&	IRTF/SpeX\tablenotemark{a}	&	T0	&	0.622	$\pm$	0.005	&	0.847	$\pm$	0.007	&	0.622	$\pm$	0.008	&	0.958	$\pm$	0.006	&	0.810	$\pm$	0.016	&	0.864	$\pm$	0.008	&	0.962	$\pm$	0.008	\\
SDSS J0151+1244	&	IRTF/SpeX\tablenotemark{a}	&	T1	&	0.654	$\pm$	0.008	&	0.714	$\pm$	0.015	&	0.628	$\pm$	0.013	&	0.954	$\pm$	0.010	&	0.856	$\pm$	0.020	&	0.864	$\pm$	0.014	&	0.955	$\pm$	0.012	\\
SDSSp J1254$-$0122	&	IRTF/SpeX\tablenotemark{a}	&	T2	&	0.477	$\pm$	0.008	&	0.756	$\pm$	0.012	&	0.476	$\pm$	0.012	&	0.909	$\pm$	0.010	&	0.752	$\pm$	0.024	&	0.778	$\pm$	0.016	&	0.930	$\pm$	0.011	\\
2MASS J1209$-$1004	&	IRTF/SpeX\tablenotemark{a}	&	T3	&	0.415	$\pm$	0.008	&	0.661	$\pm$	0.014	&	0.436	$\pm$	0.012	&	0.715	$\pm$	0.011	&	0.688	$\pm$	0.024	&	0.789	$\pm$	0.017	&	0.897	$\pm$	0.015	\\
2MASSI J2254+3123	&	IRTF/SpeX\tablenotemark{a}	&	T4	&	0.372	$\pm$	0.005	&	0.628	$\pm$	0.007	&	0.377	$\pm$	0.007	&	0.587	$\pm$	0.008	&	0.634	$\pm$	0.008	&	0.766	$\pm$	0.008	&	0.879	$\pm$	0.006	\\
\enddata																																	
\tablenotetext{a}{Spectra from Adam Burgasser's SpeX Prism Spectral Library.}																																	
\end{deluxetable}																																																																																																																																																

\clearpage

\begin{deluxetable}{ccccccccccc}																																									
\tabletypesize{\tiny}																																									
\rotate																																									
\tablewidth{8.4in}																																									
\tablenum{8}																																									
\tablecaption{Average Median-Flux Spectral Indices for Each Half Spectral Type \label{avg_indices}}																																									
\tablehead{																																									
\colhead{NIR Sp.} &																																									
\colhead{H$_2$O$-$J} & 																																									
\colhead{CH$_4$$-$J} &																																									
\colhead{H$_2$O$-$H} &																																									
\colhead{CH$_4$$-$H} &																																									
\colhead{W$_J$} &																																									
\colhead{NH$_3$$-$H} &																																									
\colhead{J-narrow} &																																									
\colhead{Y/J} &																																									
\colhead{H/J} &																																									
\colhead{K/J}\\																																									
\colhead{Type} &																																									
\colhead{} & 																																									
\colhead{} &																																									
\colhead{} &																																									
\colhead{} &																																									
\colhead{} &																																									
\colhead{} &																																									
\colhead{} &																																									
\colhead{} \\																																									
\colhead{(1)} &																																									
\colhead{(2)} &																																									
\colhead{(3)} &																																									
\colhead{(4)} &																																									
\colhead{(5)} &																																									
\colhead{(6)} &																																									
\colhead{(7)} &																																									
\colhead{(8)} &																																									
\colhead{(9)} &																																									
\colhead{(10)} &																																									
\colhead{(11)}}																																									
\startdata																																									
T0	&	0.647	$\pm$	0.054	&	0.841	$\pm$	0.063	&	0.651	$\pm$	0.036	&	0.970	$\pm$	0.054	&	0.803	$\pm$	0.030	&	0.884	$\pm$	0.021	&	0.928	$\pm$	0.024	&	0.579	$\pm$	0.033	&	0.846	$\pm$	0.051	&	0.496	$\pm$	0.060	\\
T0.5	&	0.581	$\pm$	0.087	&	0.802	$\pm$	0.050	&	0.591	$\pm$	0.074	&	1.035	$\pm$	0.100	&	0.790	$\pm$	0.036	&	0.850	$\pm$	0.039	&	0.934	$\pm$	0.028	&	0.567	$\pm$	0.040	&	0.854	$\pm$	0.037	&	0.554	$\pm$	0.121	\\
T1	&	0.572	$\pm$	0.079	&	0.772	$\pm$	0.064	&	0.582	$\pm$	0.041	&	0.912	$\pm$	0.068	&	0.766	$\pm$	0.080	&	0.853	$\pm$	0.018	&	0.922	$\pm$	0.028	&	0.557	$\pm$	0.076	&	0.814	$\pm$	0.033	&	0.445	$\pm$	0.053	\\
T1.5	&	0.537	$\pm$	0.068	&	0.728	$\pm$	0.067	&	0.518	$\pm$	0.045	&	0.924	$\pm$	0.059	&	0.777	$\pm$	0.040	&	0.840	$\pm$	0.051	&	0.937	$\pm$	0.017	&	0.585	$\pm$	0.051	&	0.734	$\pm$	0.023	&	0.344	$\pm$	0.033	\\
T2	&	0.490	$\pm$	0.071	&	0.707	$\pm$	0.077	&	0.522	$\pm$	0.052	&	0.861	$\pm$	0.064	&	0.712	$\pm$	0.055	&	0.828	$\pm$	0.045	&	0.916	$\pm$	0.030	&	0.532	$\pm$	0.039	&	0.732	$\pm$	0.035	&	0.370	$\pm$	0.060	\\
T2.5	&	0.457	$\pm$	0.037	&	0.679	$\pm$	0.055	&	0.490	$\pm$	0.033	&	0.813	$\pm$	0.059	&	0.697	$\pm$	0.040	&	0.813	$\pm$	0.032	&	0.914	$\pm$	0.007	&	0.537	$\pm$	0.023	&	0.687	$\pm$	0.030	&	0.314	$\pm$	0.026	\\
T3	&	0.410	$\pm$	0.035	&	0.625	$\pm$	0.061	&	0.485	$\pm$	0.038	&	0.722	$\pm$	0.047	&	0.685	$\pm$	0.048	&	0.802	$\pm$	0.042	&	0.911	$\pm$	0.019	&	0.563	$\pm$	0.053	&	0.660	$\pm$	0.030	&	0.281	$\pm$	0.046	\\
T3.5	&	0.368	$\pm$	0.096	&	0.604	$\pm$	0.071	&	0.390	$\pm$	0.042	&	0.702	$\pm$	0.042	&	0.621	$\pm$	0.072	&	0.759	$\pm$	0.038	&	0.900	$\pm$	0.026	&	0.510	$\pm$	0.018	&	0.611	$\pm$	0.027	&	0.233	$\pm$	0.032	\\
\enddata																																									
\end{deluxetable}																																																																																																																																																										

\clearpage

\begin{deluxetable}{llllllll}																															
\tabletypesize{\tiny}																																													
\tablewidth{5.6in}																															
\tablenum{9}																															
\tablecaption{WISE All-Sky Photometry for Interloper Sources\label{WISE_all_sky_phot_duds}}																															
\tablehead{																															
\colhead{Object Name} &																															
\colhead{$W1$} &																															
\colhead{$W2$} &																															
\colhead{$W3$} &																															
\colhead{$W4$} &																															
\colhead{$W1-W2$} &																															
\colhead{$W2-W3$} \\																															
\colhead{} &																															
\colhead{(mag)} &																															
\colhead{(mag)} &																															
\colhead{(mag)} &																															
\colhead{(mag)} &																															
\colhead{(mag)} &																															
\colhead{(mag)}\\																															
\colhead{(1)} &																															
\colhead{(2)} &																															
\colhead{(3)} &																															
\colhead{(4)} &																															
\colhead{(5)} &																															
\colhead{(6)} &																															
\colhead{(7)} 																															
}																															
\startdata																															
WISE J004733.81+671834.8   	&		17.00	$\pm$	0.15	&	14.21	$\pm$	0.06	&	$>$	12.01			&	$>$	8.82			&		2.79	$\pm$	0.16	&	$<$	2.20			\\	
WISE J014837.57$-$104803.0	&		16.54	$\pm$	0.09	&	14.56	$\pm$	0.06	&		12.67	$\pm$	0.40	&	$>$	9.38			&		1.98	$\pm$	0.11	&		1.89	$\pm$	0.41	\\	
WISE J020110.08+483633.7	&		17.41	$\pm$	0.22	&	14.67	$\pm$	0.07	&		12.05	$\pm$	0.27	&	$>$	9.13			&		2.74	$\pm$	0.23	&		2.62	$\pm$	0.28	\\	
WISE J022058.58+620236.4	&	$>$	17.42			&	13.04	$\pm$	0.03	&	$>$	11.88			&		7.18	$\pm$	0.12	&	$>$	4.39			&	$<$	1.16			\\	
WISE J022116.78+590337.7	&		16.14	$\pm$	0.09	&	14.07	$\pm$	0.06	&		11.01	$\pm$	0.10	&		8.06	$\pm$	0.17	&		2.07	$\pm$	0.10	&		3.06	$\pm$	0.11	\\	
WISE J040450.69+261825.1	&		14.23	$\pm$	0.03	&	13.32	$\pm$	0.04	&	$>$	12.65			&		8.07	$\pm$	0.22	&		0.91	$\pm$	0.05	&	$<$	0.66			\\	
WISE J042102.74+372125.2	&		13.72	$\pm$	0.03	&	13.30	$\pm$	0.04	&		11.63	$\pm$	0.27	&		8.59	$\pm$	0.45	&		0.42	$\pm$	0.05	&		1.67	$\pm$	0.27	\\	
WISE J044627.93$-$072611.6	&		16.09	$\pm$	0.06	&	14.22	$\pm$	0.05	&			---		&			---		&		1.87	$\pm$	0.08	&			---		\\
\enddata																															
\end{deluxetable}																																																																																																																									

\clearpage

\begin{deluxetable}{lcccc}													
\tabletypesize{\scriptsize}													
\tablewidth{3.5in}													
\tablenum{10}													
\tablecaption{{\it Spitzer}/IRAC Photometry for Interloper Sources \label{Spitzer_phot_duds}}													
\tablehead{													
\colhead{Object Name} &													
\colhead{$ch1$} &													
\colhead{$ch2$} &													
\colhead{$ch1-ch2$}\\													
\colhead{} &													
\colhead{(mag)} &													
\colhead{(mag)} &													
\colhead{(mag)}\\													
\colhead{(1)} &													
\colhead{(2)} &													
\colhead{(3)} &													
\colhead{(4)}}													
\startdata													
WISE J0148$-$1048	&	16.47	$\pm$	0.04	&	14.63	$\pm$	0.02	&	1.84	$\pm$	0.04	\\
WISE J0201+4836	&	16.54	$\pm$	0.04	&	14.70	$\pm$	0.02	&	1.84	$\pm$	0.04	\\
WISE J0220+6202	&	14.79	$\pm$	0.02	&	13.05	$\pm$	0.02	&	1.75	$\pm$	0.03	\\
WISE J0221+5903	&	15.06	$\pm$	0.02	&	13.93	$\pm$	0.02	&	1.13	$\pm$	0.03	\\
WISE J0446$-$0726	&	15.59	$\pm$	0.03	&	14.35	$\pm$	0.02	&	1.25	$\pm$	0.03	\\
WISE J0543+4653	&	15.91	$\pm$	0.03	&	13.64	$\pm$	0.02	&	2.27	$\pm$	0.03	\\
WISE J0546+2639  	&	12.88	$\pm$	0.02	&	12.67	$\pm$	0.02	&	0.21	$\pm$	0.02	\\
WISE J0554+2457	&	14.94	$\pm$	0.02	&	14.10	$\pm$	0.02	&	0.84	$\pm$	0.03	\\
\enddata													
\end{deluxetable}																																													

\clearpage

\begin{deluxetable}{lccl}						
\tabletypesize{\scriptsize}						
\rotate						
\tablewidth{7.6in}						
\tablenum{11}						
\tablecaption{Follow-up Near-IR Spectroscopy for Interloper Sources \label{duds}}						
\tablehead{						
\colhead{Object Name} &						
\colhead{Spectrograph} &						
\colhead{Obs.\ Date} & 						
\colhead{Description of Spectrum}\\						
\colhead{J2000 Coordinates} &						
\colhead{} &						
\colhead{(UT)} &						
\colhead{}\\						
\colhead{(1)} &						
\colhead{(2)} &						
\colhead{(3)} &						
\colhead{(4)}}						
\startdata						
WISE J0047+6718   	&	 Keck/NIRSPEC-H	&	 2010 Jul 18	&	Faint continuum throughout the $H$ band\\
WISE J0148$-$1048 	&	 Keck/NIRSPEC-J	&	 2011 Aug 09	&	Faint continuum; cluster of {\it Spitzer} sources within 0$\farcs$5 \\
WISE J0201+4836   	&	 Keck/NIRSPEC-J  	&	 2011 Oct 08/09	&	Faint continuum; two blended {\it Spitzer} sources \\
WISE J0220+6202   	&	 Palomar/TSpec 	&	 2012 Jan 07	&	Faint $K$-band continuum\\
WISE J0221+5903   	&	 Palomar/TSpec 	&	 2011 Nov 08	&	Faint $K$-band continuum; no motion between WISE and {\it Spitzer} epochs\\
WISE J0404+2618   	&	 Palomar/TSpec 	&	 2012 Jan 06	&	$K$-band and faint $H$-band continuum\\
WISE J0421+3721   	&	 IRTF/SpeX     	&	 2012 Feb 01	&	YSO with high SNR hump throughout NIR; peak flux near 1.8~$\mu$m (Figure~\ref{YSOs})\\
WISE J0446$-$0726 	&	 Palomar/TSpec 	&	 2012 Jan 07	&	Faint $H$ and $K$-band continuum; no motion between WISE and {\it Spitzer} epochs\\
\enddata						
\end{deluxetable}																			
\clearpage

\begin{deluxetable}{lcccccccc}																																			
\tabletypesize{\tiny}																																			
\rotate																																			
\tablewidth{7.5in}																																			
\tablenum{A1}																																			
\tablecaption{Appendix: Spectral Indices using Integrated (not Median) Fluxes \label{int_indices}}																																			
\tablehead{																																			
\colhead{Object Name} &																																			
\colhead{Instrument} &																																			
\colhead{NIR Sp.} &																																			
\colhead{H$_2$O$-$J} & 																																			
\colhead{CH$_4$$-$J} &																																			
\colhead{H$_2$O$-$H} &																																			
\colhead{CH$_4$$-$H} &																																			
\colhead{W$_J$} &																																			
\colhead{NH$_3$$-$H} \\																																			
\colhead{J2000 Coordinates} &																																			
\colhead{} & 																																			
\colhead{Type} &																																			
\colhead{} & 																																			
\colhead{} &																																			
\colhead{} &																																			
\colhead{} &																																			
\colhead{} &																																			
\colhead{} \\																																			
\colhead{(1)} &																																			
\colhead{(2)} &																																			
\colhead{(3)} &																																			
\colhead{(4)} &																																			
\colhead{(5)} &																																			
\colhead{(6)} &																																			
\colhead{(7)} &																																			
\colhead{(8)} &																																			
\colhead{(9)}}																																			
\startdata																																			
																																			
\\                                                                                                                                                   																																			
\multicolumn{9}{c}{T Dwarf Discoveries from \citet{kirkpatrick2011}} \\  																																			
\\ 																																			
																																			
WISE J0008$-$1739 	&	 Keck/NIRSPEC-J	&	  T6            	&	0.141	$\pm$	0.025		&	0.393	$\pm$	0.028		&		---			&		---			&	0.463	$\pm$	0.021		&		---			\\
WISE J0123+4142   	&	 Keck/NIRSPEC-J	&	  T7            	&		---			&	0.285	$\pm$	0.051		&		---			&		---			&	0.419	$\pm$	0.041		&		---			\\
WISE J0138$-$0322 	&	 IRTF/SpeX     	&	  T3            	&	0.409	$\pm$	0.006		&	0.502	$\pm$	0.004		&	0.536	$\pm$	0.006		&	0.831	$\pm$	0.007		&	0.655	$\pm$	0.005		&	0.846	$\pm$	0.008		\\
WISE J0148$-$7202 	&	 Magellan/FIRE 	&	  T9.5          	&	-0.017	$\pm$	0.011	\tablenotemark{a}	&	0.076	$\pm$	0.006	\tablenotemark{a}	&	0.070	$\pm$	0.013	\tablenotemark{a}	&	0.016	$\pm$	0.012	\tablenotemark{a}	&	0.152	$\pm$	0.006	\tablenotemark{a}	&	0.431	$\pm$	0.015	\tablenotemark{a}	\\
WISE J0150+3827   	&	 IRTF/SpeX     	&	  T0            	&	0.665	$\pm$	0.006		&	0.908	$\pm$	0.006		&	0.675	$\pm$	0.004		&	0.979	$\pm$	0.005		&	0.976	$\pm$	0.006		&	0.891	$\pm$	0.006		\\
\enddata																																			
\tablenotetext{a}{\citet{cushing2011}}																																			
\tablenotetext{b}{\citet{burgasser2011}}																																			
\end{deluxetable}																																																																																																																																			

\clearpage

\clearpage

\clearpage

\clearpage

\clearpage

\clearpage

\clearpage

\clearpage

\clearpage

\begin{figure}
\epsscale{0.9}
\figurenum{9}
\plotone{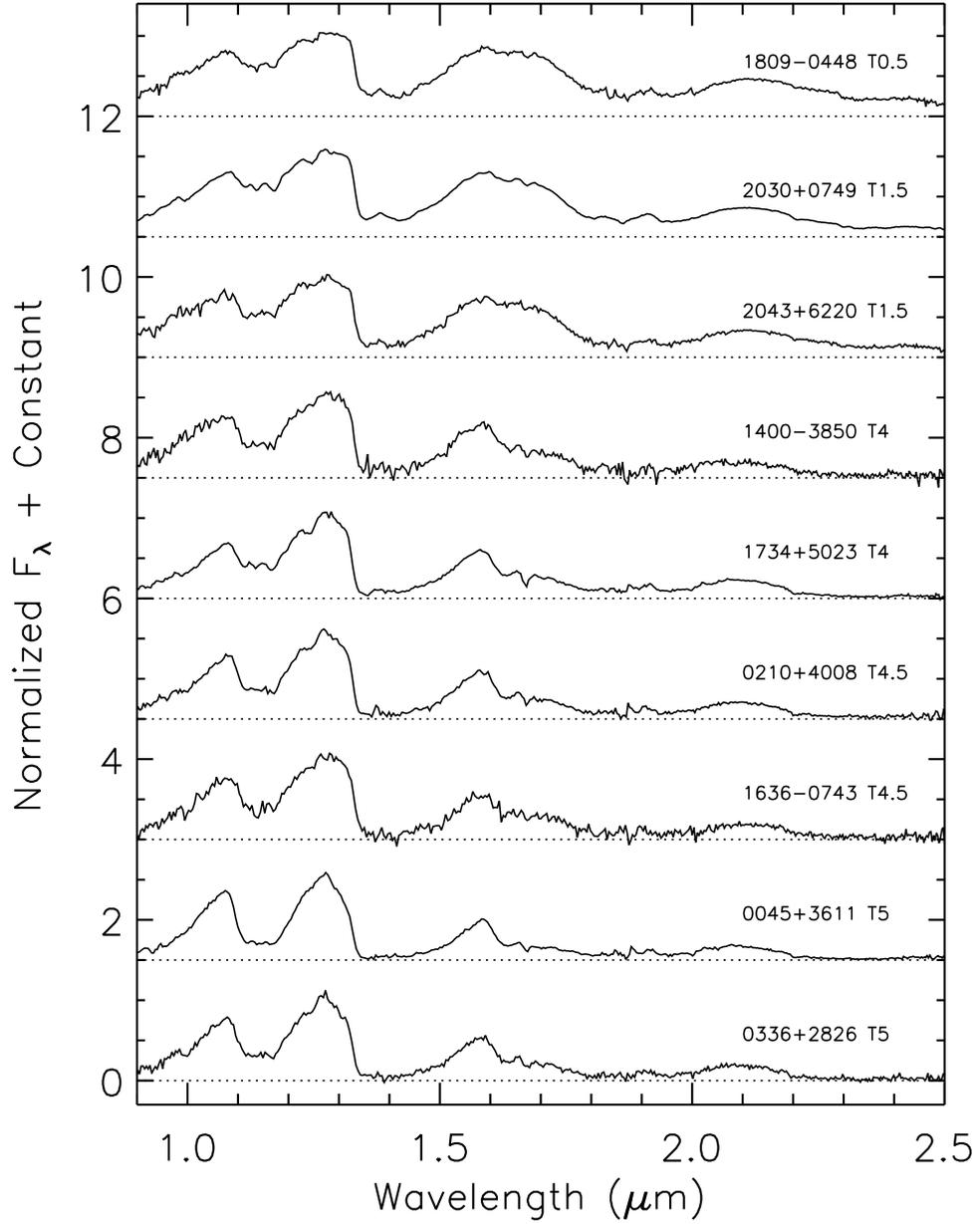}
\caption{IRTF/SpeX spectra of confirmed WISE brown dwarfs with spectral types from T0.5 to T8.5. Spectra have been normalized at 1.27$\mu$m and offset vertically.
\label{SpeX_1}}
\end{figure}

\clearpage

\begin{figure}
\epsscale{0.9}
\figurenum{9b}
\plotone{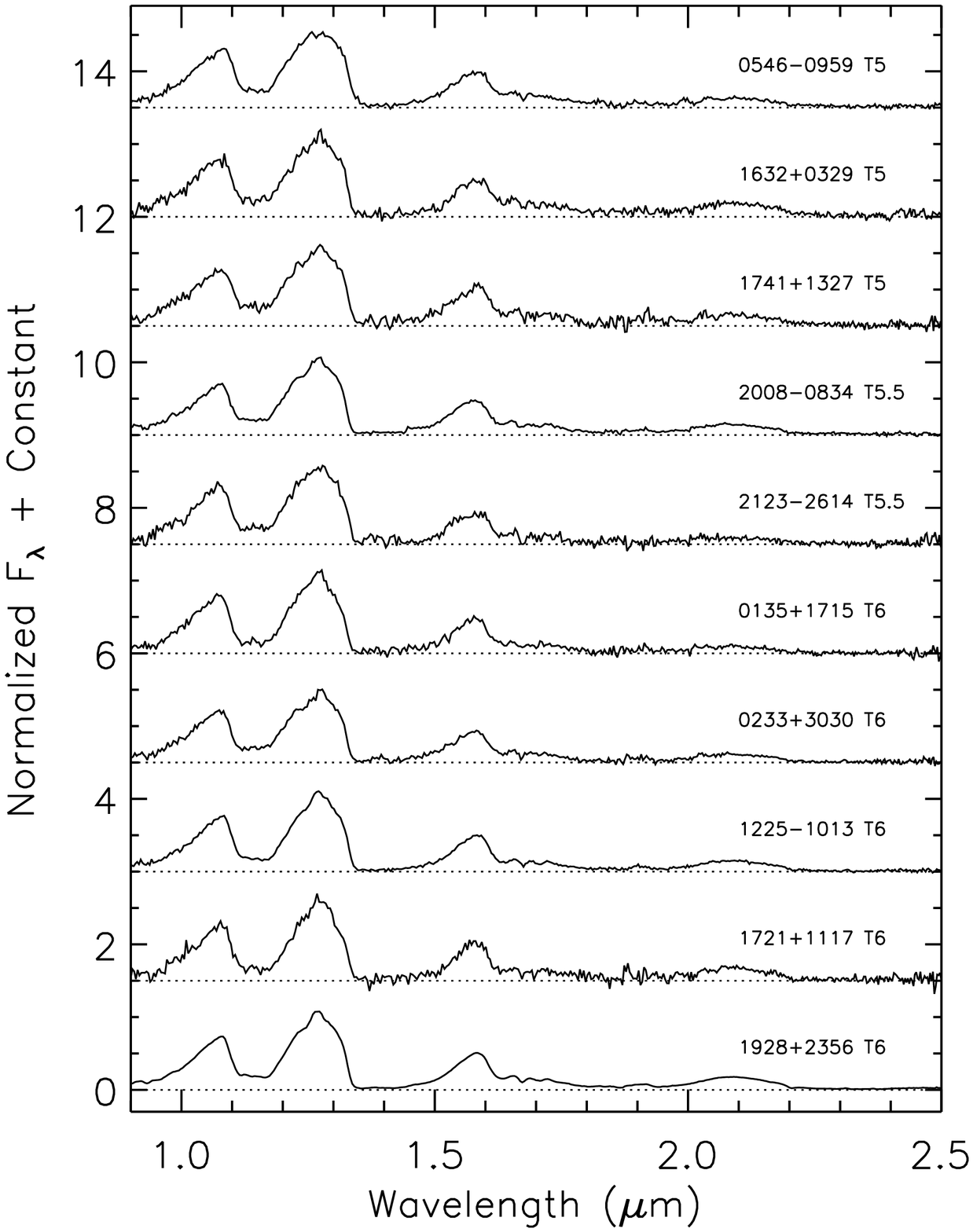}
\caption{Continued.
\label{SpeX_2}}
\end{figure}

\clearpage

\begin{figure}
\epsscale{0.9}
\figurenum{9c}
\plotone{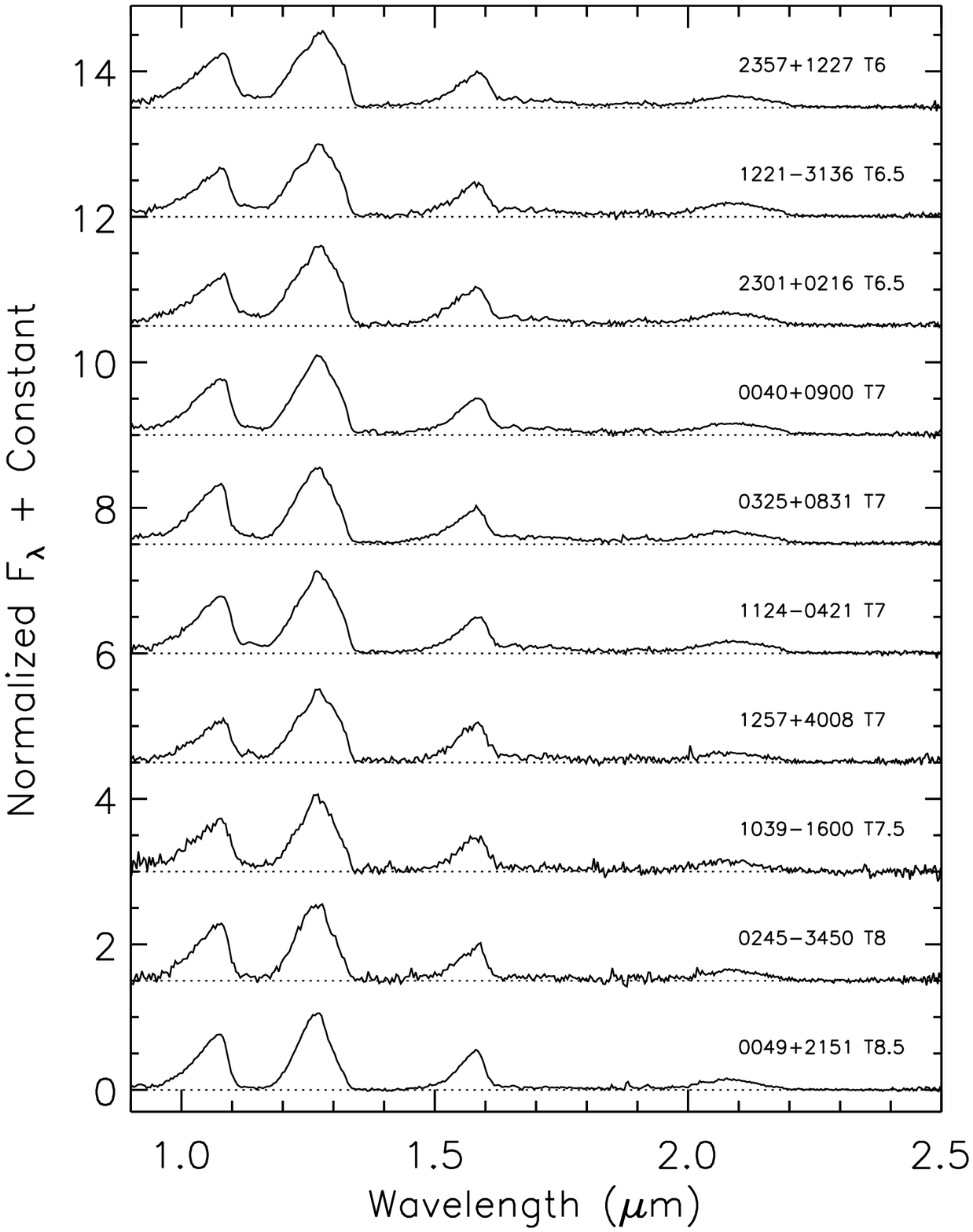}
\caption{Continued.
\label{SpeX_3}}
\end{figure}

\clearpage

\begin{figure}
\epsscale{0.9}
\figurenum{10}
\plotone{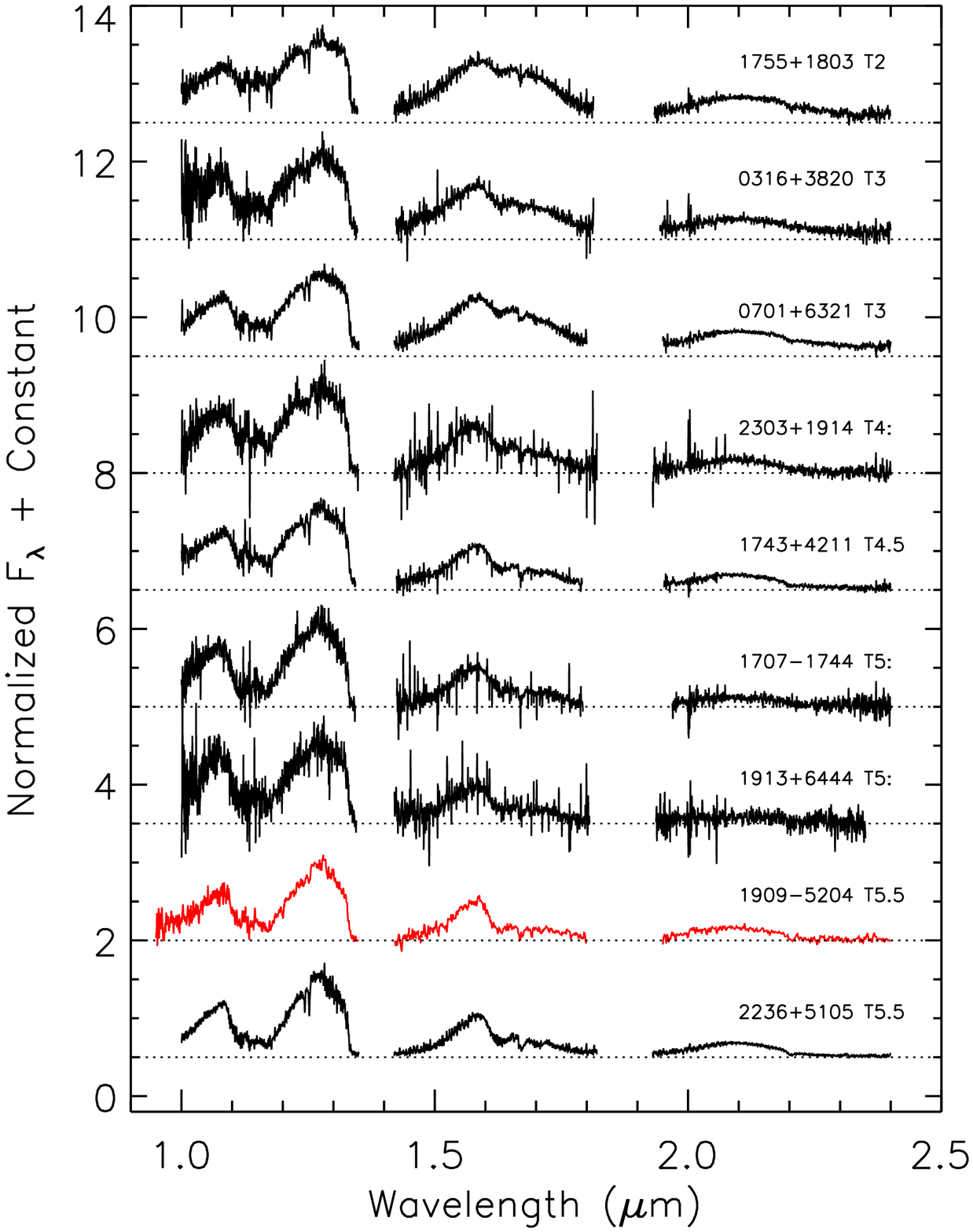}
\caption{Other spectra of confirmed WISE brown dwarfs with spectral types from T2 to T9.5. Spectra have been normalized at 1.27$\mu$m and offset vertically. Palomar/TSpec spectra are shown in black, Magellan/FIRE spectra in red, and Gemini/GNIRS spectra in blue.
\label{FGT_1}}
\end{figure}

\clearpage

\begin{figure}
\epsscale{0.9}
\figurenum{10b}
\plotone{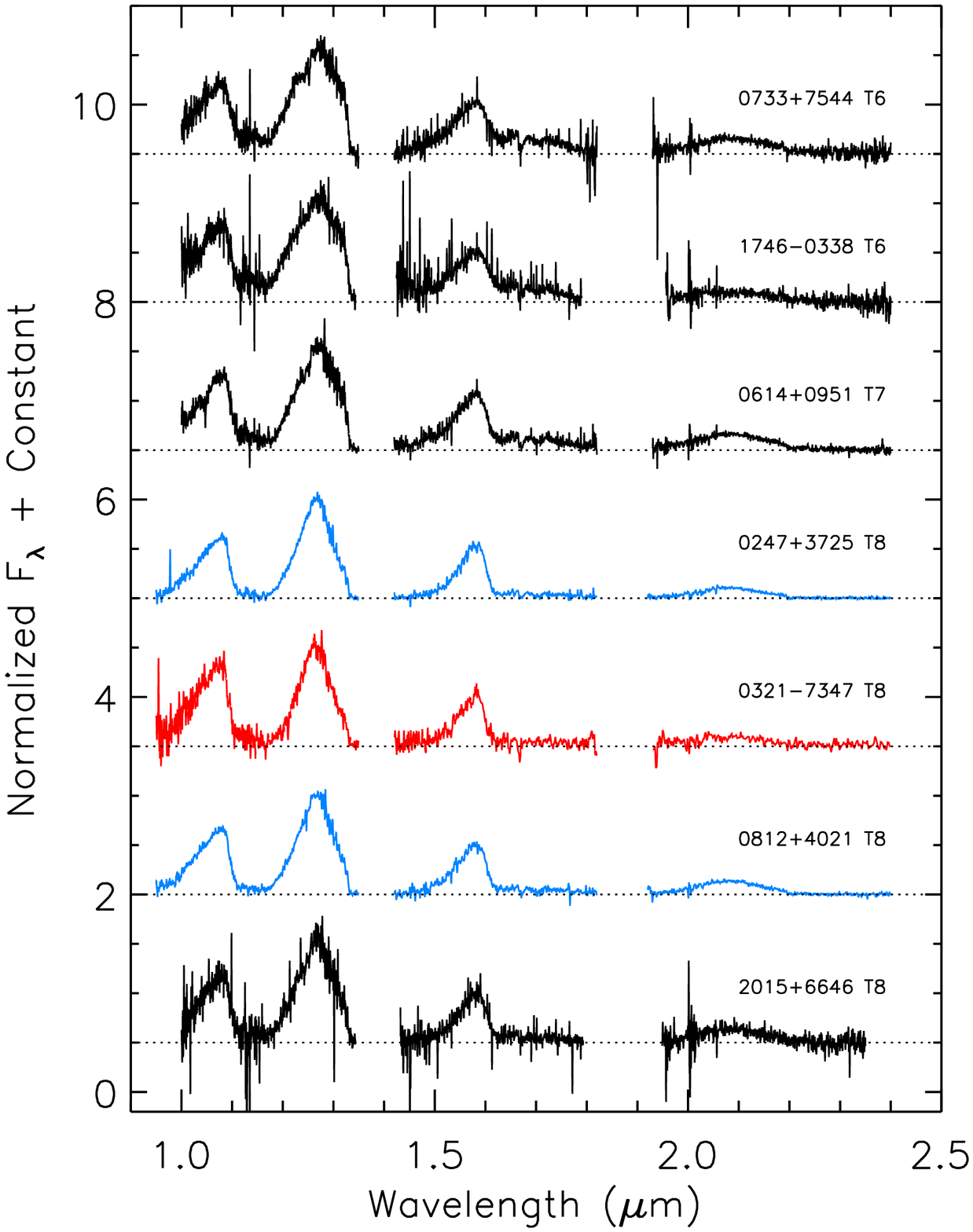}
\caption{Continued.
\label{FGT_2}}
\end{figure}

\clearpage

\begin{figure}
\epsscale{0.9}
\figurenum{10c}
\plotone{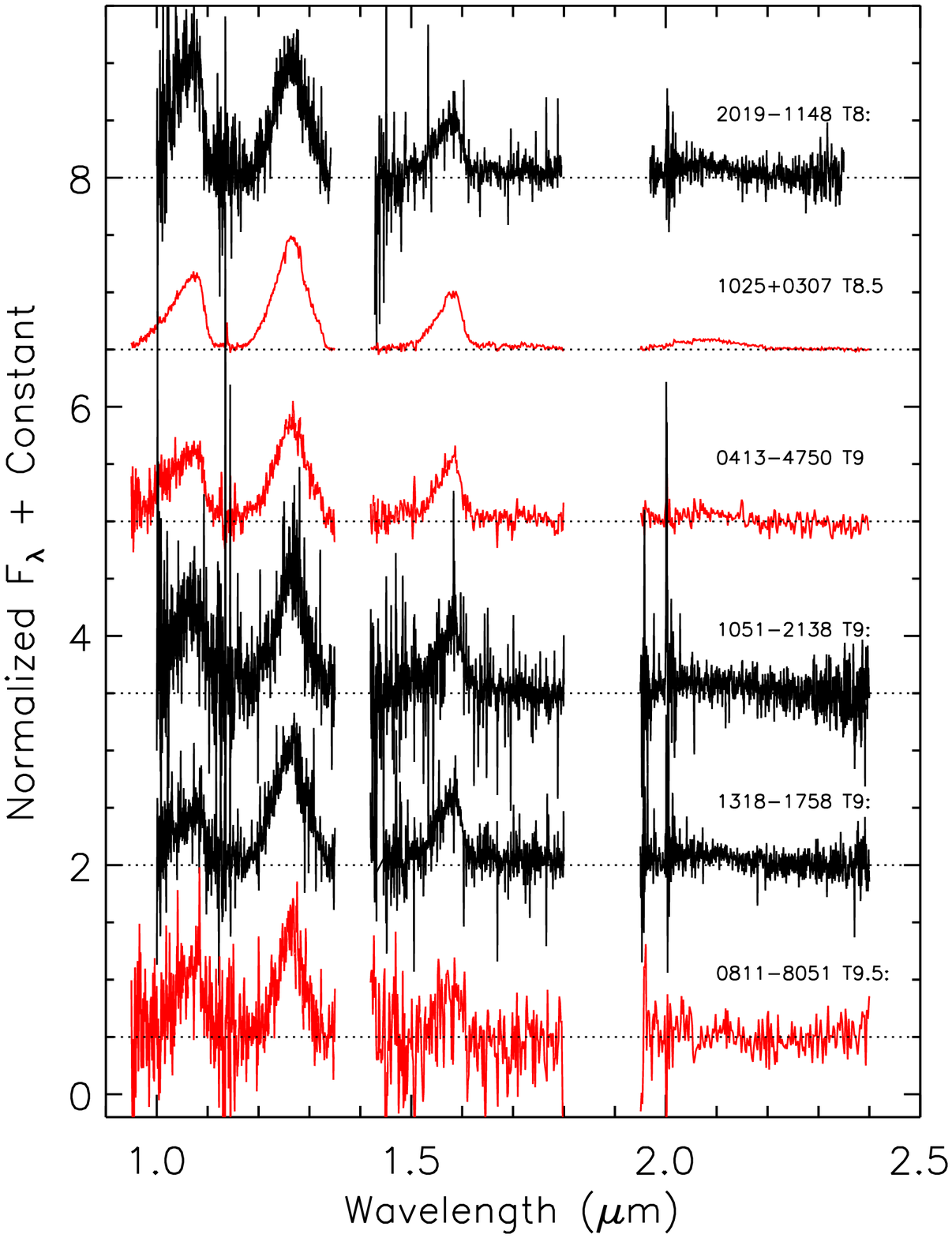}
\caption{Continued.
\label{FGT_3}}
\end{figure}

\clearpage

\begin{figure}
\epsscale{0.9}
\figurenum{11}
\plotone{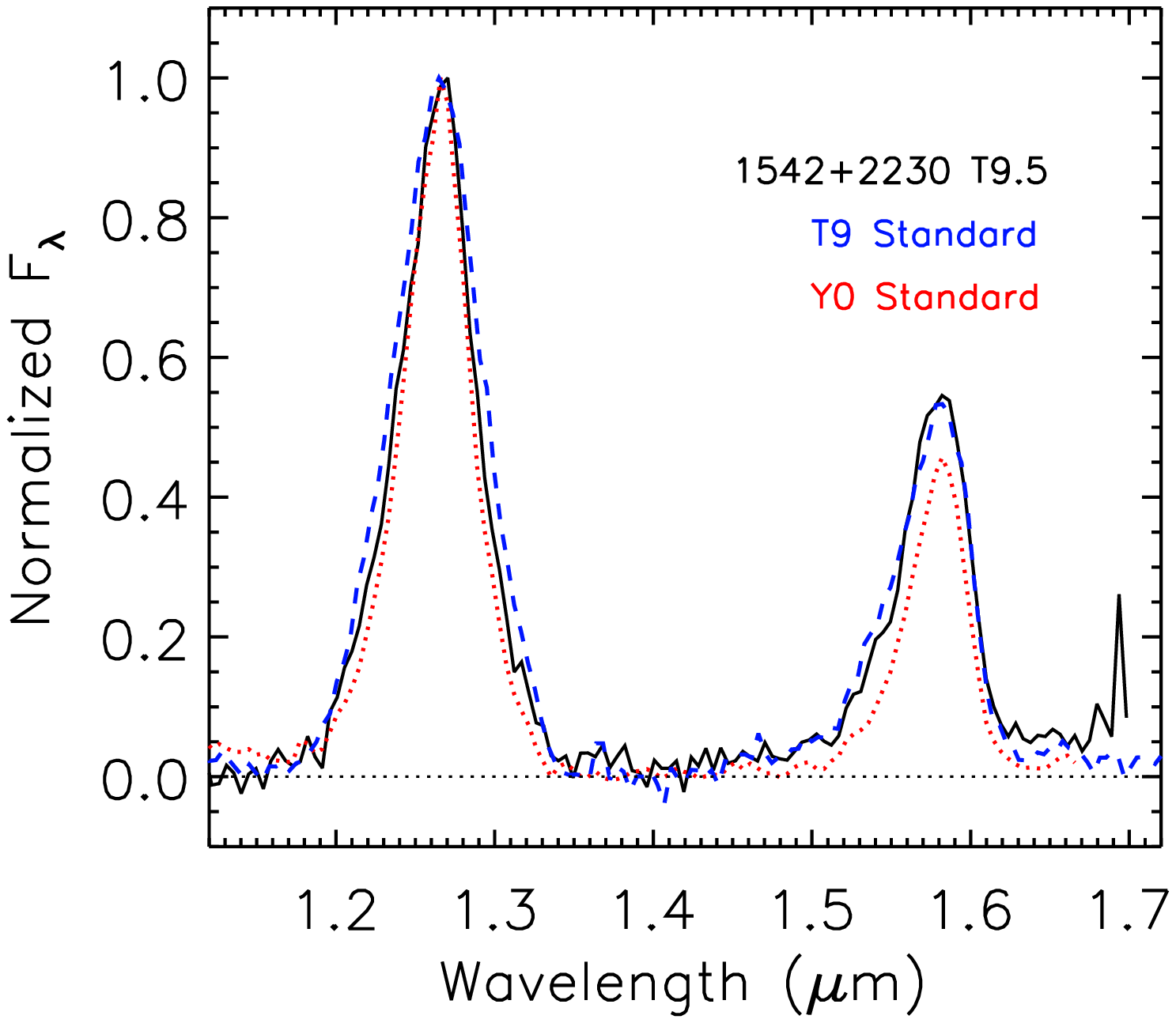}
\caption{The HST/WFC3 spectrum of WISE J154214.00+223005.2 (T9.5). The T9 (UGPS J072227.51$-$054031.2) and Y0 (WISEP J173835.52+273258.9) spectral standards \citep{cushing2011} are included for comparison. Spectra have been normalized at 1.27$\mu$m.
\label{HST}}
\end{figure}

\clearpage

\begin{figure}
\epsscale{0.9}
\figurenum{12}
\plotone{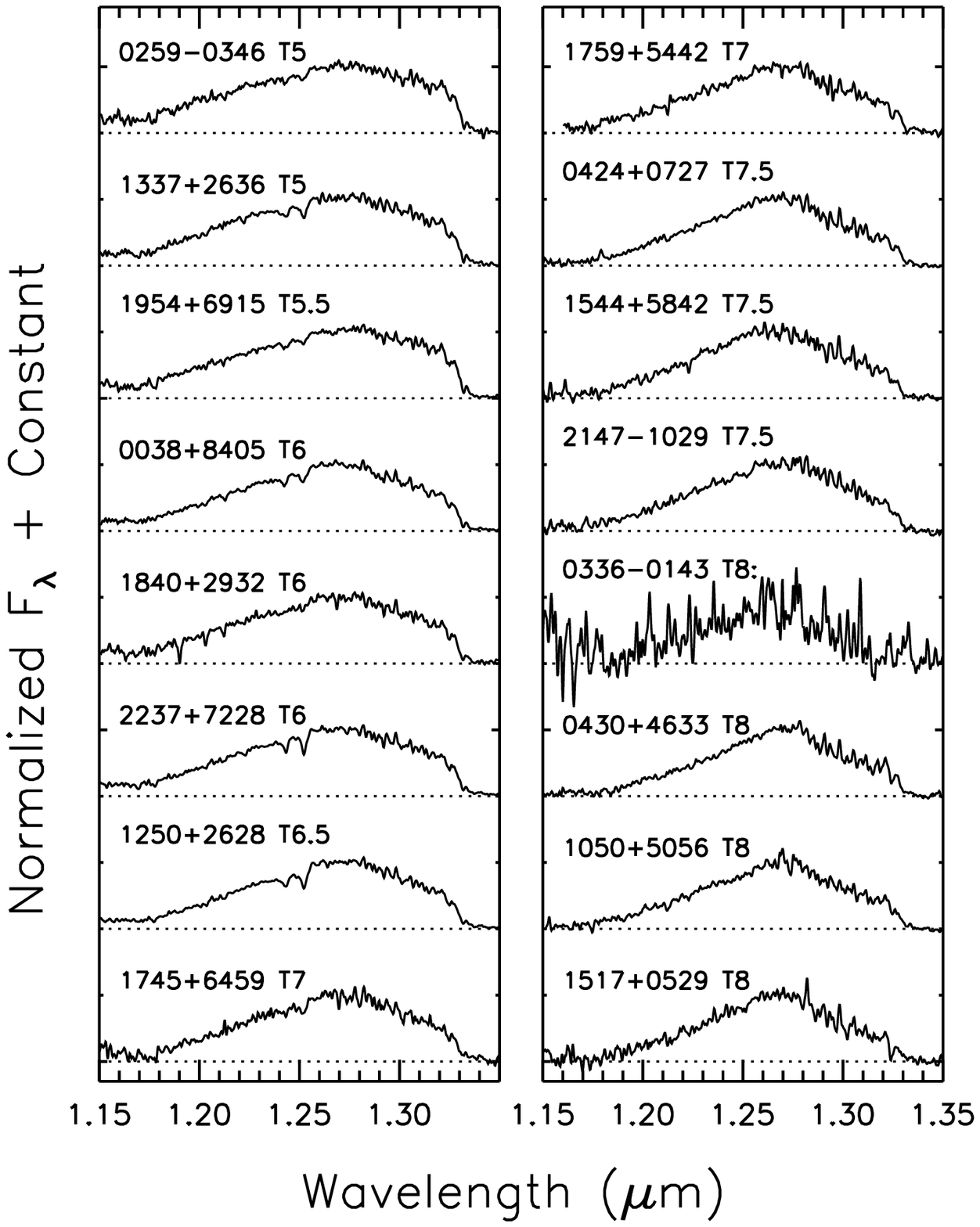}
\caption{Keck/NIRSPEC $J$-band spectra of confirmed WISE brown dwarfs with spectral types from T5 to T9. Spectra have been normalized at 1.27$\mu$m and offset vertically.
\label{NIRSPEC_N3_1}}
\end{figure}

\clearpage

\begin{figure}
\epsscale{0.9}
\figurenum{12b}
\plotone{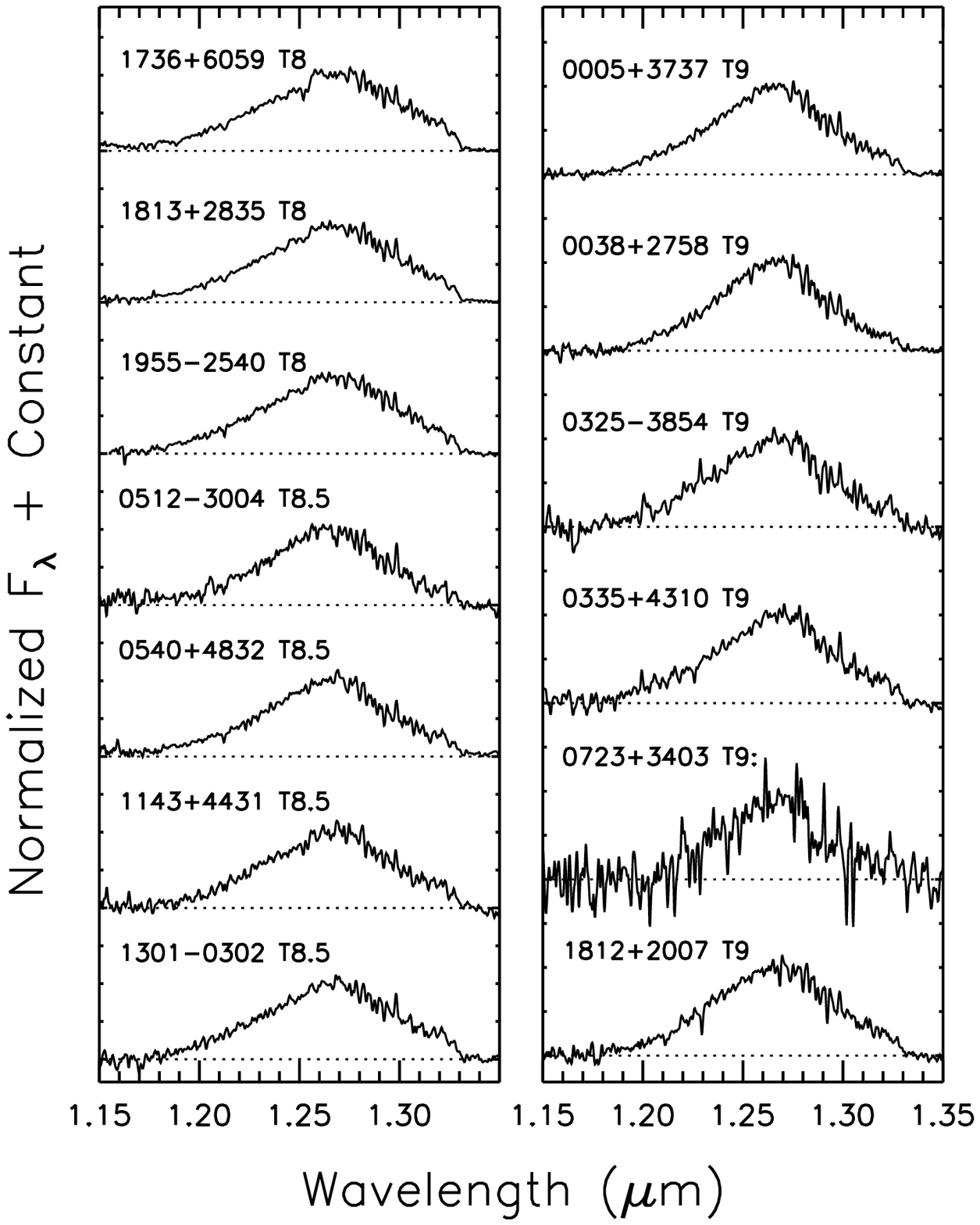}
\caption{Continued.
\label{NIRSPEC_N3_2}}
\end{figure}

\clearpage

\begin{figure}
\epsscale{0.9}
\figurenum{13}
\plotone{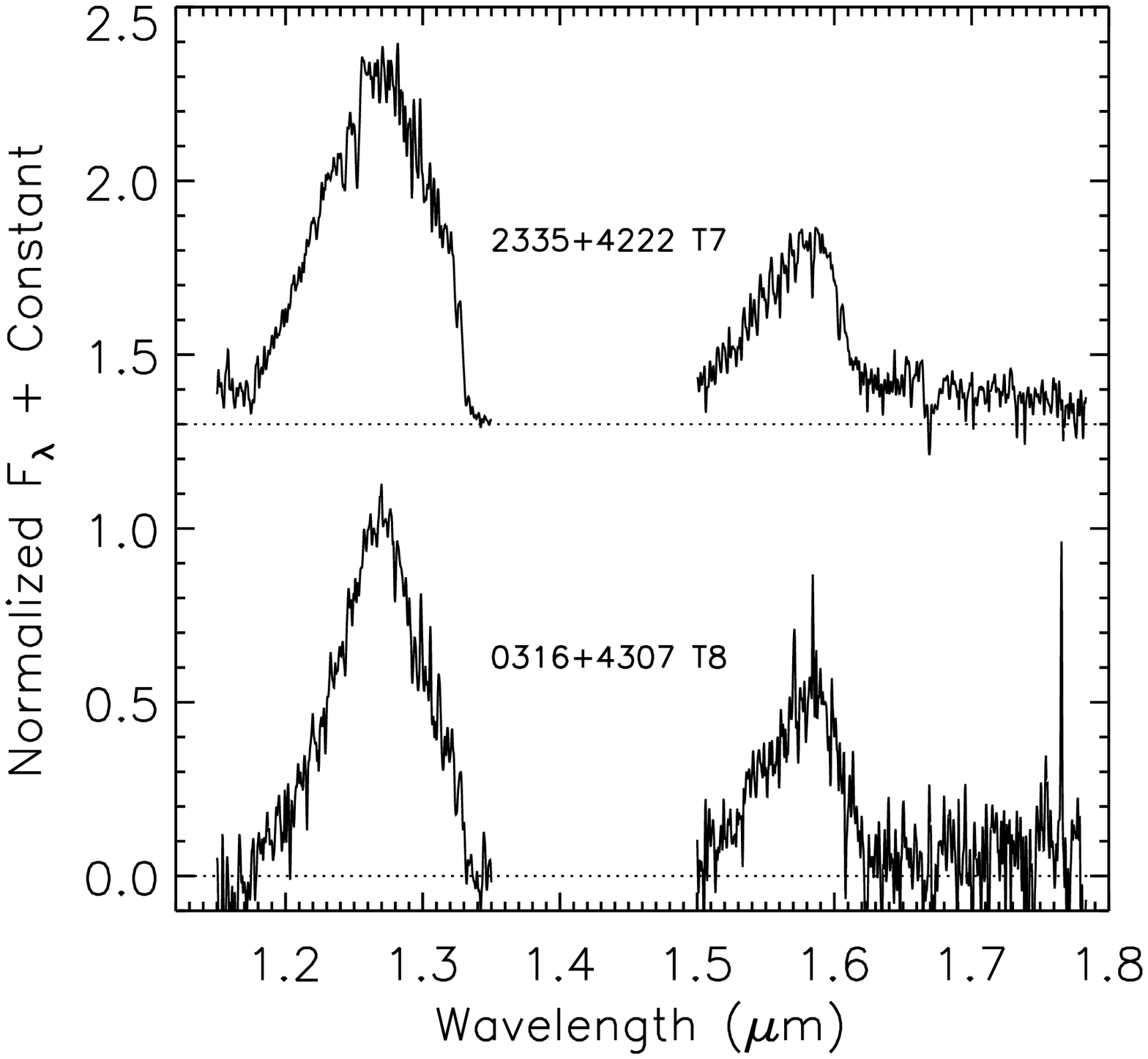}
\caption{Keck/NIRSPEC $J$- and $H$-band spectra of confirmed WISE brown dwarfs WISE J233543.79+422255.2 (T7) and WISE J031624.35+430709.1 (T8). Spectra have been normalized at 1.27$\mu$m and offset vertically. The $H$-band spectra were scaled relative to the $J$-band data using the ratio of peak fluxes produced by REDSPEC \citep{mclean2003}.
\label{NIRSPEC_N3N5}}
\end{figure}

\clearpage

\begin{figure}
\epsscale{0.9}
\figurenum{14}
\plotone{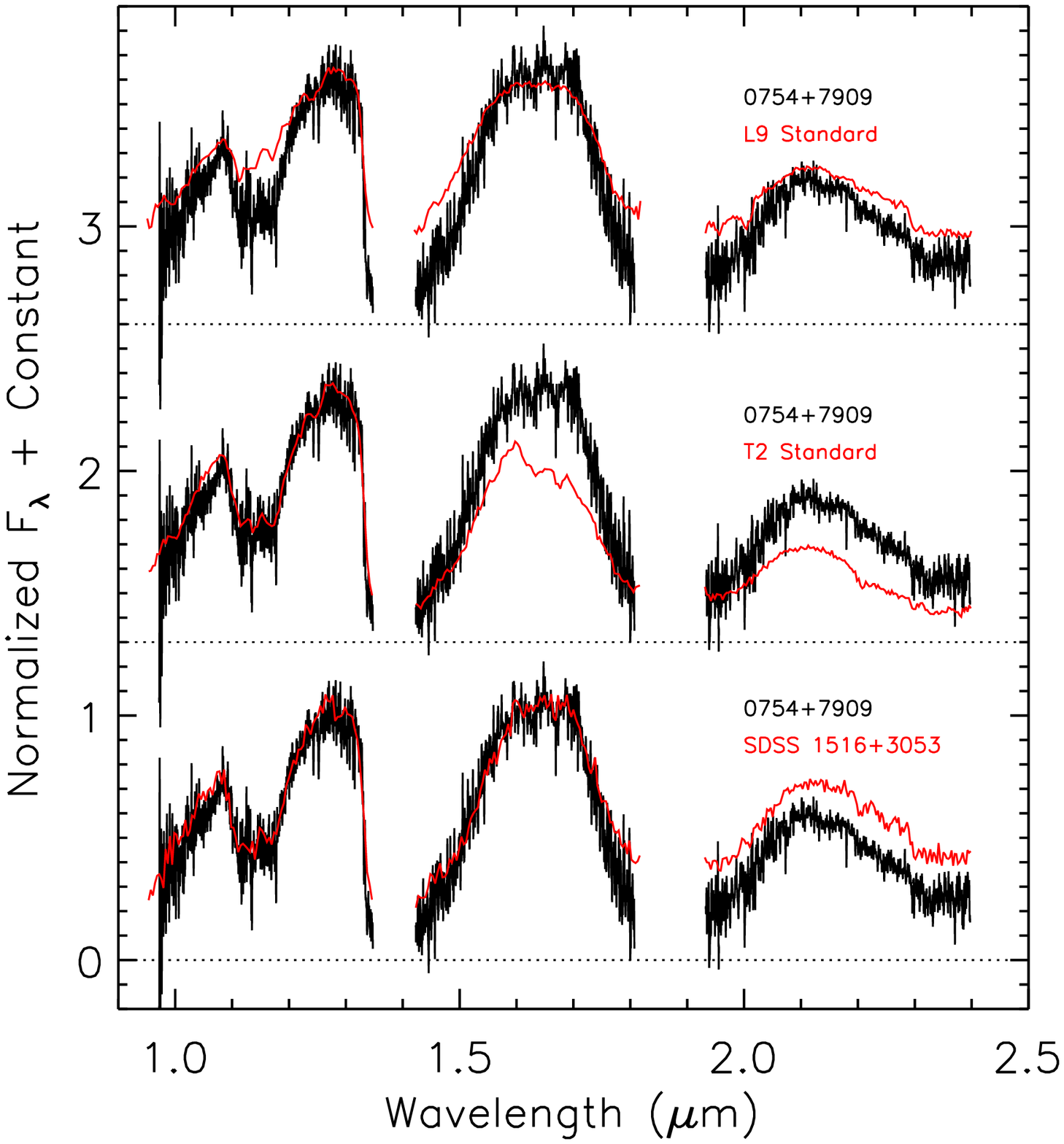}
\caption{This Palomar/TSpec spectrum of WISE 0754+7909 is not easily classified as a L or T dwarf. SDSS J1516+3053 has been classified as a T0.5$\pm$1 \citep{chiu2006} and T1.5$\pm$2 \citep{burgasser2010} and is a better match in the $J$ and $H$ bands than the spectral standards.
Spectra have been normalized at 1.27$\mu$m and offset vertically.
\label{ERGs1}}
\end{figure}

\clearpage

\begin{figure}
\epsscale{0.9}
\figurenum{15}
\plotone{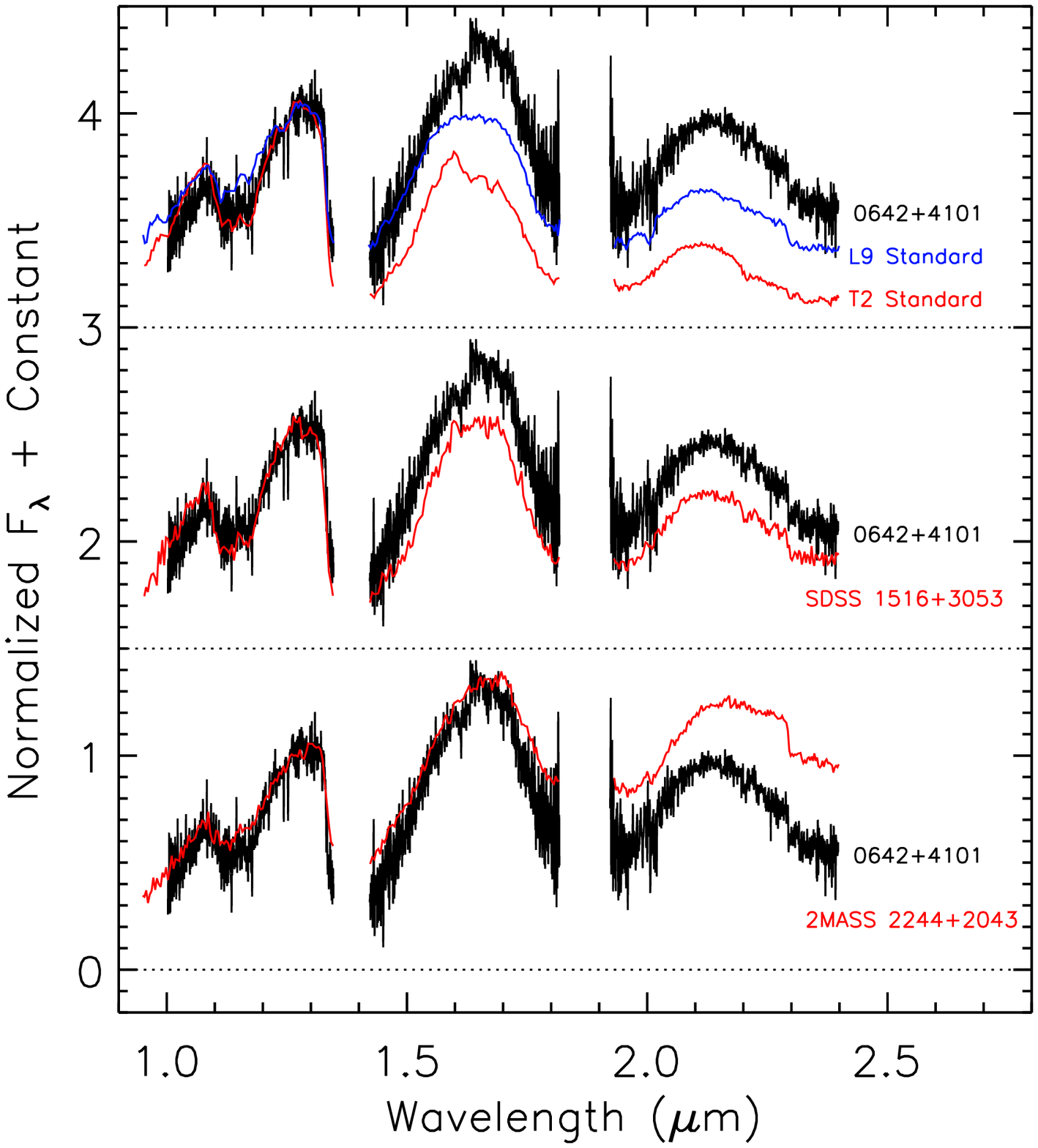}
\caption{This Palomar/TSpec spectrum of WISE 0642+4101 is not easily classified as a L or T dwarf. It is redder than SDSS J1516+3053 \citep[T1.5$\pm$2;][]{burgasser2010} and bluer than 2MASSW J2244+2043 \citep[L7.5$\pm$2;][]{looper2008}.
Spectra have been normalized at 1.27$\mu$m and offset vertically.
\label{ERGs2}}
\end{figure}

\clearpage

\begin{figure}
\epsscale{0.9}
\figurenum{16}
\plotone{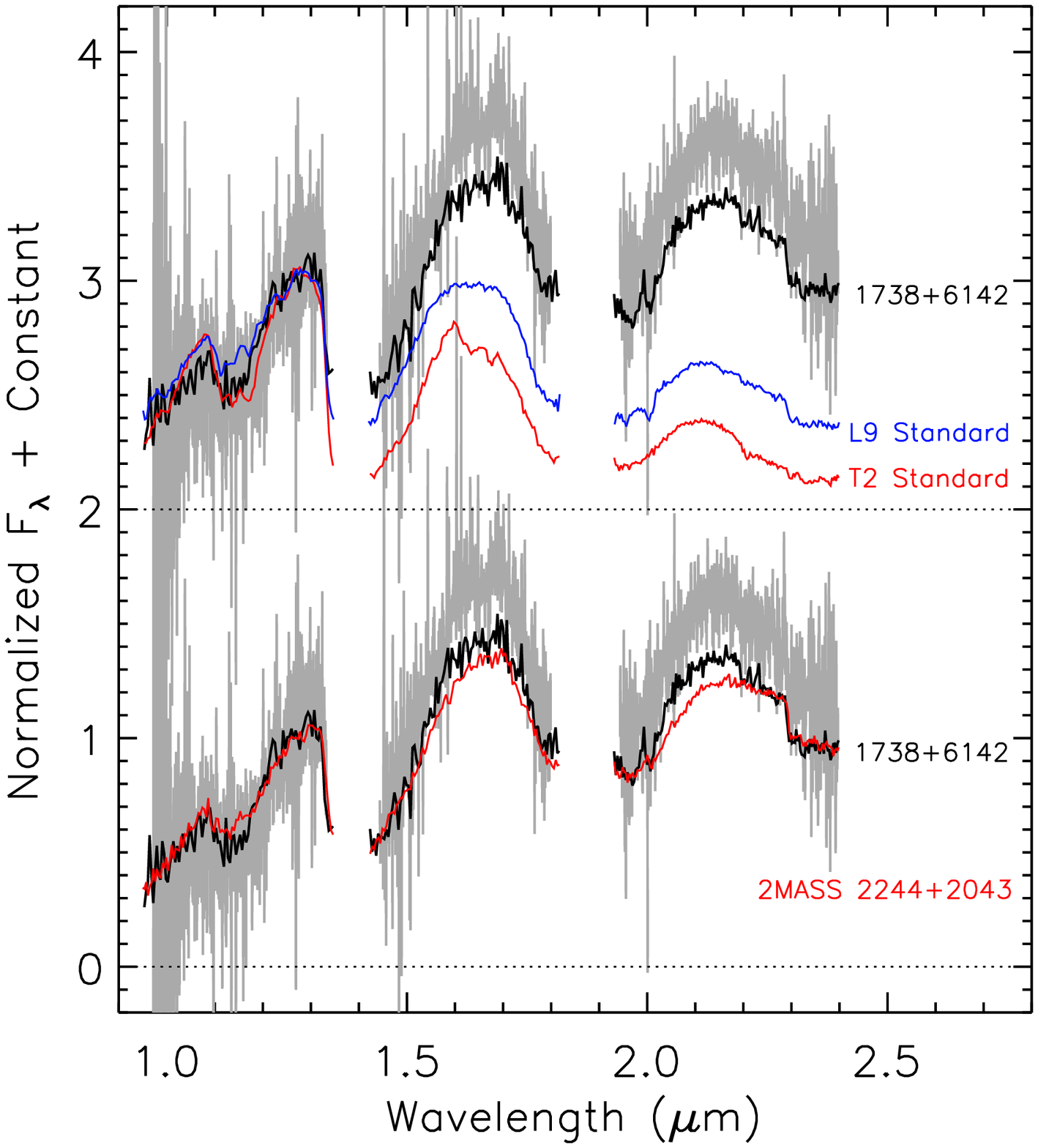}
\caption{WISE 1738+6142 was observed with both Palomar/TSpec (gray) and IRTF/SpeX (black).  If it is actually an L dwarf, it is the reddest one known and shows substantial variability. 2MASSW J2244+2043 \citep[L7.5$\pm$2;][]{looper2008} is similar to this object,
but not as red. Excess blue flux in the $H$ and $K$ bands may hint at a T dwarf companion or the presence of CH$_4$ in this extremely red object. 
Spectra have been normalized at 1.27$\mu$m and offset vertically.
\label{ERGs3}}
\end{figure}

\clearpage

\begin{figure}
\epsscale{0.9}
\figurenum{17}
\centering
\includegraphics[width=3in]{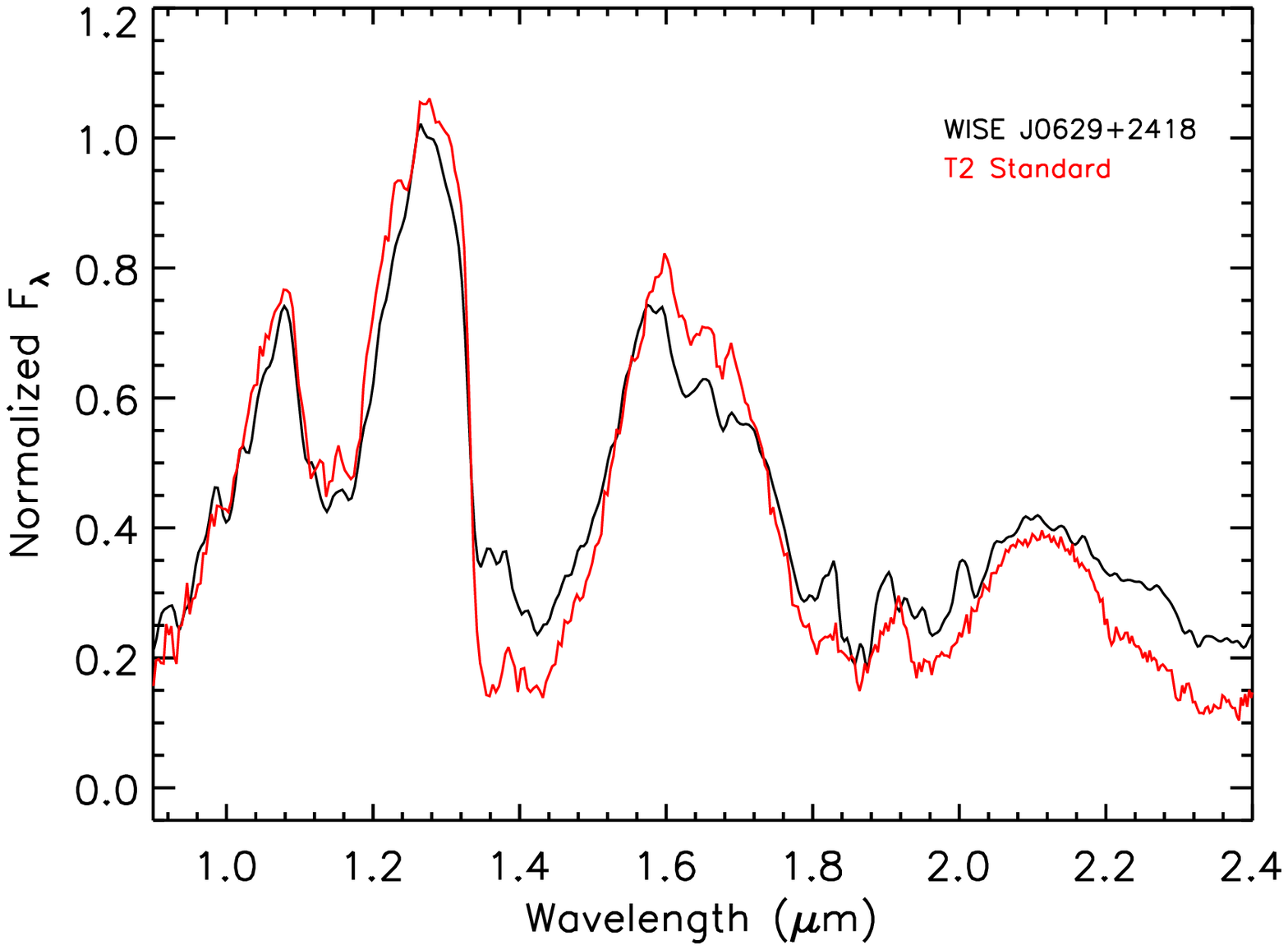}
\includegraphics[width=3in]{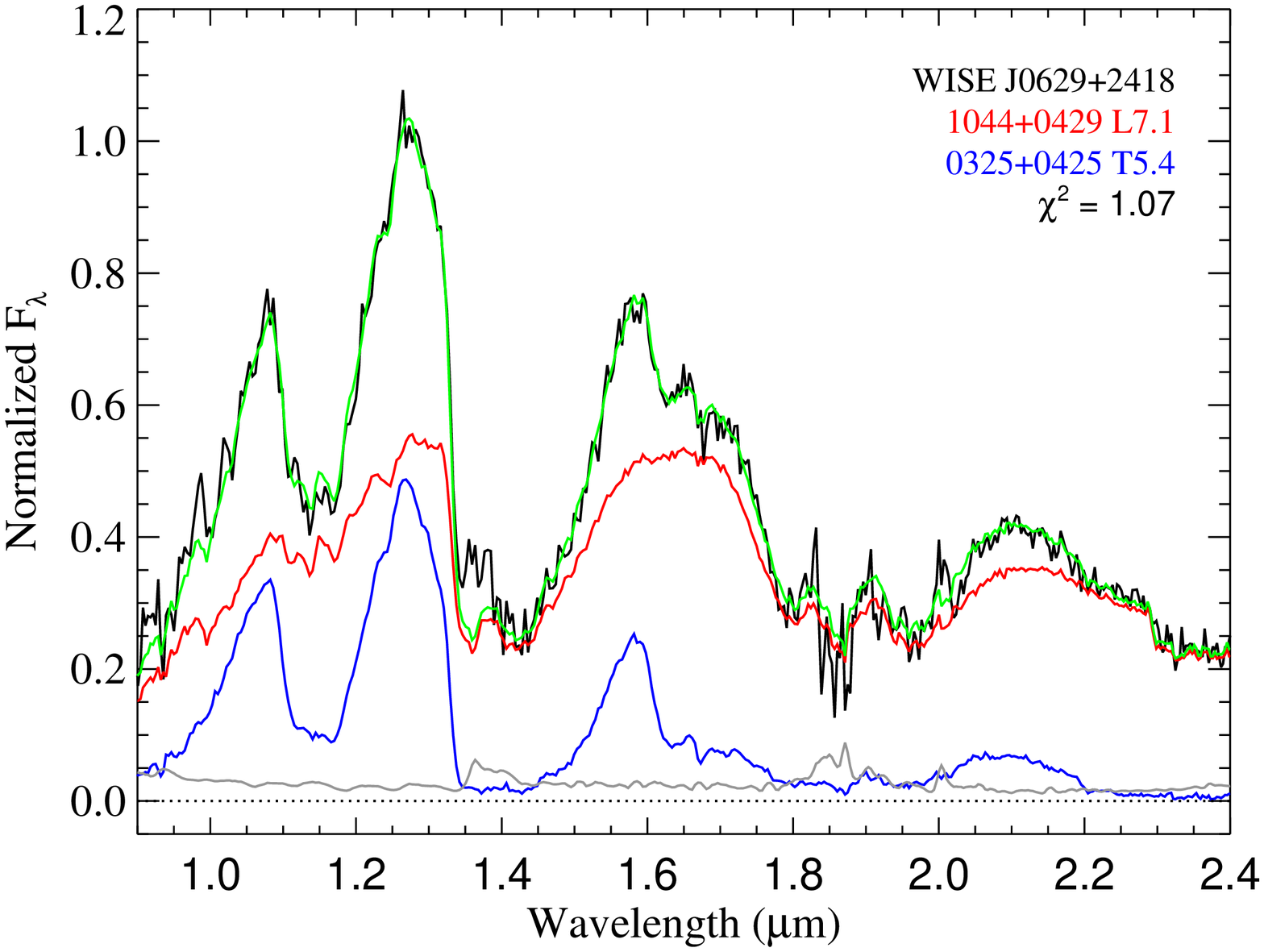}
\includegraphics[width=3in]{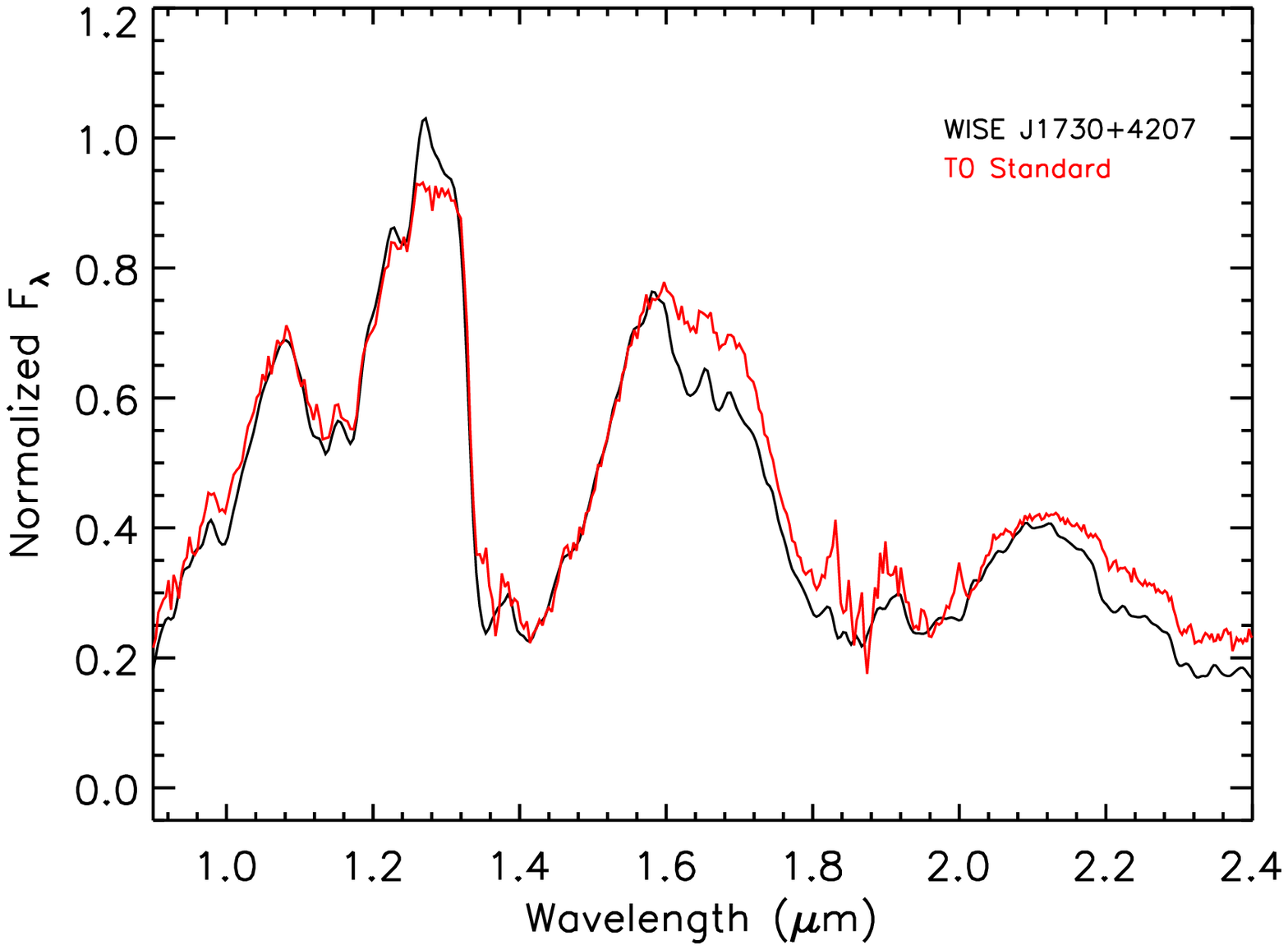}
\includegraphics[width=3in]{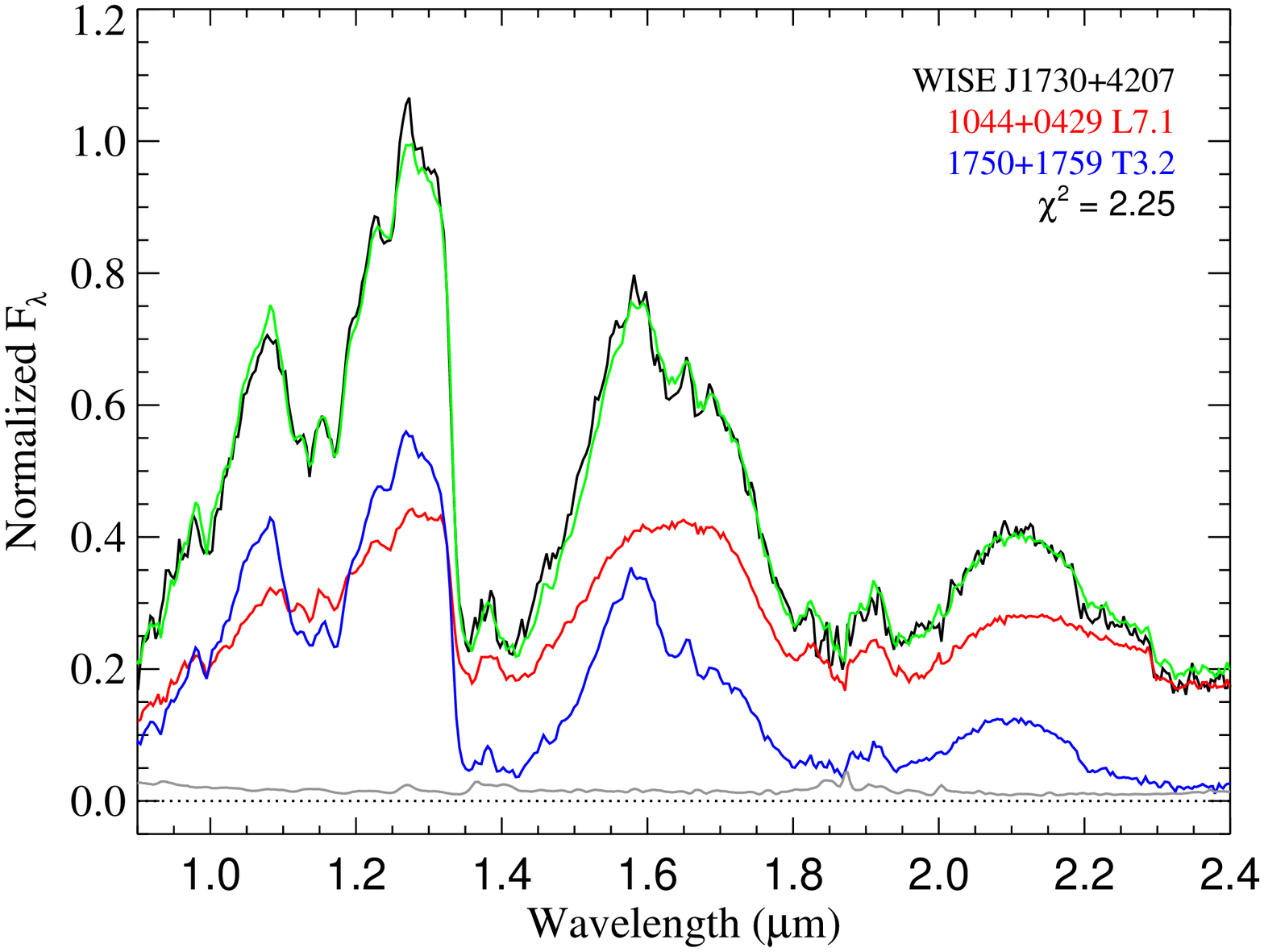}
\includegraphics[width=3in]{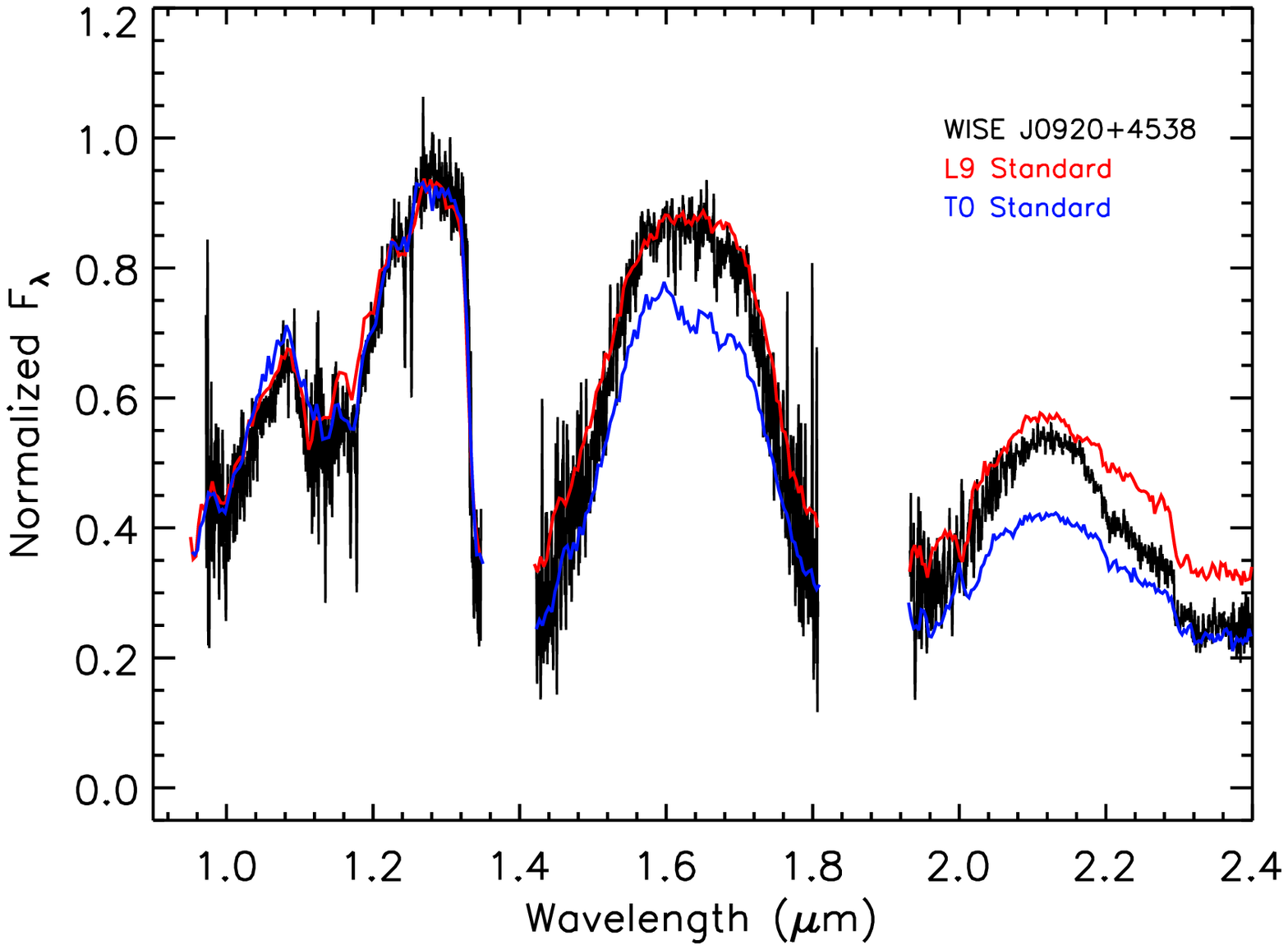}
\includegraphics[width=3in]{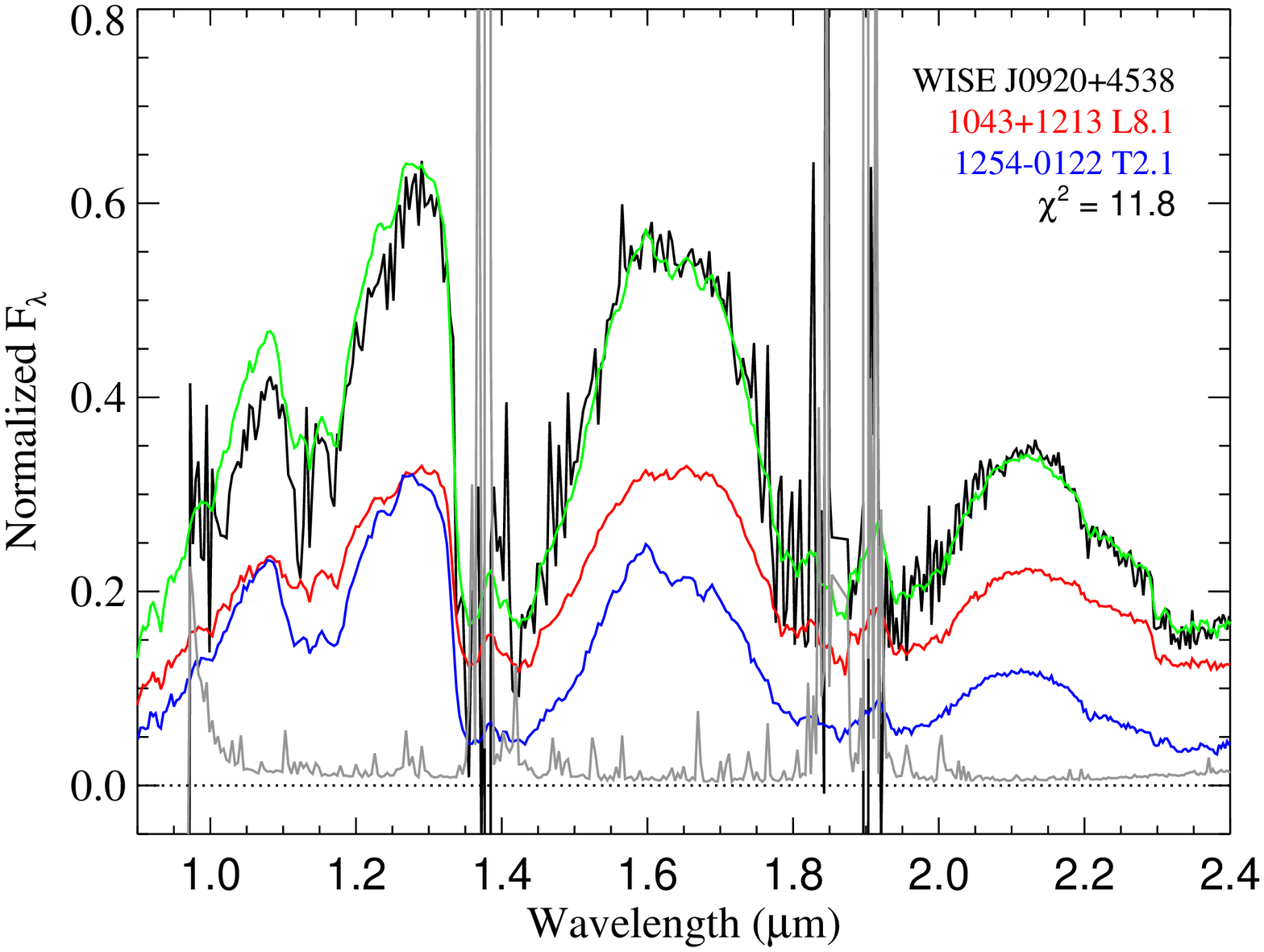}
\caption{Binary candidates identified using the method of \citet{burgasser2010}. The left side shows the best matching spectral standard and the right side shows the best composite spectrum (green), along with the constituent templates.
Spectra have been normalized at 1.27$\mu$m and offset vertically.
\label{sbsolutions}}
\end{figure}

\clearpage

\begin{figure}
\epsscale{0.9}
\figurenum{18}
\plotone{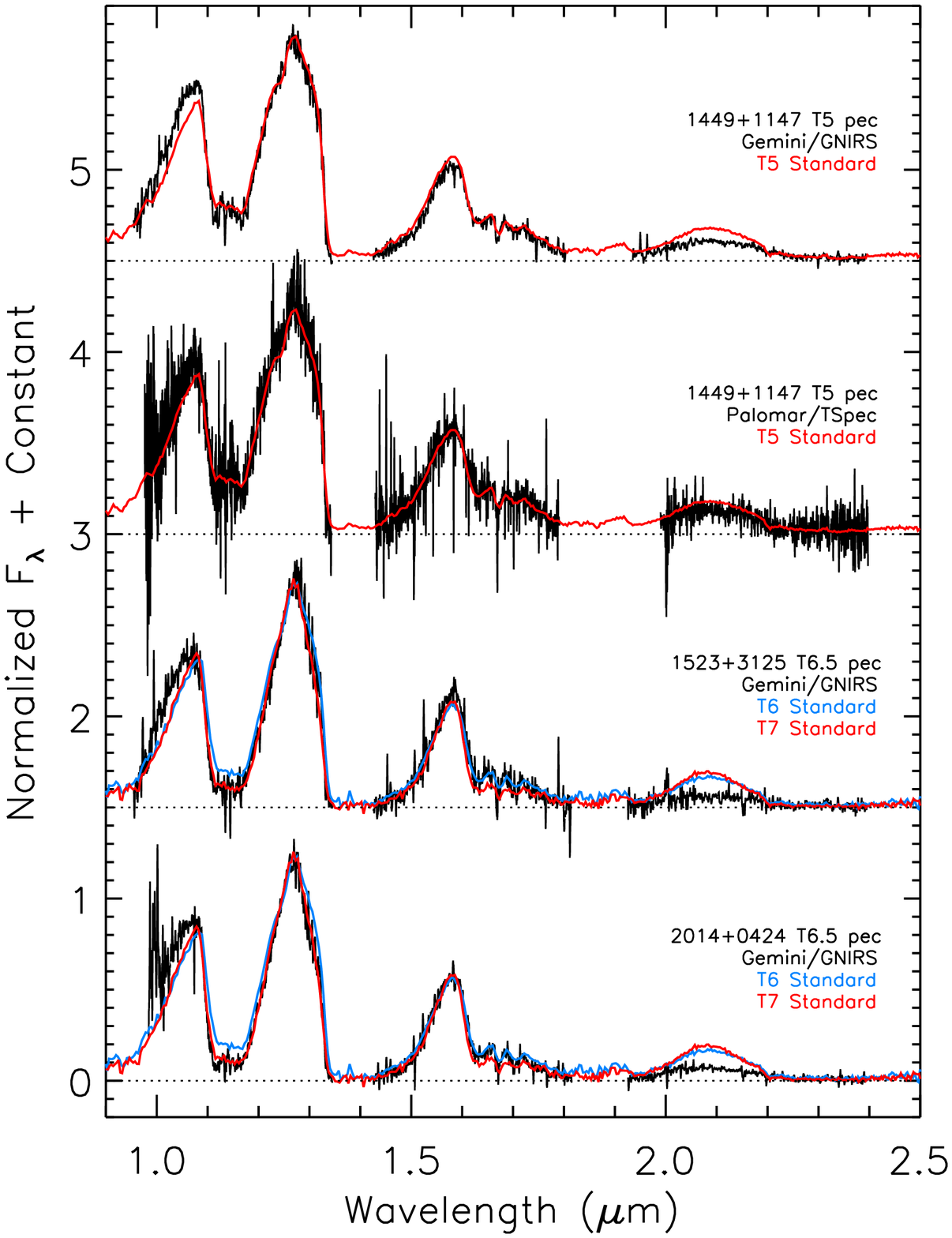}
\caption{Spectra of peculiar early-type T dwarfs, shown in black, along with spectral standards. Two different observations of the peculiar T5, WISE J1449+1147, from Gemini/GNIRS and Palomar/TSpec are shown. Spectra have been normalized at 1.27$\mu$m and offset vertically.
\label{pec}}
\end{figure}

\clearpage

\begin{figure}
\epsscale{0.9}
\figurenum{19}
\plotone{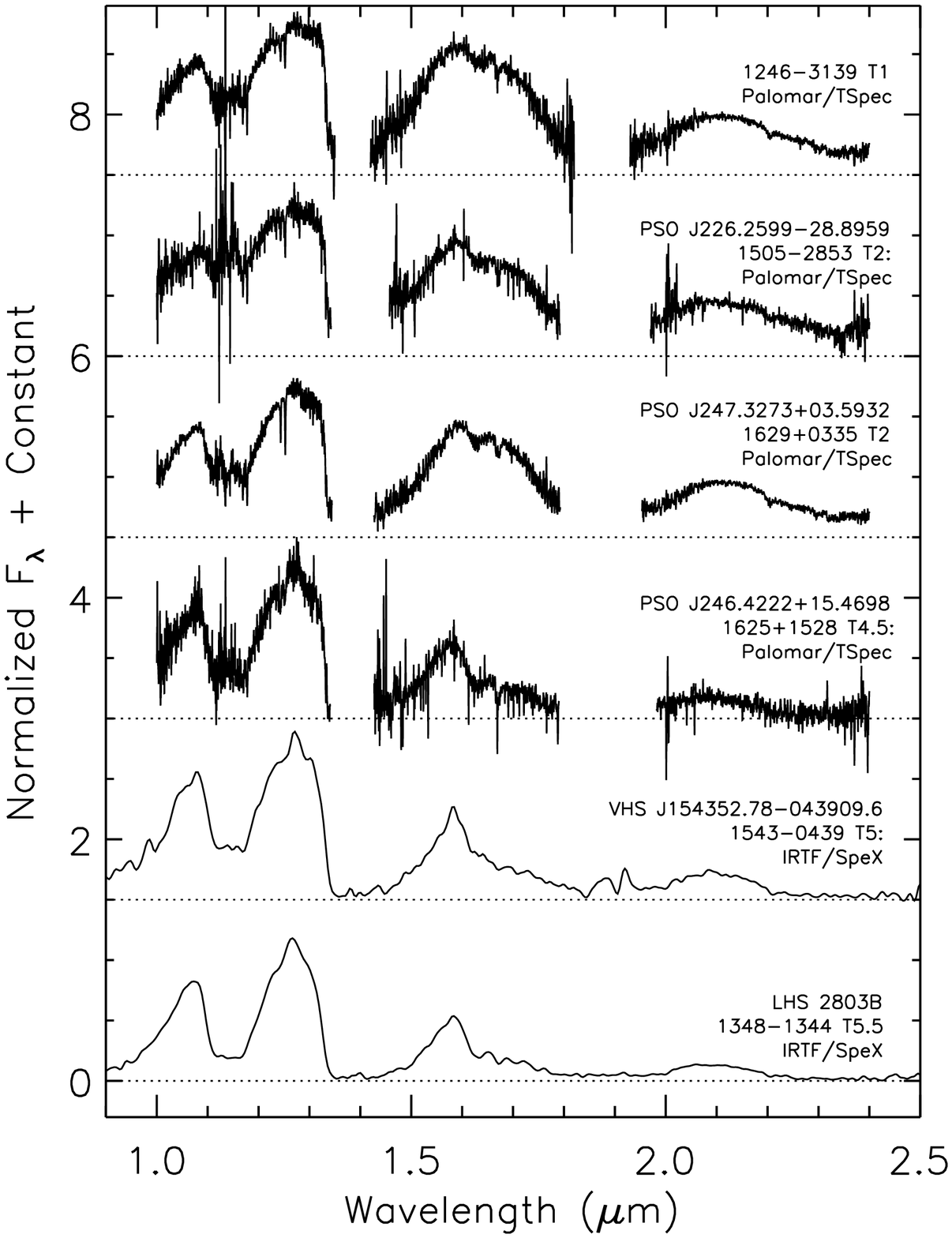}
\caption{Spectra of previously identified T dwarfs from the literature. Spectra have been normalized at 1.27$\mu$m and offset vertically.
\label{WISERedo1}}
\end{figure}

\clearpage

\begin{figure}
\epsscale{0.9}
\figurenum{19b}
\plotone{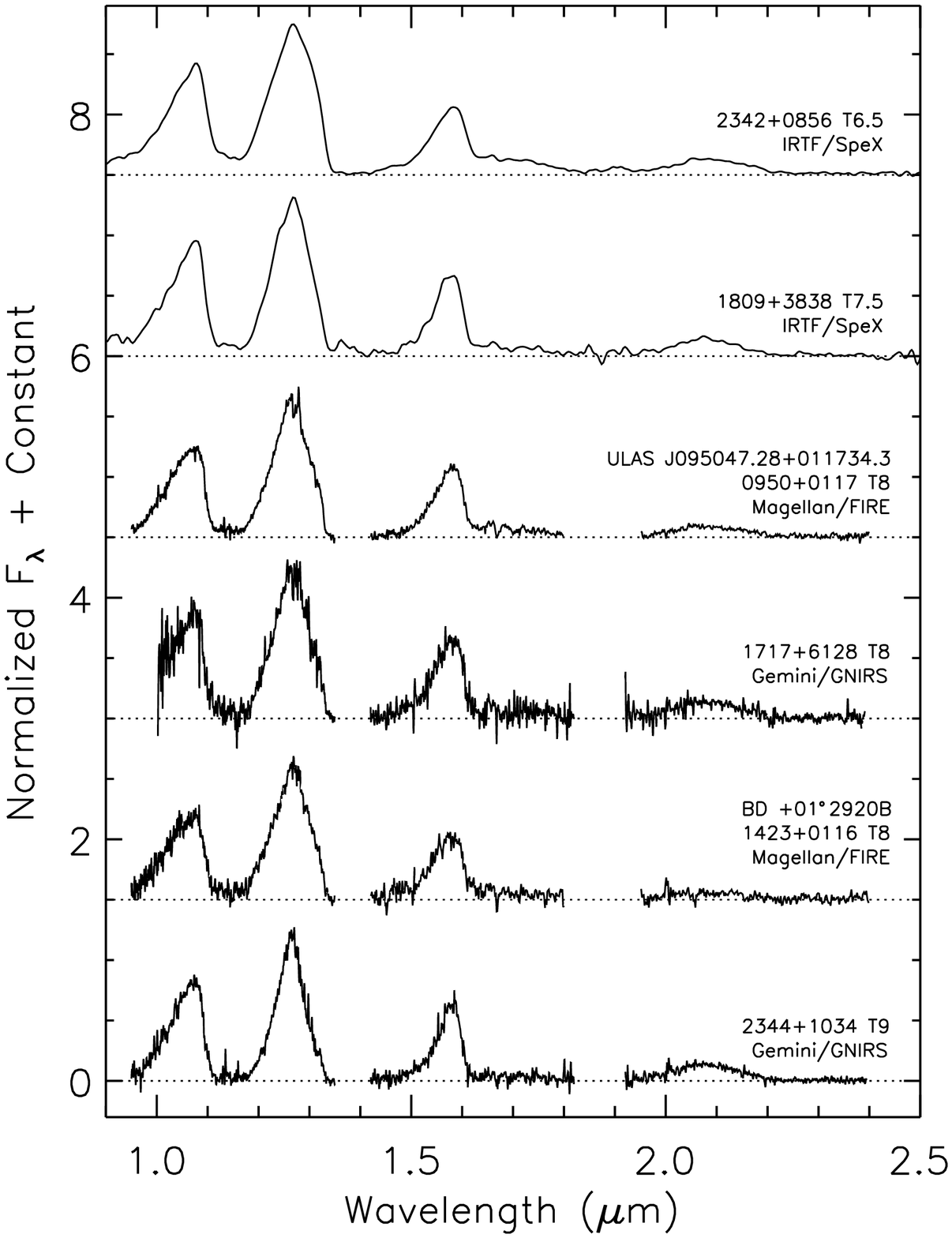}
\caption{Continued.
\label{WISERedo2}}
\end{figure}

\clearpage

\begin{figure}
\epsscale{0.9}
\figurenum{20}
\centering
\includegraphics[width=6.5in]{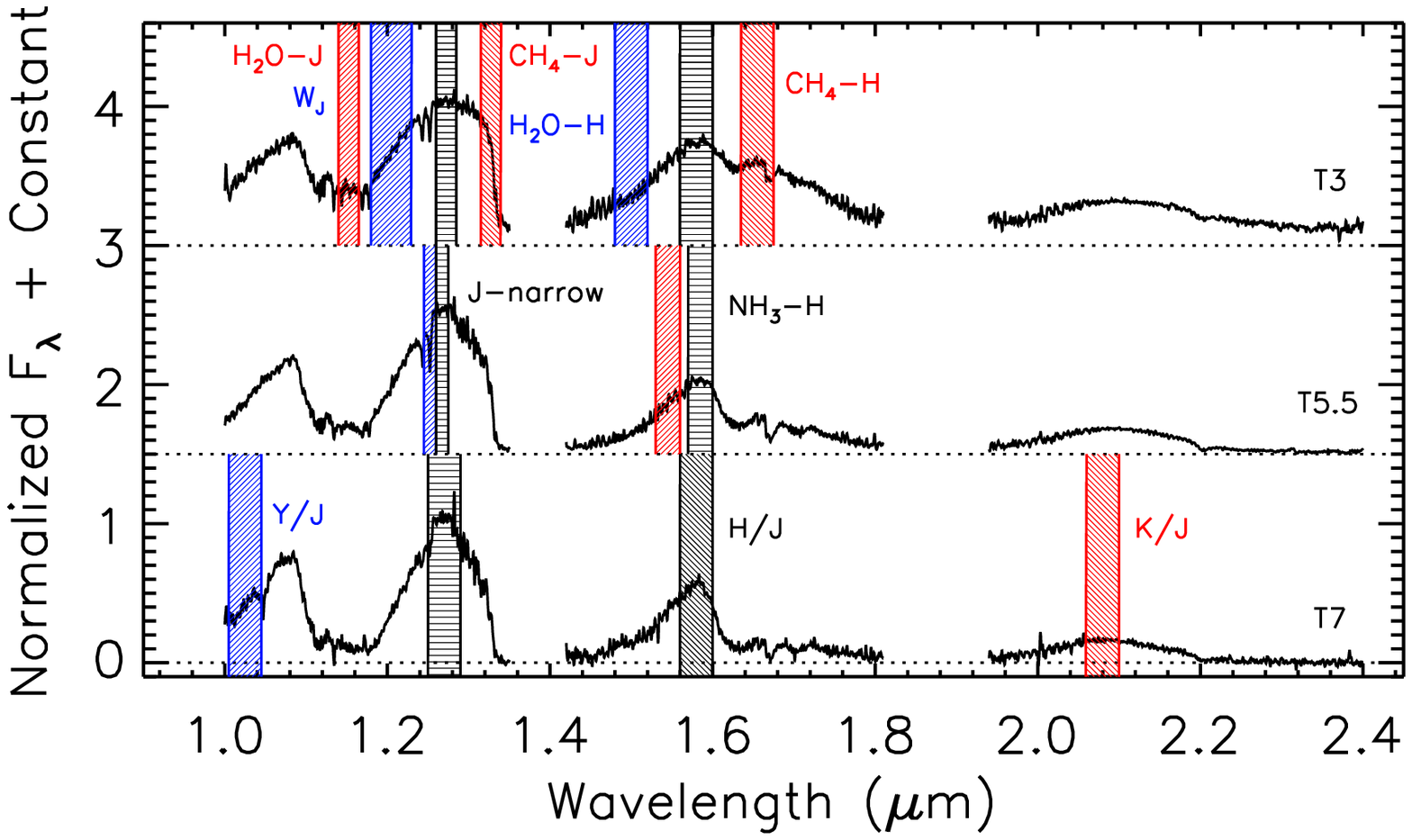}
\caption{The wavelengths that define the spectral indices derived in Equations 1-10 are depicted for a series of Palomar/TSpec T dwarf spectra. 
\label{indexwaves}}
\end{figure}

\clearpage

\begin{figure}
\epsscale{0.9}
\figurenum{21}
\centering
\includegraphics[width=5in]{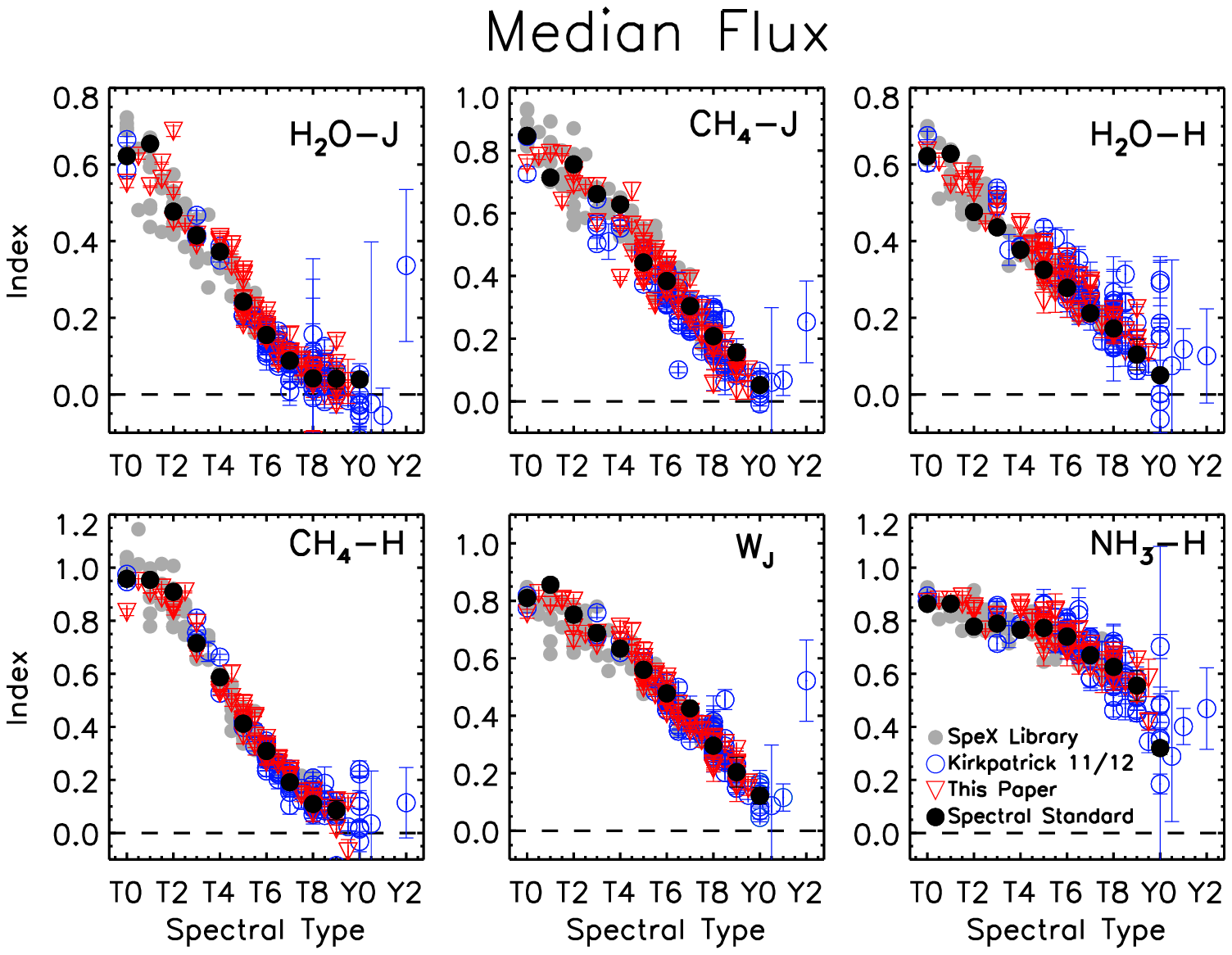}
\includegraphics[width=5in]{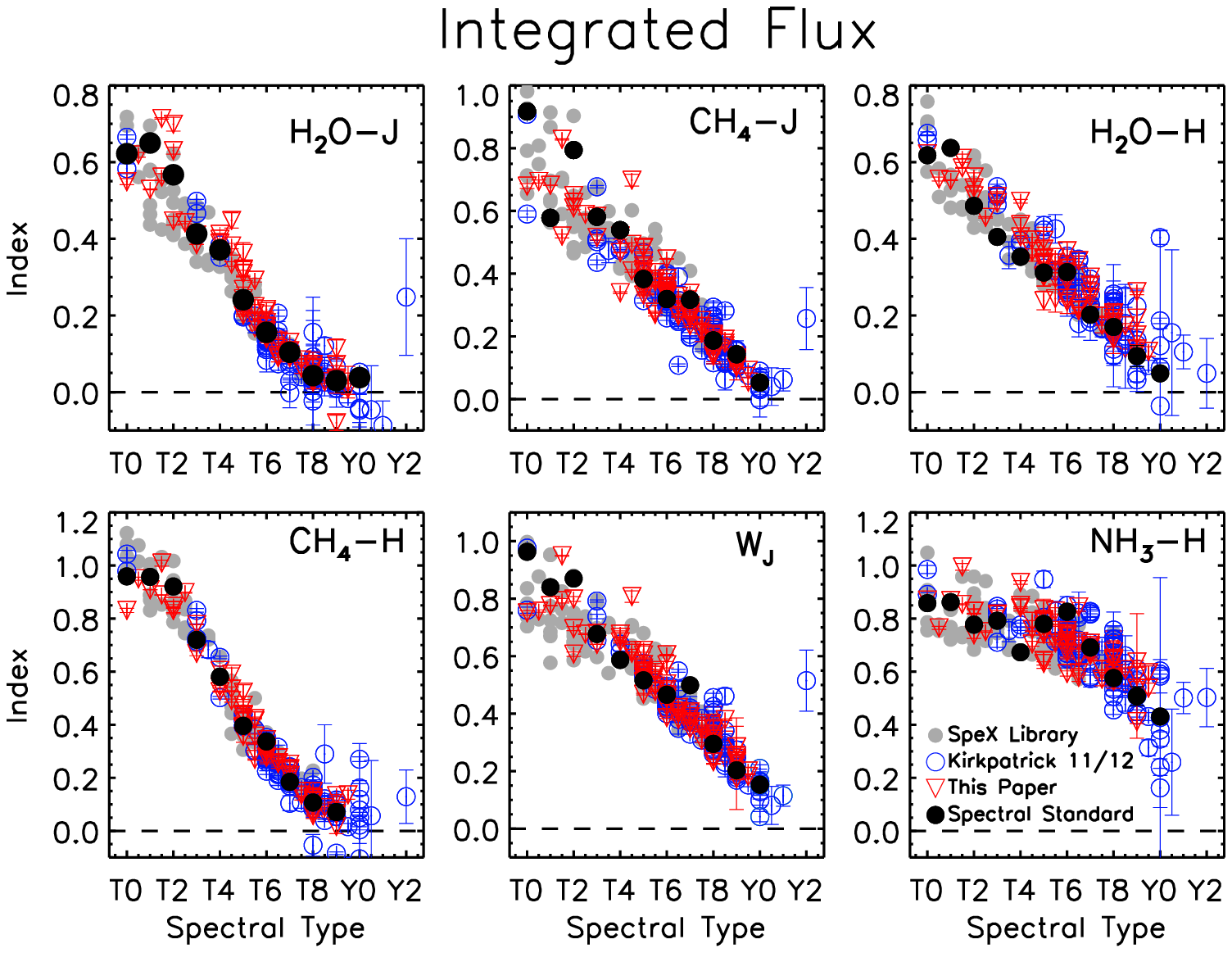}
\caption{Six spectral indices from \citet{burgasser2006}, \citet{warren2007} and \citet{delorme2008}, computed using the median flux and the integrated flux. The T dwarfs in this paper are marked by red triangles, and the T and Y dwarfs of 
\citet{kirkpatrick2011,kirkpatrick2012} and \citet{cushing2011} are marked as blue hollow circles. Also included are the T dwarfs from the SpeX Prism Library as gray filled circles, and spectral standards as black filled circles.
\label{Six_Indices_Ref}}
\end{figure}

\clearpage

\begin{figure}
\epsscale{0.9}
\figurenum{22}
\plotone{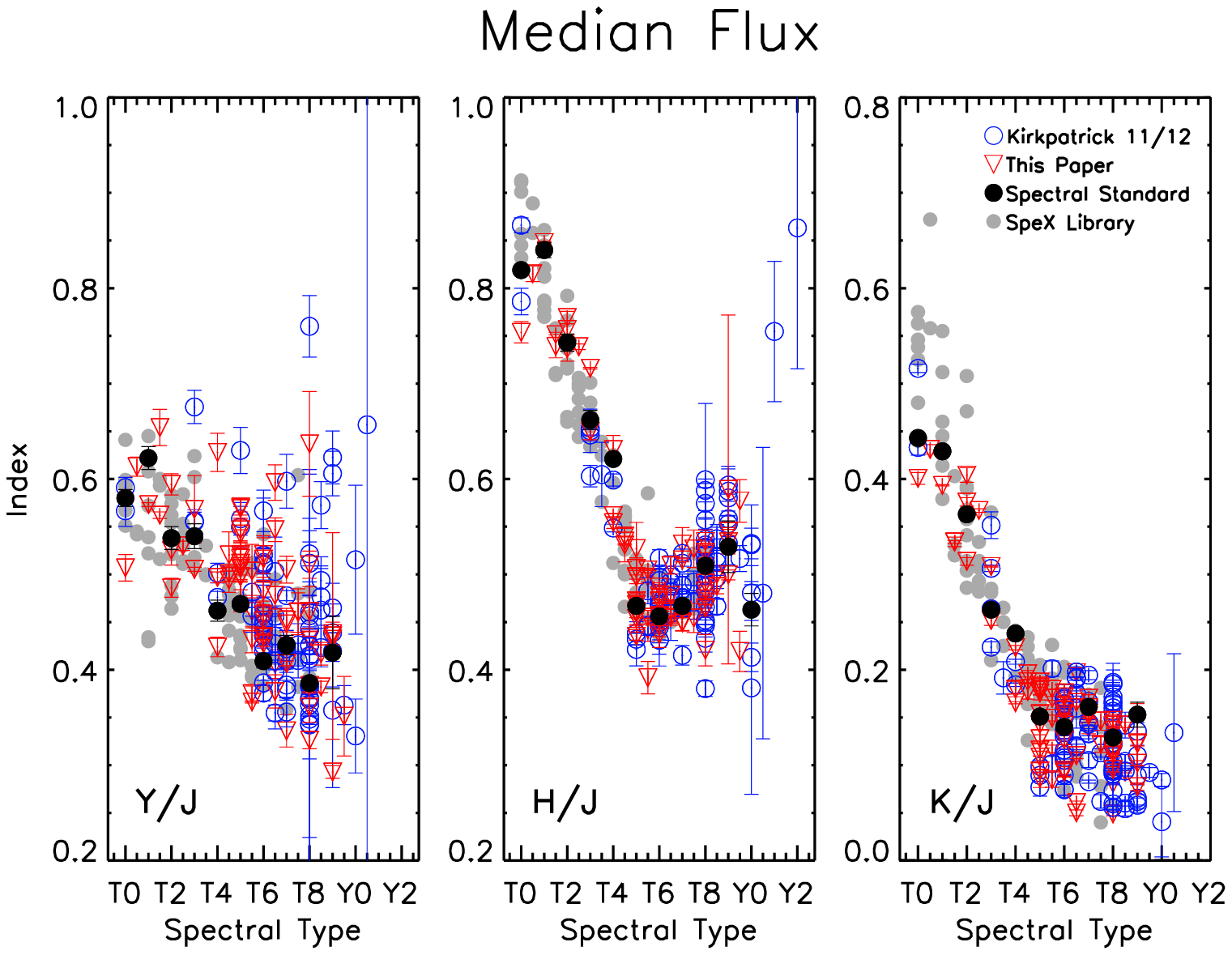}
\caption{The peak flux ratios defined by \citet{burgasser2004b,burgasser2006}, computed using the median flux instead of the integrated flux.  Same symbols as in Figure~\ref{Six_Indices_Ref}.
\label{Indices_Bands_Ref_Med}}
\end{figure}

\clearpage

\begin{figure}
\epsscale{0.9}
\figurenum{23}
\plotone{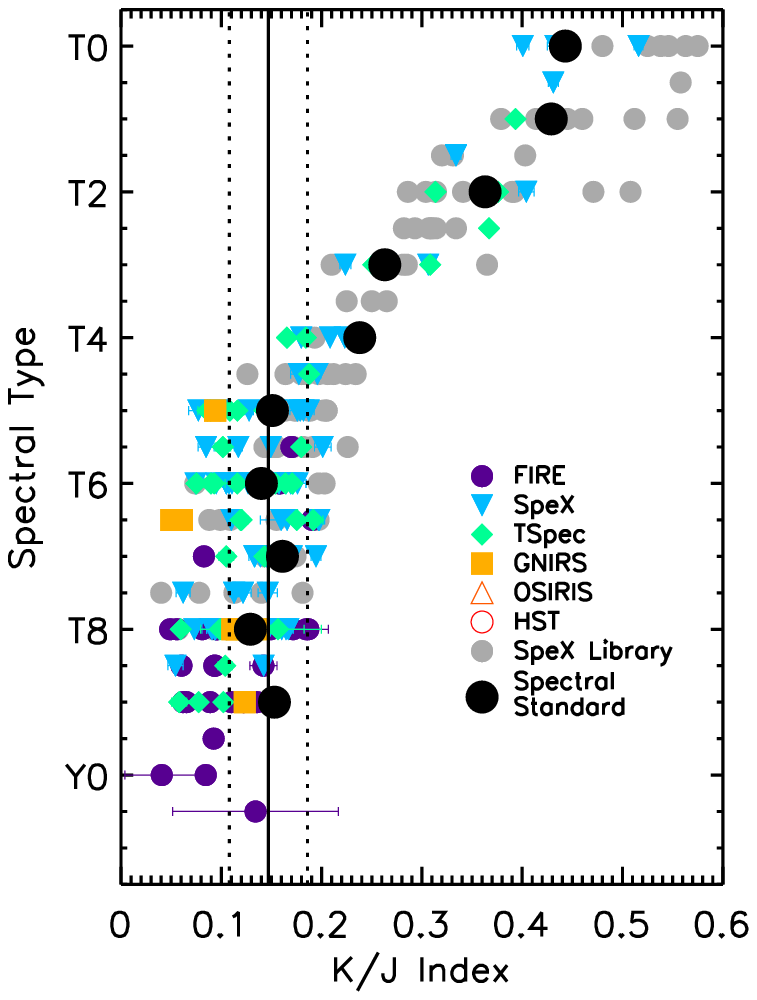}
\caption{Spectral type versus K/J ratio for the same objects in Figure~\ref{Indices_Bands_Ref_Med}, sorted by spectrograph. The average K/J index for the T5-T9 spectral standards is 0.147$\pm$0.013, and the 3$\sigma$ limits are shown here. We find that 54 out of 170 (about one third) of T5 or later dwarfs are below this limit, while only 18 out of 170 are above this limit.
\label{SpTKJ}}
\end{figure}

\clearpage

\begin{figure}
\epsscale{0.9}
\figurenum{24}
\plotone{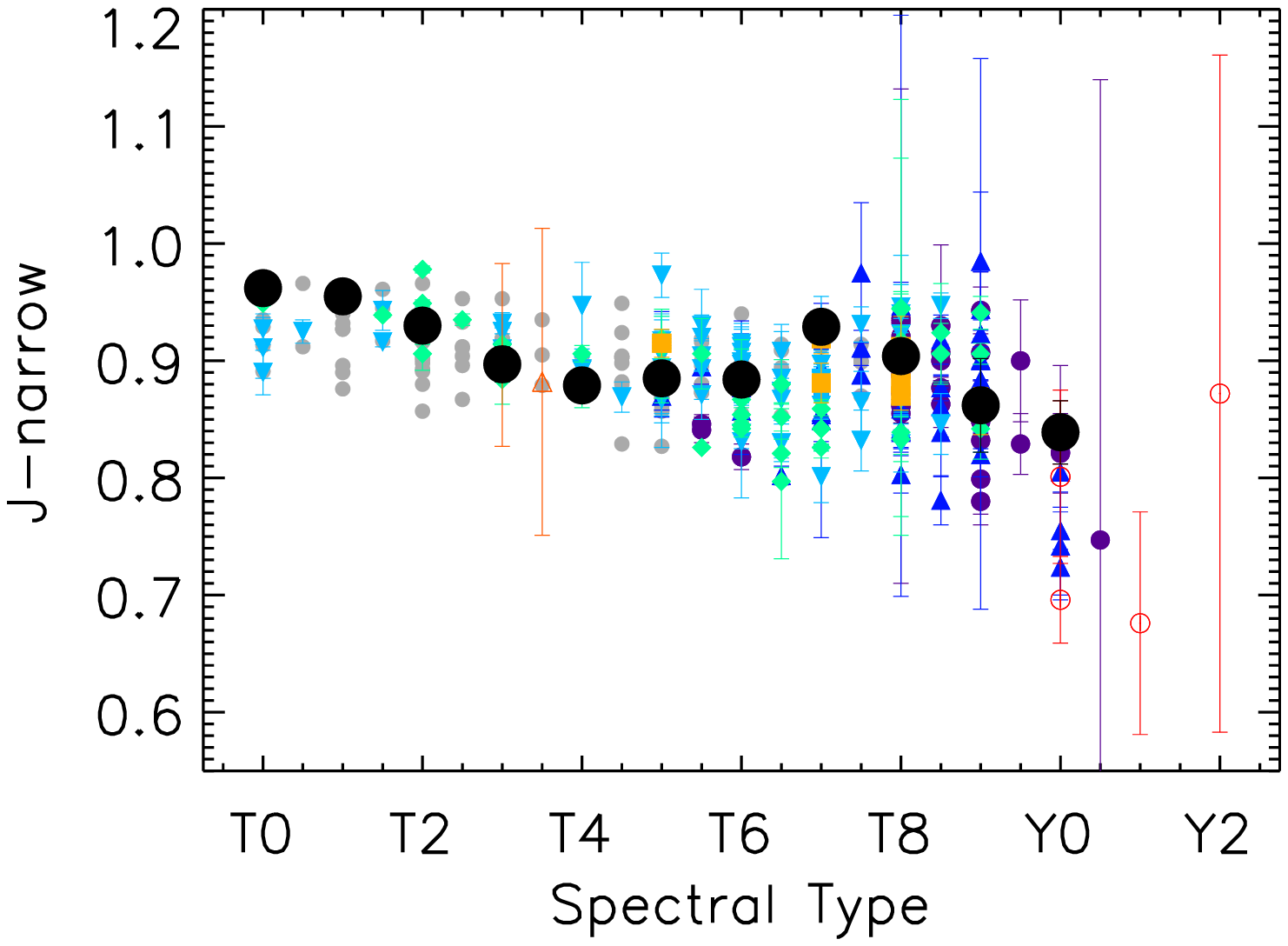}
\caption{The J-narrow index for T and Y dwarfs in Figure~\ref{Indices_Bands_Ref_Med}. Undulation in the T dwarfs is due to changes in the 1.254$\mu$m KI line. Y dwarfs have smaller indices because of a narrowing of the $J$-band flux peak.
\label{Jnarrow}}
\end{figure}

\clearpage

\begin{figure}
\epsscale{0.9}
\figurenum{25}
\plotone{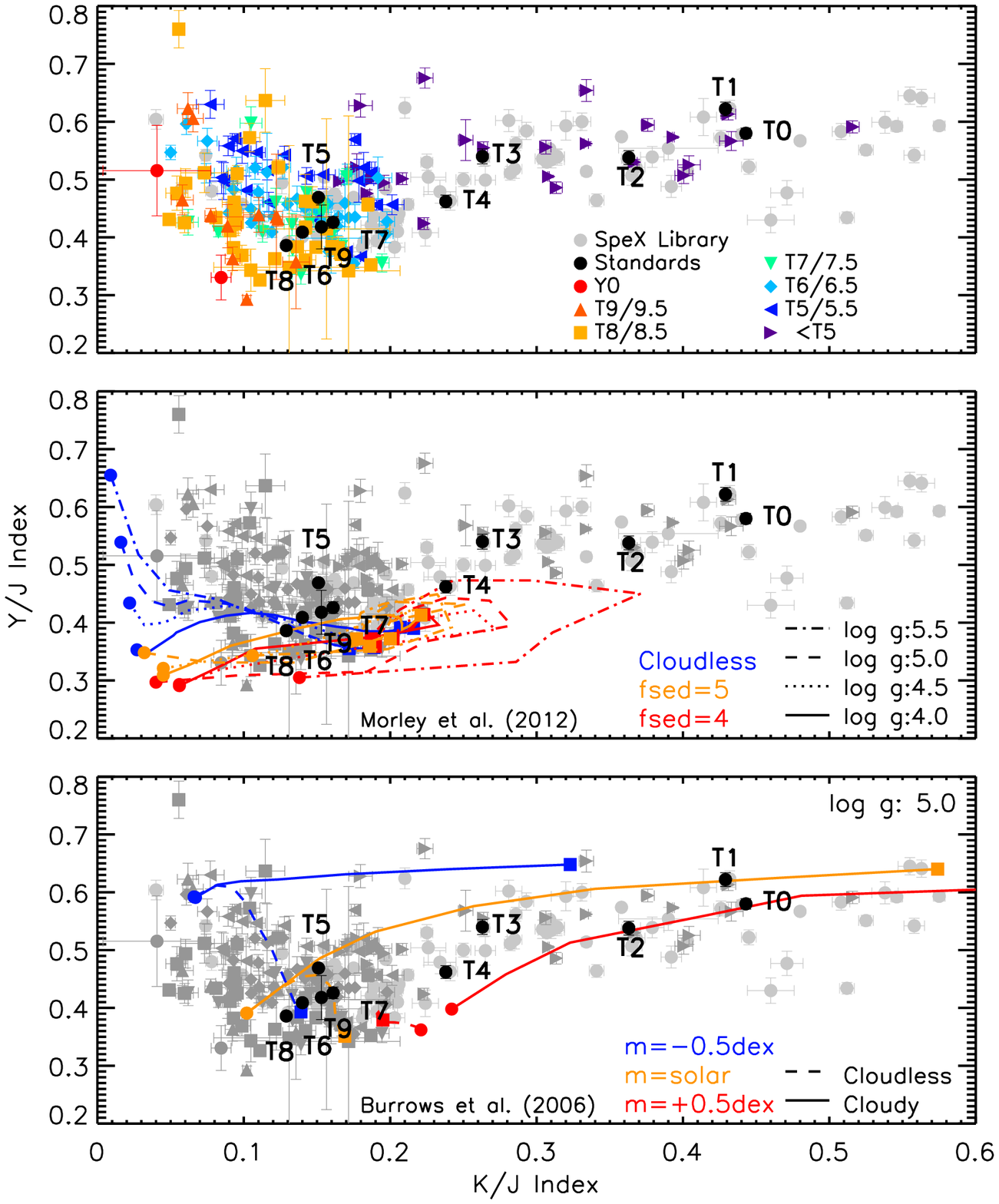}
\caption{The Y/J versus K/J indices of late-type T dwarfs compared to models. Top: Index values sorted by spectral type. Middle: The \citet{morley2012} sulfide cloud models (which do not include silicate clouds expected in early-type T dwarfs) are shown for log g =4, 4.5, 5, 5.5 (solid,dotted,dashed,dot-dash) and T$_{eff}$=400K to 1300K. Cloudless models are shown in blue, $f_{sed}$=5 in orange, and $f_{sed}$=4 in red. Bottom: The cloudy (solid lines) and clear (dashed lines) models from \citet{burrows2006} with log g =5 and metallicities of solar and $\pm$0.5dex with T$_{eff}$= 800K to 1700K. The highest temperature is marked with a square and the lowest temperature is marked with a circle.
\label{model_indices}}
\end{figure}

\clearpage

\begin{figure}
\epsscale{0.9}
\figurenum{26}
\plotone{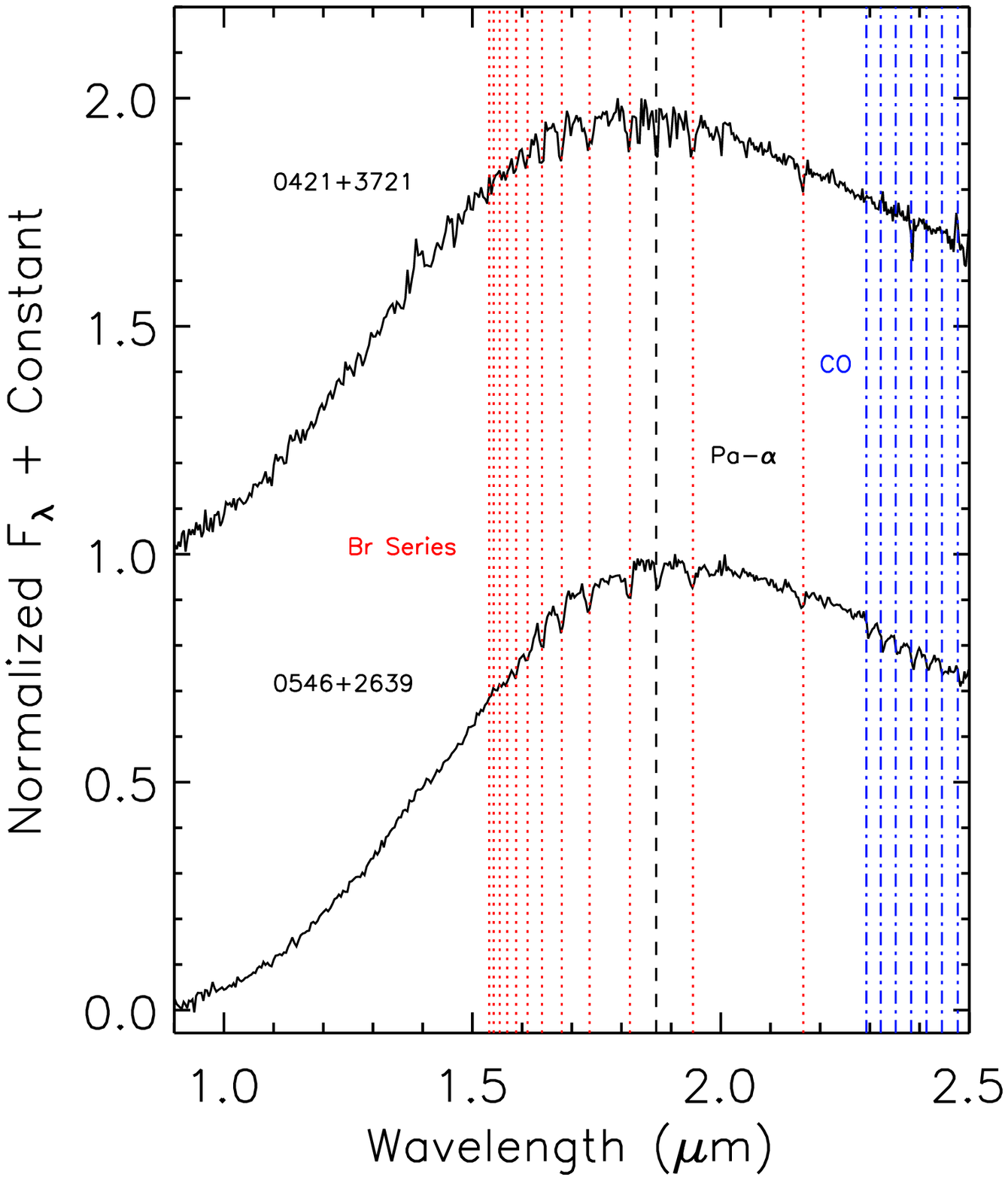}
\caption{IRTF/SpeX spectra of two YSOs found in the same WISE color space as T dwarfs. Both sources lack clear emission features and display photospheric H(dotted and dashed lines) and CO(dash-dotted lines) absorption.
\label{YSOs}}
\end{figure}

\clearpage

\begin{figure}
\epsscale{0.9}
\figurenum{27}
\plotone{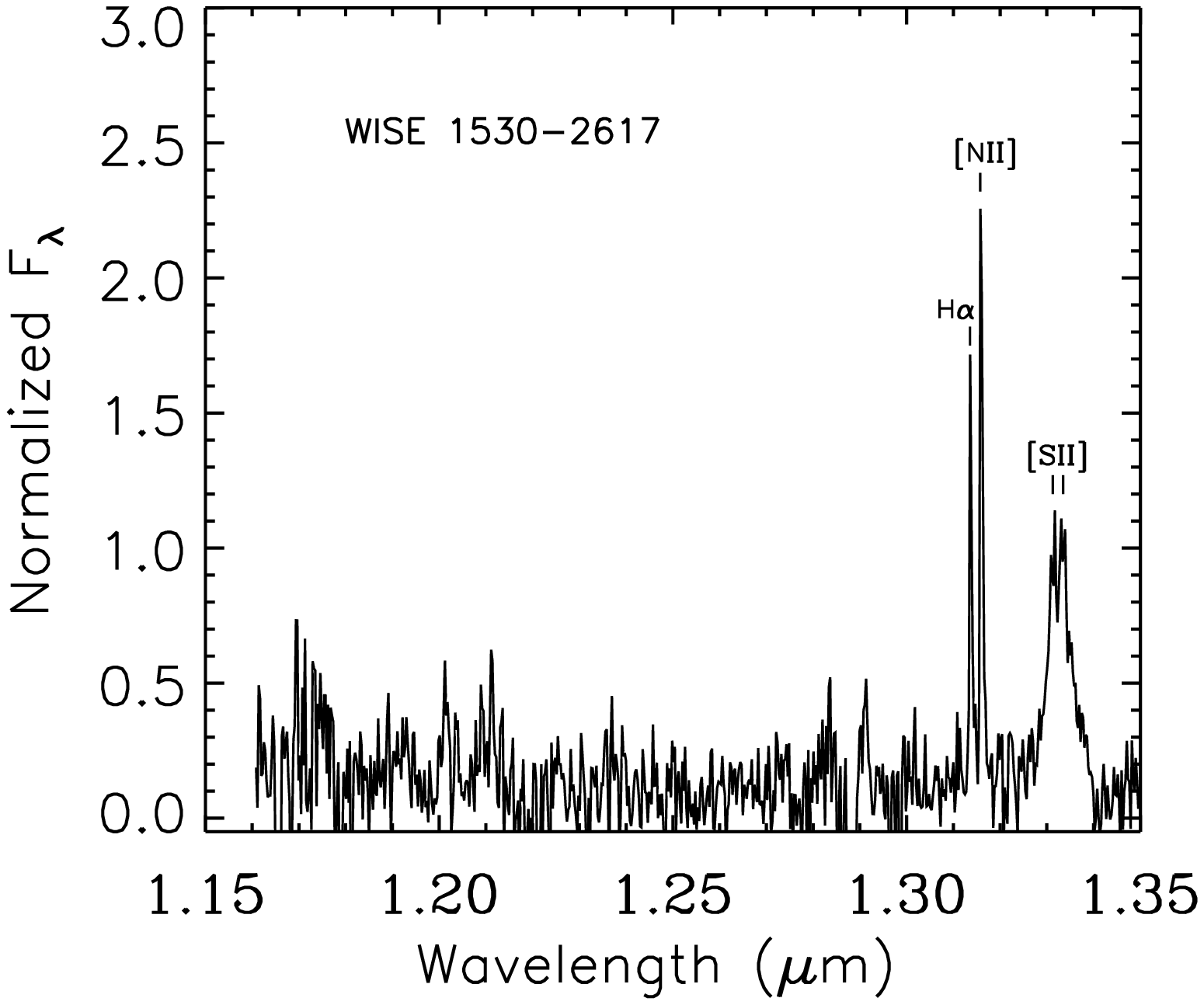}
\caption{Keck/NIRSPEC spectrum of WISE J153015.18$-$261753.6, an AGN that is on the reddest edges of WISE brown dwarf color space. The H$\alpha$, [NII]$\lambda$6583, and broad double-peaked [SII]$\lambda$$\lambda$6716, 6731 lines in the $J$ band place this object at z=1.00.
\label{WISE1530_NIRSPEC}}
\end{figure}

\clearpage

\begin{figure}
\epsscale{0.9}
\figurenum{28}
\plotone{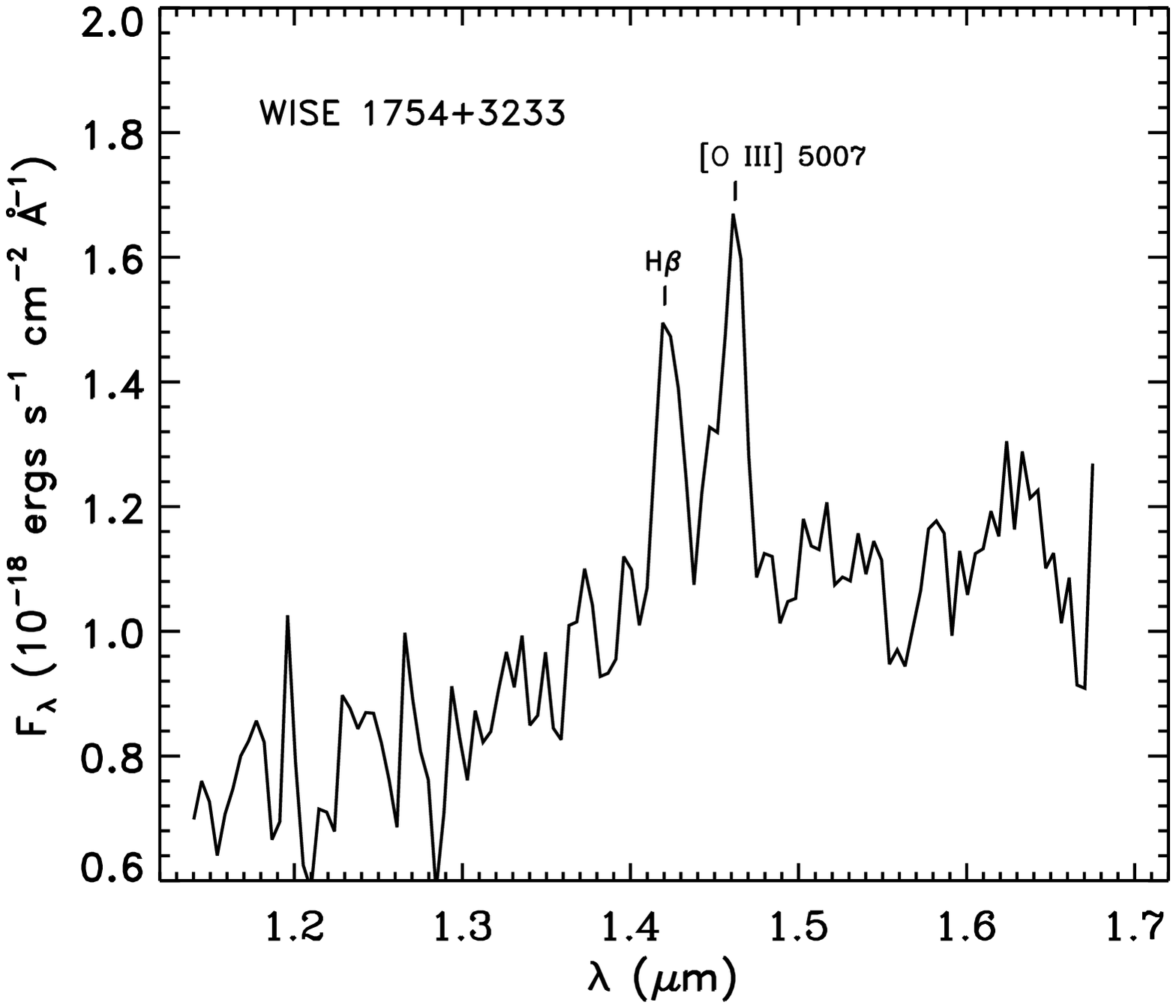}
\caption{HST/WFC3 spectrum of WISE J175443.93+323332.5, an AGN interloper. The H$\beta$ and [O III]$\lambda$$\lambda$4959, 5007 lines in the $H$ band place this object at z=1.92.
\label{WISE1754_HST}}
\end{figure}

\clearpage

\begin{figure}
\epsscale{0.9}
\figurenum{A1}
\centering
\includegraphics[width=5in]{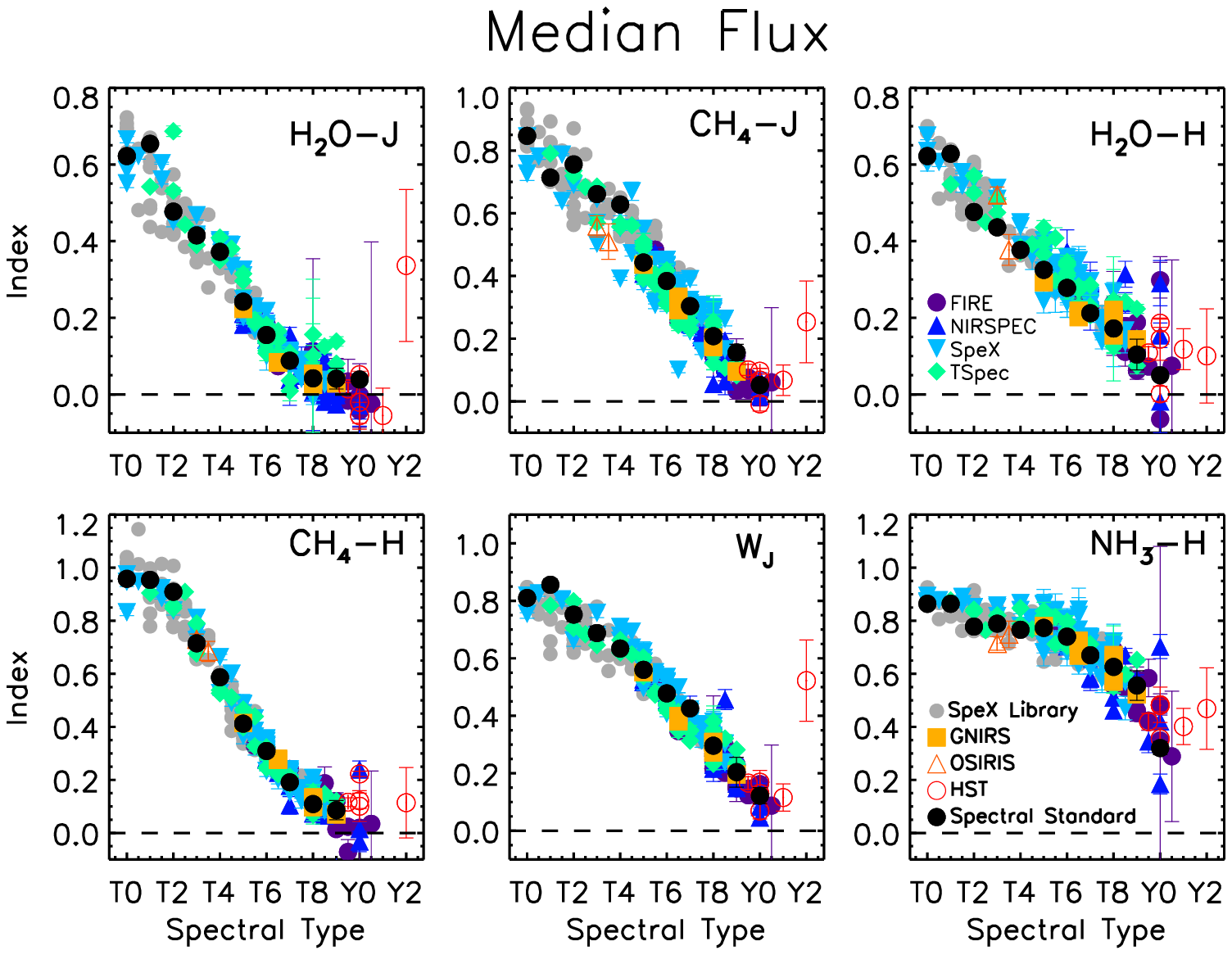}
\includegraphics[width=5in]{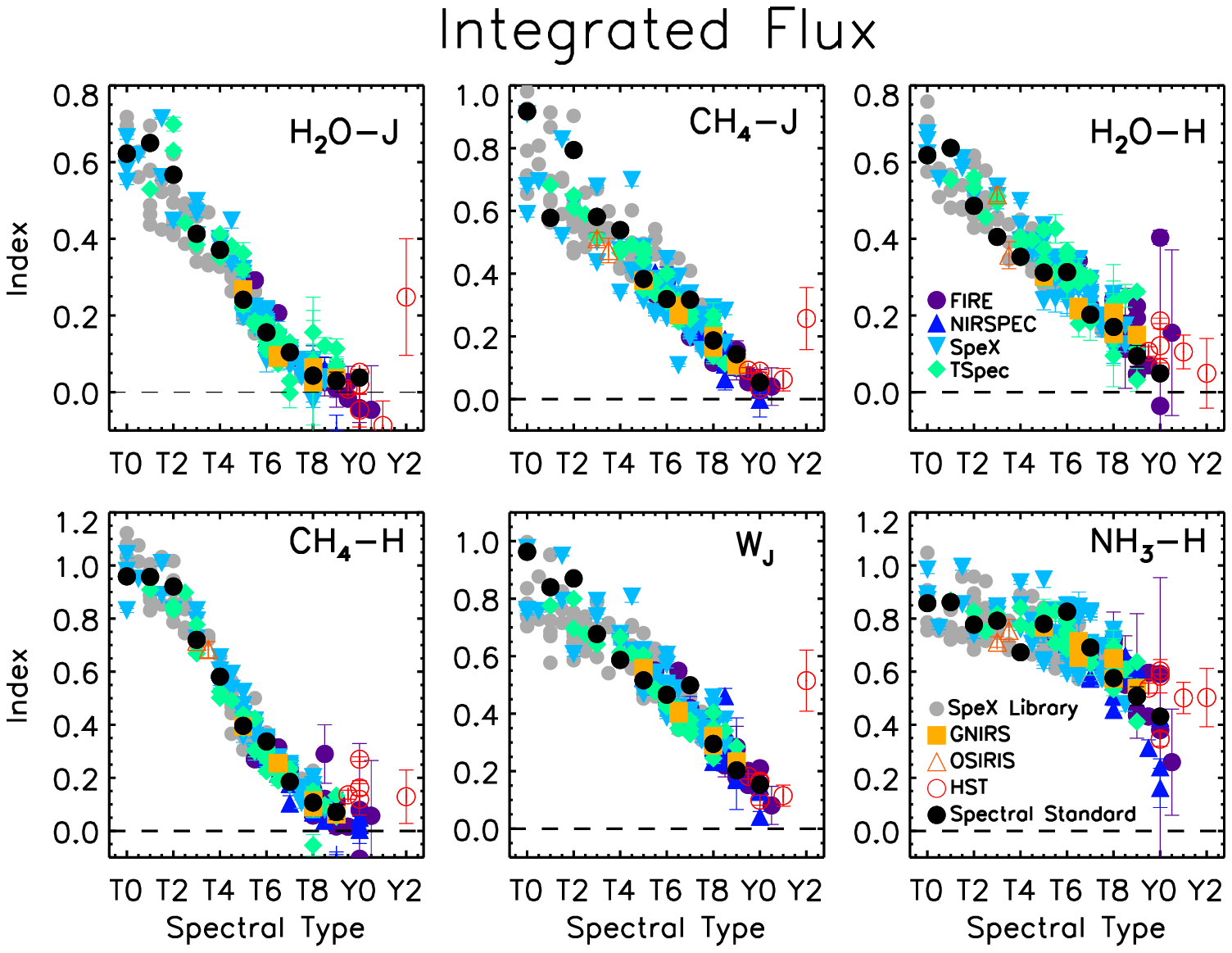}
\caption{The six spectral indices from \citet{burgasser2006}, \citet{warren2007} and \citet{delorme2008}, computed using the median flux and integrated flux for the T dwarfs in this paper and the T and Y dwarfs of \citet{kirkpatrick2011,kirkpatrick2012} and \citet{cushing2011}. Objects are marked by instrument with the symbols and colors in the legend. Also included are the spectral standards as black filled circles and T dwarfs from the SpeX Prism Library as gray circles.  
\label{Six_Indices_Inst_A}}
\end{figure}

\clearpage

\begin{figure}
\epsscale{0.9}
\figurenum{A2}
\centering
\includegraphics[width=5in]{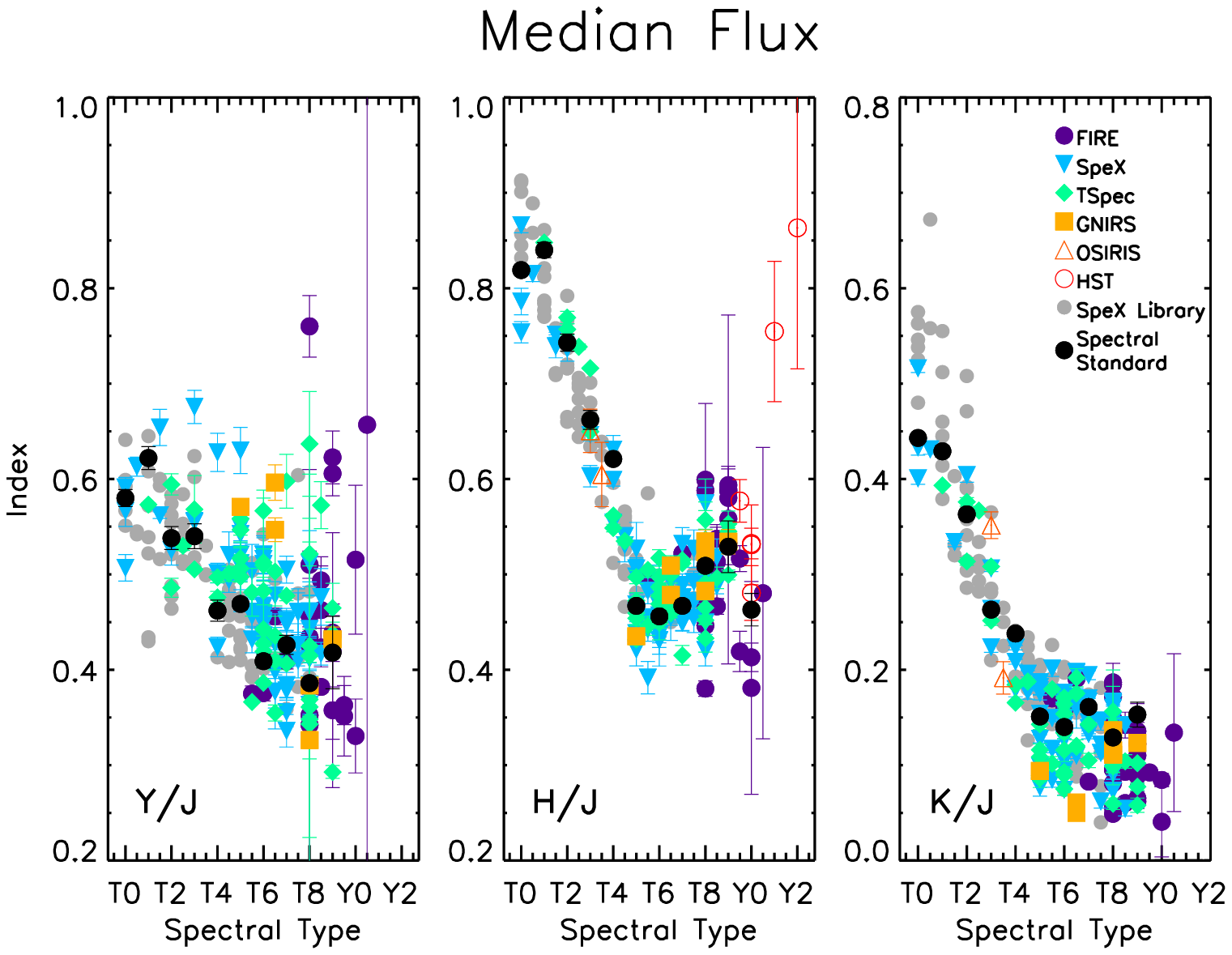}
\includegraphics[width=5in]{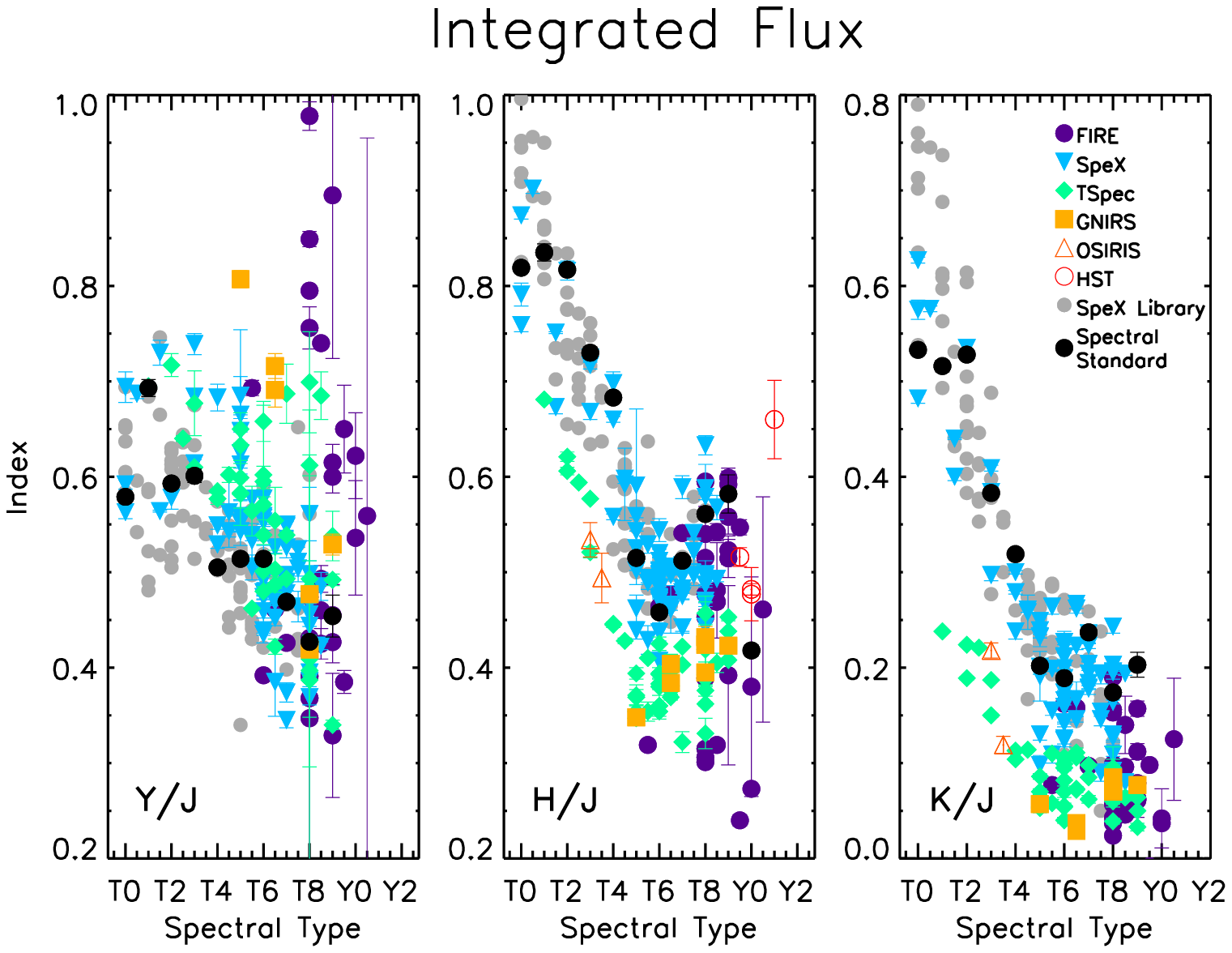}
\caption{The peak flux ratios defined by \citet{burgasser2004b,burgasser2006}, computed using the median flux and the integrated flux.  Same symbols as in Figure~\ref{Six_Indices_Inst_A}.
\label{Indices_Bands_Inst_A}}
\end{figure}

\clearpage

\end{document}